\documentclass[11pt, a4paper]{article}
\usepackage{jheppub}

\title{Component reduction in $\cN=2$ supergravity:\\
the vector, tensor, and vector-tensor multiplets}
\author{Daniel Butter and}
\author{Joseph Novak}
\affiliation{School of Physics M013, The University of Western Australia \\
35 Stirling Highway, Crawley, W.A. 6009, Australia}
\emailAdd{daniel.butter@uwa.edu.au, joseph.novak@uwa.edu.au}

\abstract{
Recent advances in curved $\cN=2$ superspace methods have rendered the
component reduction of superspace actions more feasible than
in the past. In this paper, we consider models involving both vector
and tensor multiplets coupled to supergravity and demonstrate explicitly
how component actions may be efficiently obtained.
In addition, tensor multiplets coupled to conformal supergravity
are considered directly within projective superspace, where their
formulation is most natural.
We then demonstrate how the inverse procedure -- the lifting of
component results to superspace -- can simplify the analysis of
complicated multiplets. We address the off-shell $\cN=2$ vector-tensor
multiplet coupled to conformal supergravity with a central charge
and demonstrate explicitly how its constraints and Lagrangian
can be written in a simpler way using superfields.
}

\usepackage{latexsym}
\usepackage{amssymb,amsfonts,amsmath}
\usepackage{graphicx} 
\usepackage{indentfirst}
\usepackage{tensor}
\usepackage{bbm}
\usepackage{slashed, tensor}

\usepackage[mathscr]{euscript} % alternate script style

\newcommand {\cA}{{\cal A}}
\newcommand {\cB}{{\cal B}}
\newcommand {\cC}{{\cal C}}
\newcommand {\cD}{{\cal D}}
\newcommand {\cE}{{\cal E}}
\newcommand {\cF}{{\cal F}}
\newcommand {\cG}{{\cal G}}

\newcommand {\cL}{{\cal L}}
\newcommand {\cM}{{\cal M}}
\newcommand {\cN}{{\cal N}}
\newcommand {\cO}{{\cal O}}

\newcommand {\cQ}{{\cal Q}}
\newcommand {\cR}{{\cal R}}

\newcommand {\cT}{{\cal T}}

\newcommand {\cV}{{\cal V}}
\newcommand {\cW}{{\cal W}}

\newcommand {\cZ}{{\cal Z}}
\def\a{\alpha}
\def\b{\beta}
\def\c{\chi}
\def\d{\delta}

\def\g{\gamma}
\def\G{\Gamma}

\def\k{\kappa}
\def\l{\lambda}

\def\q{\theta}
\def\r{\rho}
\def\s{\sigma}

\def\x{\xi}

\def\D{\Delta}

\def\S{\Sigma}
\def\U{\Upsilon}

\def\ri{{\rm i}}

\newcommand{\gd}{{\dot\g}}
\newcommand{\dd}{{\dot\d}}
\newcommand{\ad}{{\dot{\alpha}}}
\newcommand{\bd}{{\dot{\beta}}}
\newcommand{\dalpha}{{\dot{\alpha}}}
\newcommand{\dbeta}{{\dot{\beta}}}
\newcommand{\dgamma}{{\dot{\gamma}}}
\newcommand{\ddelta}{{\dot{\delta}}}

\newcommand{\dsC}{{\mathbb C}}

\newcommand{\1}{{\underline{1}}}
\newcommand{\2}{{\underline{2}}}

\newcommand{\ve}{\varepsilon}

\newcommand{\pa}{\partial}
\newcommand{\hf}{\frac12}

\newcommand{\psib}{\bar{\psi}}

\newcommand{\be}{\begin{equation}}
\newcommand{\ee}{\end{equation}}
\newcommand{\bea}{\begin{eqnarray}}
\newcommand{\eea}{\end{eqnarray}}
\newcommand{\non}{\nonumber}
\newcommand{\ba}{\begin{array}}
\newcommand{\ea}{\end{array}}

\newcommand{\bm}[1]{\mbox{\boldmath$#1$}}

\def\double #1{#1{\hbox{\kern-2pt $#1$}}}

\newcommand{\ts}{{\tilde{\s}}}

\newcommand{\bsubeq}{\begin{subequations}}
\newcommand{\esubeq}{\end{subequations}}

\newcommand{\N}{{\mathcal N}}
\newcommand{\rd}{\mathrm d}
\newcommand{\HC}{{\mathrm{c.c.}}}
\newcommand{\veps}{\varepsilon}

\newcommand{\bphi}{{\bar\phi}}
\newcommand{\bpsi}{{\bar\psi}}

\newcommand{\eps}{{\epsilon}}

\newcommand{\bsigma}{\bar{\sigma}}

\newcommand{\eol}{\notag \\}
\newcommand{\lc}{{\vert}}

\def\tr{{\rm tr}}

\newcommand{\poin}[1]{\hat{#1}}
\renewcommand{\eps}{\ve}

\begin{document}
%%%%%%%%%%%%%%%%
%%%%%%%%%%%%%%%%
\maketitle

\section{Introduction}

Supergravity-matter model building has attracted substantial interest ever since 
the first models for supergravity were found over thirty years ago.
Much of this interest has been due to its intimate relationship with complex 
geometry. For globally supersymmetric $\cN=1$ theories, the K\"ahler structure of
the target space is most apparent when cast in the language of superfields in
superspace. The same may be said of the Hodge-K\"ahler geometry of locally
supersymmetric theories. The most general matter Lagrangians,
when written in terms of chiral superfields in superspace, naturally lead to the required
geometries with the Lagrangian playing the role of the generating function
for the target space geometry. This is due to the relative simplicity
of $\cN=1$ superspace, its corresponding action principle, and the relative
ease of moving between $\cN=1$ superspace and component actions.
Both approaches are important. Superspace methods economically
incorporate supersymmetry and provide generating formulations, while component methods
are usually given directly by higher dimensional supergravity
and string computations. Therefore, it is advantageous to be able to move
deftly between them.

However, for the case of $\cN=2$ supersymmetric theories, especially locally supersymmetric
theories, the relationship between superspace and component methods has remained distant.
$\cN=2$ component methods, {\it e.g.} those based on the $\cN=2$ superconformal tensor
calculus \cite{sct_rules, BdeRdeW, sct_structure},
are quite mature and have been used to describe off-shell vector and
tensor multiplets and on-shell hypermultiplets and their couplings to conformal
supergravity. However, they cannot describe off-shell hypermultiplets
without a central charge and do not provide a generating formulation
for the most general couplings. 

In superspace, this last problem was overcome with the development of
harmonic \cite{GIKOS, GIOS} and projective \cite{KLR, GHR, LR:Proj, LR:SYM}
superspace, in which general supersymmetric models involving
off-shell hypermultiplets could be constructed.
These models naturally lead to the hyperk\"ahler target space geometries required
by rigid $\cN=2$ supersymmetry upon elimination of the infinite numbers of auxiliary
fields. However, coupling such theories to supergravity remained problematic
for some time, at least for projective superspace.\footnote{Models in harmonic
superspace can be coupled directly to the supergravity prepotentials \cite{GIKOS,GIOS}.
However, the relation between this and the differential geometric structure of
superspace -- that is, the full algebra of covariant derivatives -- remains
unclear.}

A recent advance in superspace has resolved this last problem.
In 2008, a projective superspace version of
general $\cN=2$ supergravity-matter couplings was constructed
by Kuzenko, Lindstr\"om, Ro\v cek, and Tartaglino-Mazzucchelli
\cite{KLRT-M08}.\footnote{The construction was
inspired by Kuzenko and Tartaglino-Mazzucchelli's solution to five dimensional
supergravity and its matter couplings, which also made use of projective superspace
\cite{KT-M:5DSugra, KT-M:5DSugraProj, KT-M:5DSuperWeyl}.}
This new formulation allows the most general couplings of
off-shell $\cN=2$ matter to conformal supergravity and
(at least in principle) provides a generating formalism to
produce the most general on-shell supergravity-matter component
actions.

This approach, like all conventional superspace methods,
was based on a superspace geometry where superconformal transformations
were not part of the structure group; rather, they were manifested
as nonlinear super-Weyl transformations. For this reason, matching
the component structure of these general actions to prior work
involving the superconformal tensor calculus remained nontrivial.
Recently, this issue has also been resolved by the construction
of a curved $\cN = 2$ superspace, which has been called conformal superspace \cite{Butter4D},
which possesses the full superconformal algebra in its structure group and reproduces
exactly in its component structure the $\cN=2$ superconformal tensor
calculus. Though the structure group is more complicated, the algebra
of covariant derivatives is markedly simpler than more conventional superspace
approaches. Since conformal superspace can be ``degauged'' to the conventional
superspaces, it provides a useful bridge between existing superspace
constructions and the superconformal tensor calculus.

The goal of this paper is to make use of both these recent advances
in curved $\cN=2$ superspace techniques to bridge some of the gap which remains
between superspace and component methods.
The layout of the paper is as follows.
In section \ref{SuperspaceGeometry}, we briefly review the superspace geometry for
conformal supergravity as well as for the one-form and two-form potentials corresponding 
to the vector and tensor multiplets. Section \ref{ComponentActions} is devoted to the
component reduction of some simple superspace actions. Section \ref{ProjectiveComponents}
presents a component reduction in the projective superspace formulation
of conformal supergravity, which is a novel calculation. We choose a relatively
simple action for which we can easily verify the result.

In section \ref{VectorTensorSection}, we reverse the process of component reduction and
demonstrate how certain very complicated multiplets can simplify dramatically
when recast in a superfield setting. The multiplet which we address is
the off-shell vector-tensor multiplet \cite{SSW:VT} coupled to supergravity with
a central charge, which received a great deal
of interest in the late 1990s due to its appearance in $\cN = 2$ supersymmetric
vacua of heterotic string theory \cite{deWKLL}. Although several off-shell formulations of
this multiplet have been constructed in flat superspace before
\cite{DKT, DK, IS, DIKST, DT, HOW, GHH, BHO},
the intricate off-shell formulation of Claus \textit{et al.} \cite{Claus1, Claus2, Claus3},
which is coupled to conformal supergravity with a central charge, has to our knowledge
only been fully understood at the component level. We will show how the use of superfields can
simplify a number of its features and demonstrate that it is 
equivalent to the superspace formulations of the vector-tensor
multiplet coupled to supergravity recently considered in \cite{KN} as
a curved generalization of previous results.\footnote{See also
the alternative formulations of \cite{Theis1, Theis2} and
\cite{ADS, ADST}.}

In addition, we include several technical appendices. Appendix \ref{NC} summarizes
the notations and conventions which we will be using throughout this paper.
Appendix \ref{TensorBI} gives the technical details of the solution to the
Bianchi identities for the $\cN=2$ tensor multiplet in conformal superspace.
Appendix \ref{CompIdentities} gives a number of useful identities and
relations for the superspace geometries which we will be using.

%%%%%%%%%%%%%%%%%%%%%%%%%%%%%%%%%%%%%%%%%%%%%%%%%%%%%%
%%%%%%%%%%%%%%%%%%%%%%%%%%%%%%%%%%%%%%%%%%%%%%%%%%%%%%

\section{Superspace geometry and conformal supergravity}\label{SuperspaceGeometry}

The Weyl multiplet of 4D $\cN=2$ conformal supergravity
\cite{sct_rules,  BdeRdeW, sct_structure}
can be realized in a number of different ways in superspace. The comprehensive
analysis due to Howe \cite{Howe} addressed the general case of $\cN$-extended
superspace (with $\cN \leq 4$) and chose the structure group $\rm SL(2, \mathbb C) \times U(\cN)_R$.
The geometry uncovered there may be understood (at least for $\cN \leq 2$) at the component
level as conformal supergravity coupled to a real scalar supermultiplet of non-vanishing Weyl weight.
This scalar supermultiplet is a compensator, and a change in the choice
of compensator field manifests as a nonlinear super-Weyl transformation in superspace.
When coupled to a superconformal action, the compensator becomes a pure gauge degree
of freedom.
A simpler version of $\cN=2$ conformal supergravity in superspace was constructed
in \cite{KLRT-M08} using the superspace geometry derived earlier by Grimm \cite{Grimm}
using the structure group $\rm SL(2, \mathbb C) \times SU(2)_R$.
This $\rm SU(2)$ superspace may be understood as a gauge-fixed version of $\rm U(2)$ superspace
({\it i.e.} Howe's geometry for the case $\cN=2$) \cite{KLRT-M09}.
Recently, $\rm SU(2)$ superspace has proven useful in the construction of a projective
superspace formulation of conformal supergravity \cite{KLRT-M08}.\footnote{This approach was later
extended to $\rm U(2)$ superspace \cite{KLRT-M09}.}

However, the superspace geometry which we will utilize most in this paper is neither
$\rm U(2)$ nor $\rm SU(2)$ superspace, but a more general one which
gauges not just $\rm SL(2, \mathbb C) \times U(2)_R$ but the entire
superconformal algebra $\rm SU(2,2|2)$. This formulation, known as
$\cN=2$ conformal superspace \cite{Butter4D} has the advantage that it reduces
\emph{precisely} to conformal supergravity in components, without any
additional compensating multiplet. A drawback (or feature) is that
only superconformal actions can be described with it; but
if the covariant derivatives are ``degauged'', $\rm U(2)$ superspace
results.

In this section we review $\cN = 2$ conformal superspace \cite{Butter4D},
adapted to the notation used in \cite{KLRT-M08}.
We give a brief discussion of both the one-form (vector) and
two-form (tensor) superspace geometries. Then we end with a discussion of the formulation
of $\cN=2$ superspace due to Grimm (elaborated upon in \cite{KLRT-M08}),
explaining how it may be recovered from the formulation of \cite{Butter4D} using
the method of compensators.

%%%%%%%%%%%%%%%%%%%%%%%%%%%%%%%%%%%%%%%%%%%%%%%%%%%%%%

\subsection{Conformal superspace}
We begin with a curved 4D $\cN = 2$ superspace $\cM^{4|8}$ parametrized by
local bosonic $(x)$ and fermionic $(\q, \bar{\q})$ coordinates 
$z^M = (x^m, \ \q^\mu_\imath, \ \bar{\q}_{\dot{\mu}}^\imath)$, where
$m = 0, 1, \cdots, 3,$ $\mu = 1, 2$, $\dot{\mu} = 1, 2$ and $\imath = \1, \2$. The
Grassmann variables $\q^\mu_\imath$ and $\bar{\q}_{\dot{\mu}}^\imath$ are related
to each other by complex conjugation: $\overline{\q^\mu_\imath} = \bar{\q}^{\dot{\mu} \imath}$.
The structure group is chosen to be $\rm SU(2, 2|2)$ and the covariant derivatives
$\nabla_A = (\nabla_a, \nabla_\a^i , \bar{\nabla}^\ad_i)$ have the form
\begin{align} \nabla_A &= E_A + \hf \Omega_A{}^{ab} M_{ab} + \Phi_A{}^{ij} J_{ij} + \ri \Phi_A Y 
+ B_A \mathbb{D} + \frak{F}_{A}{}^B K_B \non\\
&= E_A + \Omega_A{}^{\b\g} M_{\b\g} + \bar{\Omega}_A{}^{\bd\gd} \bar{M}_{\bd\gd} + \Phi_A{}^{ij} J_{ij} + \ri \Phi_A Y 
+ B_A \mathbb{D} + \frak{F}_{A}{}^B K_B \ .
\end{align}
Here $E_A = E_A{}^M(z) \partial_M$ is the supervielbein, with $\partial_M = \partial/\partial z^M$,
$J_{kl} = J_{lk}$ are generators of the group $\rm SU(2)_R$,
$M_{ab}$ are the Lorentz generators, $Y$ is the generator of the chiral rotation
group $\rm U(1)_R$, and $K^A = (K^a, S^\a_i, \bar{S}_\ad^i)$ are the special
superconformal generators.\footnote{Following common usage, we will refer to
$K^a$ as the special conformal generator and $S^\a_i$ as the $S$-supersymmetry generator.}
The one-forms $\Omega_A{}^{bc}$, $\Phi_A{}^{kl}$, $\Phi_A$, $B_A$ and
$\frak{F}_A{}^B$ are the corresponding connections.
The conventions we use here differ in numerous ways from those used originally
in \cite{Butter4D}. For the most part, they follow the conventions
of \cite{Ideas} and \cite{KLRT-M08, KLRT-M09} and are summarized in appendix \ref{NC}.

The Lorentz generators obey
\begin{align}
[M_{ab}, M_{cd} ] &= 2 \eta_{c [a} M_{b] d} - 2 \eta_{d[a} M_{b] c}~,\quad
[M_{ab}, \nabla_c ] = 2 \eta_{c [a} \nabla_{b]}~, \non \\
[M_{ab}, \nabla_\a^i] &= (\s_{ab})_\a{}^\b \nabla_\b^i ~,\quad
[M_{ab}, \bar\nabla^\ad_i] = (\tilde{\s}_{ab})^\ad{}_\bd \bar\nabla^\bd_i~.
\end{align}
The $\rm SU(2)_R$, $\rm U(1)_R$ and dilatation generators obey
\begin{align}
[J_{ij}, J_{kl}] &= -\ve_{k(i} J_{j)l} - \ve_{l(i} J_{j) k}~, \quad
[J_{ij}, \nabla_\a^k] = - \d^k_{(i} \nabla_{\a j)} ~,\quad
[J_{ij}, \bar\nabla^\ad_k] = - \ve_{k (i} \bar\nabla^{\ad}_{j)}~, \non \\
[Y, \nabla_\a^i] &= \nabla_\a^i ~,\quad [Y, \bar\nabla^\ad_i] = - \bar\nabla^\ad_i~,  \non \\
[\mathbb{D}, \nabla_a] &= \nabla_a ~, \quad
[\mathbb{D}, \nabla_\a^i] = \hf \nabla_\a^i ~, \quad
[\mathbb{D}, \bar\nabla^\ad_i] = \hf \bar\nabla^\ad_i ~.
\end{align}
The special superconformal generators $K^A$ transform in the obvious
way under Lorentz and $\rm SU(2)_R$ rotations,
\begin{align}
[M_{ab}, K_c] &= 2 \eta_{c [a} K_{b]} ~, \quad
[M_{ab} , S^\g_i] = - (\s_{ab})_\b{}^\g S^\b_i ~, \quad
[M_{ab} , \bar S_\gd^i] = - (\ts_{ab})^\bd{}_\gd \bar S_\bd^i~, \non \\
[J_{ij}, S^\g_k] &= - \ve_{k (i} S^\g_{j)} ~, \quad
[J_{ij}, \bar{S}^k_\gd] = - \d^k_{(i} \bar{S}_{\gd j)}~,
\end{align}
while their transformation under $\rm U(1)_R$ and dilatations is opposite
that of $\nabla_A$:
\begin{align}
[Y, S^\a_i] &= - S^\a_i ~, \quad
[Y, \bar{S}^i_\ad] = \bar{S}^i_\ad~, \non \\
[\mathbb{D}, K_a] &= - K_a ~, \quad
[\mathbb{D}, S^\a_i] = - \hf S^\a_i ~, \quad
[\mathbb{D}, \bar{S}_\ad^i] = - \hf \bar{S}_\ad^i ~.
\end{align}
Among themselves, the generators $K^A$ obey the algebra
\begin{align}
\{ S^\a_i , \bar{S}^j_\ad \} &= 2 \ri \d^j_i (\s^a)^\a{}_\ad K_a~.
\end{align}

Finally, the algebra of $K^A$ with $\nabla_B$ is given by
\begin{align}
[K^a, \nabla_b] &= 2 \delta^a_b \mathbb{D} + 2 M^{a}{}_b ~,\non \\
\{ S^\a_i , \nabla_\b^j \} &= 2 \d^j_i \d^\a_\b \mathbb{D} - 4 \d^j_i M^\a{}_\b 
- \d^j_i \d^\a_\b Y + 4 \d^\a_\b J_i{}^j ~,\non \\
\{ \bar{S}^i_\ad , \bar{\nabla}^\bd_j \} &= 2 \d^i_j \d^\bd_\ad \mathbb{D} 
+ 4 \d^i_j \bar{M}_\ad{}^\bd + \d^i_j \d_\ad^\bd Y - 4 \d_\ad^\bd J^i{}_j ~,\non \\
[K^a, \nabla_\b^j] &= -\ri (\s^a)_\b{}^\bd \bar{S}_\bd^j \ , \quad [K^a, \bar{\nabla}^\bd_j] = 
-\ri ({\s}^a)^\bd{}_\b S^\b_j ~, \non \\
[S^\a_i , \nabla_b] &= \ri (\s_b)^\a{}_\bd \bar{\nabla}^\bd_i \ , \quad [\bar{S}^i_\ad , \nabla_b] = 
\ri ({\s}_b)_\ad{}^\b \nabla_\b^i \ ,
\end{align}
where all other (anti-)commutations vanish.

The covariant derivatives obey (anti-)commutation relations of the form
\begin{align} [\nabla_A, \nabla_B\} 
         &= T_{AB}{}^C \nabla_C + \hf R_{AB}{}^{cd} M_{cd} + R_{AB}{}^{kl} J_{kl}
	\eol & \quad
	+ \ri R_{AB}(Y) Y + R_{AB} (\mathbb{D}) \mathbb{D} + R_{AB}{}^C K_C \ , \label{CDA1}
\end{align}
where $T_{AB}{}^C$ is the torsion, and $R_{AB}{}^{cd}$, $R_{AB}{}^{kl}$, $R_{AB}(Y)$, $R_{AB} (\mathbb{D})$ 
and $R_{AB}{}^C$ are the curvatures. Some of the components of the torsion and curvature must be constrained. Following 
\cite{Butter4D}, the spinor derivative torsions and curvatures are chosen to obey
\begin{align}\label{fullConstraints}
\{ \nabla_\a^i , \nabla_\b^j \} = - 2 \ve^{ij} \ve_{\a\b} \bar{\cW}~,\quad
\{ \bar{\nabla}^\ad_i , \bar{\nabla}^\bd_j \} = 2 \ve_{ij} \ve^{\ad\bd} \cW~,\quad
\{ \nabla_\a^i , \bar{\nabla}^\bd_j \} = -2 \ri \d^i_j \nabla_\a{}^\bd ~,
\end{align}
where $\cW$ is some operator valued in the superconformal algebra.
In \cite{Butter4D}, it was shown how to constrain $\cW$ entirely in terms
of a superfield $W_{\a\b}$ so that the component structure reproduces
$\cN=2$ conformal supergravity. In our notation, the constraints lead to
\begin{subequations}\label{CSGAlgebra}
\begin{align}
\{ \nabla_\a^i , \nabla_\b^j \} &= 2 \ve^{ij} \ve_{\a\b} \bar{W}_{\gd\dd} \bar{M}^{\gd\dd} + \hf \ve^{ij} \ve_{\a\b} \bar{\nabla}_{\gd k} \bar{W}^{\gd\dd} \bar{S}^k_\dd - \hf \ve^{ij} \ve_{\a\b} \nabla_{\g\dd} \bar{W}^\dd{}_\gd K^{\g \gd}~, \\
\{ \bar{\nabla}^\ad_i , \bar{\nabla}^\bd_j \} &= - 2 \ve_{ij} \ve^{\ad\bd} W^{\g\d} M_{\g\d} + \frac{1}{2} \ve_{ij} \ve^{\ad\bd} \nabla^{\g k} W_{\g\d} S^\d_k - \frac{1}{2} \ve_{ij} \ve^{\ad\bd} \nabla^{\g\gd} W_{\g}{}^\d K_{\d \gd}~, \\
\{ \nabla_\a^i , \bar{\nabla}^\bd_j \} &= - 2 \ri \d_j^i \nabla_\a{}^\bd~, \\
[\nabla_{\a\ad} , \nabla_\b^i ] &= - \ri \ve_{\a\b} \bar{W}_{\ad\bd} \bar{\nabla}^{\bd i} - \frac{\ri}{2} \ve_{\a\b} \bar{\nabla}^{\bd i} \bar{W}_{\ad\bd} \mathbb{D} - \frac{\ri}{4} \ve_{\a\b} \bar{\nabla}^{\bd i} \bar{W}_{\ad\bd} Y + \ri \ve_{\a\b} \bar{\nabla}^\bd_j \bar{W}_{\ad\bd} J^{ij}
	\eol & \quad
	- \ri \ve_{\a\b} \bar{\nabla}_\bd^i \bar{W}_{\gd\ad} \bar{M}^{\bd \gd} - \frac{\ri}{4} \ve_{\a\b} \bar{\nabla}_\ad^i \bar{\nabla}^\bd_k \bar{W}_{\bd\gd} \bar{S}^{\gd k} + \frac{1}{2} \ve_{\a\b} \nabla^{\g \bd} \bar{W}_{\ad\bd} S^i_\g
	\eol & \quad
	+ \frac{\ri}{4} \ve_{\a\b} \bar{\nabla}_\ad^i \nabla^\g{}_\gd \bar{W}^{\gd \bd} K_{\g \bd}~, \\
[ \nabla_{\a\ad} , \bar{\nabla}^\bd_i ] &=  \ri \d^\bd_\ad W_{\a\b} \nabla^{\b}_i + \frac{\ri}{2} \d^\bd_\ad \nabla^{\b}_i W_{\a\b} \mathbb{D} - \frac{\ri}{4} \d^\bd_\ad \nabla^{\b}_i W_{\a\b} Y + \ri \d^\bd_\ad \nabla^{\b j} W_{\a\b} J_{ij}
	\eol & \quad
	+ \ri \d^\bd_\ad \nabla^{\b}_i W^\g{}_\a M_{\b\g} + \frac{\ri}{4} \d^\bd_\ad \nabla_{\a i} \nabla^{\b j} W_\b{}^\g S_{\g j} - \hf \d^\bd_\ad \nabla^\b{}_\gd W_{\a\b} \bar{S}^{\gd}_i
	\eol & \quad
	+ \frac{\ri}{4} \d^\bd_\ad \nabla_{\a i} \nabla^\g{}_\gd W_{\b\g} K^{\b\gd} ~.
\end{align}
\end{subequations}
The complex superfield $W_{\a\b} = W_{\b\a}$ and its complex conjugate
${\bar{W}}_{\ad \bd} := \overline{W_{\a\b}}$ are superconformally primary,
$K_A W_{\a\b} = 0$, and obey the additional constraints
\begin{align}
\bar{\nabla}^\ad_i W_{\b\g} = 0~,\qquad
\nabla_{\a\b} W^{\a\b} &= \bar{\nabla}^{\ad\bd} \bar{W}_{\ad\bd} ~,
\end{align}
where we introduce the notation
\begin{align}
\nabla_{\a\b} := \nabla_{(\a}^k \nabla_{\b) k} \ , \quad
\bar{\nabla}^{\ad\bd} := \bar\nabla^{(\ad}_k \bar\nabla^{\bd) k} \ .
\end{align}
Despite the appearance of the $S$-supersymmetry and special conformal $K_a$ generators,
the algebra of covariant derivatives \eqref{CSGAlgebra} is significantly
simpler to work with than the corresponding algebras of $\rm SU(2)$ \cite{Grimm, KLRT-M08}
or $\rm U(2)$ superspace \cite{Howe, KLRT-M09}.

%%%%%%%%%%%%%%%%%%%%%%%%%%%%%%%%%%%%%%%%%%%%%%%%%%%%%%

\subsection{One-form geometry of the abelian vector multiplet}\label{VectorMultipletGeometry}

It is possible to introduce a one-form gauge potential, describing the $\cN =2$
vector multiplet, in the formulation of conformal supergravity presented in the
previous section, as a generalization of the flat superspace solution \cite{GSW:ExtendedGauge}.
We shall consider the abelian vector multiplet, since it will
be used extensively in subsequent sections and its generalization to the
non-abelian case is straightforward. 

The field strength two-form $F$ is given in terms of its one-form potential
$V = \rd z^A V_A$ by $F = \rd V$, or equivalently,
\begin{align}
F_{AB} &= 2 \nabla_{[A} V_{B\}} - T_{AB}{}^C V_C ~.
\end{align}
Due to the existence of the one-form potential the field strength must satisfy the Bianchi identity
\be
\rd F = 0 \implies \nabla_{[A} F_{BC\}} - T_{[AB}{}^{D} F_{|D| C\}} = 0 ~.
\ee

At mass dimension-1 we impose the constraints
\begin{align}\label{2formConstraint}
F_\a^i{}_\b^j = - 2 \ve^{ij}\ve_{\a\b} \bar{\cW} \ , \quad F^\ad_i{}^\bd_j = 2 \ve_{ij} \ve^{\ad\bd} \cW \ , \quad F_\a^i{}^\bd_j = 0 \ ,
\end{align}
where $\cW$ is a primary superfield with dimension 1 and $\rm U(1)$ weight $-2$,
\be \label{Wprimary}
K_A \cW = 0 \ , \quad \mathbb{D} \cW = \cW \ , \quad Y \cW = -2 \cW \ .
\ee
Then the Bianchi identities may be solved giving
\begin{subequations}\label{2formCurvs}
\begin{align}
F_a{}_\b^j =& \frac{\ri}{2} (\s_a)_\b{}^\gd \bar{\nabla}_\gd^j \bar{\cW} \ , \qquad 
F_a{}^\bd_j = - \frac{\ri}{2} (\s_a)_\g{}^\bd \nabla^\g_j \cW \ , \\  \label{2formCurvsb}
F_{ab} =& - \frac{1}{8} (\s_{ab})_{\a\b} ( \nabla^{\a \b} \cW + 4 W^{\a\b} \bar{\cW})
+ \frac{1}{8} (\tilde{\s}_{ab})_{\ad\bd}  (\bar{\nabla}^{\ad \bd} \bar{\cW} + 4 \bar{W}^{\ad\bd} \cW) \ .
\end{align}
\end{subequations}
$\cW$ is further required to be a reduced chiral superfield,
\begin{align}\label{BianchiYM}
\bar{\nabla}_{\ad}^i \cW = 0~,\qquad
\nabla^{ij} \cW = \bar{\nabla}^{ij} \bar{\cW} ~.
\end{align}
Here we have introduced the notation
\be \nabla^{ij} := \nabla^{\g (i}  \nabla_\g^{j)} \ , \quad \bar{\nabla}^{ij} := \bar{\nabla}_\gd^{(i}  \bar{\nabla}^{\gd j)} \ .
\ee

As a straightforward extension of the results found in \cite{Butter4D} we can
introduce a gauged central charge using the off-shell vector multiplet,
generalizing the superspace formulation in \cite{KLRT-M08}. First, we introduce
a modified covariant  derivative:
\be \bm \nabla_A := \nabla_A + V_A \D \ ,
\ee
where $V_A(z)$ is the gauge connection and $\D$ is a real central charge. Its (anti-)commutation relations are
\begin{align} [\bm \nabla_A, \bm \nabla_B\} &= T_{AB}{}^C \bm \nabla_C + \hf R_{AB}{}^{cd} M_{cd} + R_{AB}{}^{kl} J_{kl}
	\eol & \quad
	+ \ri R_{AB}(Y) Y + R_{AB} (\mathbb{D}) \mathbb{D} + R_{AB}{}^C K_C + F_{AB} \D~,
\end{align}
where the torsion and curvature remain the same as in \eqref{CDA1}.
The central charge abelian field strength $F_{AB}$ obeys
\eqref{2formConstraint} and \eqref{2formCurvs}, with
$\cZ$ (instead of $\cW$) denoting the corresponding reduced
chiral superfield, which is annihilated by the
central charge.

%%%%%%%%%%%%%%%%%%%%%%%%%%%%%%%%%%%%%%%%%%%%%%%%%%%%%%

\subsection{Two-form geometry of the tensor multiplet}\label{TensorMultipletGeometry}

The $\cN = 2$ tensor multiplet \cite{Wess, de Wit:1979pq} can be described by a two-form gauge potential
in supergravity. Its formulation in $\rm U(2)$ superspace was given in, {\it e.g.},
\cite{Muller:ChiralActions, Muller}.  The extension to conformal superspace is entirely
straightforward and we summarize it briefly here.

The field strength three-form $H$ is given in terms of its two-form gauge
potential $B = \frac{1}{2} E^B E^A B_{AB}$ by
\begin{align}
H = \rd B = \frac{1}{3!} E^C E^B E^A H_{ABC}~,\qquad
H_{ABC} = 3 \nabla_{[A} B_{BC\}} - 3 T_{[AB}{}^{D} B_{|D|C\}}~. \label{defH}
\end{align}
The field strength remains invariant under gauge transformations
$\d B = \rd V$ with $V$ a one-form gauge parameter.
The existence of the gauge potential requires that the Bianchi identity
\be
\rd H = 0 \implies \nabla_{[A} H_{BCD\}} - \frac{3}{2} T_{[AB}{}^{E} H_{|E|CD\}} = 0
\ee
be satisfied. As with the one-form, we must impose constraints.
At mass dimension-$\frac{3}{2}$ they consist of
\be \label{3formConstraint}
H_\a^i{}_\b^j{}_\g^k = H^\ad_i{}^\bd_j{}^\gd_k
= H_\a^i{}_\b^j{}^\gd_k = H^\ad_i{}^\bd_j{}_\g^k = 0 \ .
\ee
The Bianchi identities for $H$ can then be solved (see appendix \ref{TensorBI}).
The solution is
\begin{subequations}\label{3formCurvs}
\begin{align}
H_a{}_\a^i{}_\b^j &= 0 \ , \quad H_a{}^\ad_i{}^\bd_j = 0 \ , \quad 
H_a{}^i_\a{}^\ad_j = 2 (\s_a)_\a{}^\ad \cG^i{}_j \ , \quad \\
H_{ab}{}_\a^i &= \frac{2 \ri}{3} (\s_{ab})_\a{}^\b \nabla_\b^k \cG^i{}_k \ , \quad 
H_{ab}{}^\ad_i = \frac{2 \ri}{3} (\tilde{\s}_{ab})^\ad{}_\bd \bar{\nabla}^\bd_k \cG^k{}_i \ , \\
H_{abc} &= \frac{\ri}{24} \ve_{abcd} (\s^d)^\a{}_\bd [\nabla_\a^i , 
\bar{\nabla}^\bd_j] \cG^j{}_i = \ve_{abcd} \tilde{H}^d \ ,
\end{align}
\end{subequations}
where $\cG^{ij}$ is a real symmetric conformally primary dimension-2 superfield, {\it i.e.}
\begin{align}\label{Gweight}
K_A \cG^{ij} = 0 \ , \quad \mathbb{D} \cG^{ij} = 2 \cG^{ij} \ , \quad Y \cG^{ij} = 0 \ , \quad
(\cG^{ij})^* = \cG_{ij} = \veps_{ik} \veps_{jl} \cG^{kl}~,
\end{align}
obeying the constraint
\begin{align}\label{Gconstraint}
\bar{\nabla}^{(i}_\ad \cG^{jk)} = \nabla^{(i}_\a \cG^{jk)} = 0~.
\end{align}
Such a superfield $\cG^{ij}$ contains the $\cN=2$ tensor multiplet
\cite{BS, SSW}.

It is possible to construct a superfield which automatically obeys the
above constraints \eqref{Gconstraint} by imposing constraints on the
two-form $B_{AB}$ itself \cite{Muller:ChiralActions, Muller}. One chooses
\begin{align}
B_\a^i{}_\b^j = 4 \ri \ve^{ij} \ve_{\a\b} \bar{\Psi} \ , \quad
B^\ad_i{}^\bd_j = 4 \ri \ve_{ij} \ve^{\ad\bd} \Psi \ , \quad
B_\a^i{}^\bd_j = 0 \ ,
\end{align}
where $\Psi$ is a chiral superfield, $\bar{\nabla}^\ad_i \Psi = 0$,
of dimension 1 and $\rm U(1)_R$ weight 2, but otherwise arbitrary.
Such constraints are quite natural since the gauge transformation
$\delta B = \rd V = F$ amounts to
\be\label{gaugeT}
\d \Psi = - \frac{\ri}{2} \cW ~.
\ee
We can then proceed to solve \eqref{defH} for the full two-form $B$ (see appendix \ref{TensorBI}): 
\begin{subequations}
\begin{align}
B_a{}_\a^i &= (\s_a)_\a{}^\ad \bar{\nabla}_\ad^i \bar{\Psi} \ , \ \ B_a{}^\ad_i = (\s_a)_\a{}^\ad \nabla^\a_i \Psi \ , \\
B_{ab} &= - \frac{\ri}{4} (\s_{ab})^{\a\b} (\nabla_{\a\b} \Psi - 4 W_{\a\b} \bar{\Psi}) 
- \frac{\ri}{4} (\tilde{\s}_{ab})_{\ad\bd} (\bar{\nabla}^{\ad\bd} \bar{\Psi} - 4 \bar{W}^{\ad\bd} \Psi) \ .
\end{align}
\end{subequations}
Inserting this solution into the definition of $H$ leads to an expression for
the tensor multiplet in terms of an unconstrained chiral prepotential \cite{HST,GS82,Siegel83},
\begin{align}
\cG^{ij} &= \frac{1}{4} \nabla^{ij} \Psi + \frac{1}{4} \bar{\nabla}^{ij} \bar{\Psi}~.
\end{align}
One can check that $\cG^{ij}$ indeed obeys \eqref{Gweight} and \eqref{Gconstraint}
and is invariant under \eqref{gaugeT}.

%%%%%%%%%%%%%%%%%%%%%%%%%%%%%%%%%%%%%%%%%%%%%%%%%%%%%%

\subsection{$\rm SU(2)$ superspace geometry}\label{GrimmGeometry}

The conformal superspace geometry presented earlier realizes
the superconformal algebra covariantly. Such a formulation proves quite
advantageous in finding exact agreement with results derived from $\cN = 2$
superconformal tensor calculus. However, one can derive other equally valid 
formulations via gauge-fixing the additional superspace symmetries \cite{Butter4D, KLRT-M08}. 

In this section we present the superspace formulation for $\cN=2$ conformal
supergravity developed in \cite{KLRT-M08} as a gauge fixed version of the formulation
of conformal supergravity in the previous section. In \cite{Butter4D}, it
was described how to do this by the method of degauging, resulting in the
superspace geometry with structure group $\rm SL(2, \dsC) \times \rm U(2)_R$;
the analysis of \cite{KLRT-M09} could then be applied to reduce to $\rm SU(2)$
superspace. Here we give an alternative approach which emphasizes how both
$\rm U(2)$ and $\rm SU(2)$ superspace may be understood as conformal supergravity
coupled to a compensator field at the superspace level.

We begin by introducing a real primary superfield $X$ of dimension 2,
\begin{align}
\mathbb D X = 2 X~, \qquad Y X = 0~, \qquad K^A X = 0~.
\end{align}
Using this superfield, it is possible to deform the curvatures in
a way as to render new covariant derivatives which are completely
\emph{inert} under dilatation, conformal, and $S$-supersymmetry
transformations. The new covariant derivatives are given by
\begin{subequations}
\begin{align}
\cD_\alpha^i &= e^{-U/4}
	\left(\nabla_\alpha^i
	- \nabla^{\beta i} U M_{\beta \alpha}
	+ \frac{1}{4} \nabla_\alpha^i U Y
	- \nabla_\alpha^j U J_j{}^i\right)~, \\
\bar\cD^\dalpha_i &= e^{-U/4}
	\left(\bar\nabla^\dalpha_i
	+ \bar\nabla_{\dbeta i} U \bar M^{\dbeta \dalpha}
	- \frac{1}{4} \bar\nabla^\dalpha_i U Y
	+ \bar\nabla^\dalpha_j U J^j{}_i \right)~,
\end{align}
\end{subequations}
where $U := \log X$. We refer to these as the $X$-associated
derivatives.\footnote{The $\cN=1$ analogue of this construction
first appeared in \cite{KU:STC} and is related to a similar
construction in \cite{Ideas}. See \cite{BK:DualForm} for a recent discussion
of the analogous construction for $\cN=1$ superspace.}
These are constructed so that if $\Psi$ is some conformally primary
tensor superfield of vanishing dilatation weight, then $\cD_\alpha^i \Psi$
and $\bar\cD^\dalpha_i \Psi$ are as well.

When acting on a conformally primary dimensionless tensor, the algebra of
the covariant derivatives becomes
\begin{subequations}
\begin{align}
\{\cD_\alpha^i, \cD_\beta^j\}
	&=
	4 S^{ij} M_{\alpha \beta}
	+ 2 \eps^{ij} \eps_{\alpha \beta} Y^{\gamma \delta} M_{\gamma\delta}
	+ 2 \eps^{ij} \eps_{\alpha \beta} \bar W'_{\dgamma \ddelta} \bar{M}^{\dgamma \ddelta}
	\eol & \quad
	+ 2 \eps^{ij} \eps_{\alpha \beta} S^{kl} J_{kl}
	+ 4 Y_{\alpha \beta} J^{ij}~, \\
\{\bar\cD^\dalpha_i, \bar\cD^\dbeta_j\}
	&=
	-4 \bar S_{ij} \bar M^{\dalpha \dbeta}
	- 2 \eps_{ij} \eps^{\dalpha \dbeta} \bar Y_{\dgamma \ddelta} \bar M^{\dgamma\ddelta}
	- 2 \eps_{ij} \eps^{\dalpha \dbeta} W'^{\gamma \delta} M_{\gamma \delta}
	\eol & \quad
	- 2 \eps_{ij} \eps^{\dalpha \dbeta} \bar S^{kl} J_{kl}
	- 4 \bar Y^{\dalpha \dbeta} J_{ij}~,
\end{align}
\end{subequations}
where the curvature superfields are defined by
\begin{subequations}
\begin{alignat}{2}
S^{ij} &:= \frac{1}{4X^{3/2}} \nabla^{ij} X~, &\qquad
\bar S_{ij} &:= \frac{1}{4X^{3/2}} \bar \nabla_{ij} X~, \\
Y_{\alpha \beta} &:= -\frac{X^{1/2}}{4} \nabla_{\alpha \beta} X^{-1}~, &
\bar Y_{\dalpha \dbeta} &:= -\frac{X^{1/2}}{4} \bar\nabla_{\dalpha \dbeta} X^{-1}~, \\
W'_{\a\b} &:= X^{-1/2} W_{\a\b}~,&
\bar W'_{\ad\bd} &:= X^{-1/2} \bar W_{\ad\bd}~.
\end{alignat}
\end{subequations}
We have redefined $W_{\a \b}$ and its conjugate to absorb factors of the
compensator field and render them inert under dilatations. Henceforth, we
will drop the primes. The superfields $S^{ij}$ and $Y_{\a\b}$ appearing
above are the only two conformally invariant combinations of two like-chirality
spinor derivatives acting on $X$. For two derivatives of opposite chirality,
there are two conformally invariant combinations:
\begin{align}
G_{\alpha \dalpha} := -\frac{1}{16} X^{1/2} [\nabla_\alpha^k, \bar\nabla_{\dalpha k}] X^{-1}~, \quad
G_{\alpha \dalpha}{}^{ij} := -\frac{\ri}{8} X^{-1/2} [\nabla_\alpha^{(i}, \bar\nabla_\dalpha^{j)}] U~.
\end{align}

Now we construct $\cD_{\alpha \dalpha} = (\s^a)_{\a\ad} \cD_a$. The precise definition of
$\cD_{\alpha \dalpha}$ depends, as usual, on conventional constraints.
For the usual constraints in $\rm U(2)$ superspace, one takes
\begin{align}
 \cD_\alpha{}^\dalpha &:=
	X^{-1/2} \nabla_\alpha{}^\dalpha
	- \frac{\ri}{2} X^{-1/4} \nabla_\alpha^k U \bar\cD^\dalpha_k
	- \frac{\ri}{2} X^{-1/4} \bar \nabla^\dalpha_k U \cD_\alpha^k
 	\eol & \quad
	- \left(\frac{\ri}{4} X^{-1/2} \bar\nabla^\dalpha_k \nabla^{\beta k} U + 2 \ri G^{\dalpha \beta} \right)
		M_{\beta \alpha}
	+ \left(\frac{\ri}{4} X^{-1/2} \nabla_\alpha^k \bar\nabla_{\dbeta k} U - 2 \ri G_{\alpha \dbeta} \right)
		\bar M^{\dbeta \dalpha}
	\eol & \quad
	- \ri \left(\frac{1}{16} X^{-1/2} [\nabla_\alpha^k, \bar\nabla^\dalpha_k] U - G_\alpha{}^\dalpha \right) Y
	+ \frac{\ri}{2} X^{-1/2} \nabla_\alpha^k U \bar\nabla^\dalpha_{j} U  J^j{}_k~.
\end{align}
This leads to the anti-commutator
\begin{align}
\{\cD_\alpha^i, \bar \cD^\dalpha_j\} &=
	-2\ri \delta^i_j  \cD_\alpha{}^\dalpha
	- 2 (G_\alpha{}^\dalpha \delta^i_j + \ri G_\alpha{}^\dalpha{}^i{}_j) Y
	\eol & \quad
	+ 4 (G_{\alpha \dbeta} \delta^i_j + \ri G_{\alpha \dbeta}{}^i{}_j) \bar M^{\dbeta \dalpha}
	+ 4 (G^{\dalpha \beta} \delta^i{}_j + \ri G^{\dalpha \beta}{}^i{}_j) M_{\beta \alpha}
	\eol & \quad
	+ 8 G_\alpha{}^\dalpha J^i{}_j
	+ 4 \ri \delta^i_j G_\alpha{}^\dalpha{}^k{}_l J^l{}_k ~.
\end{align}

To recover $\rm SU(2)$ superspace, we choose $X = \Phi \Phi^\dag$
where $\Phi$ is a chiral superfield of unit dimension. This implies that
$G_{\alpha \dalpha}{}^{ij}$ must vanish. Then modifying slightly the
definition of the vector derivative \cite{KLRT-M09}
\begin{align}
\cD_a' = \cD_a - \ri G_a Y~,
\end{align}
leads to a new algebra
\begin{align}
\{\cD_\alpha^i, \bar \cD^\dalpha_j\} &=
	-2\ri \delta^i_j \cD'_\alpha{}^\dalpha
	+ 4 G_{\alpha \dbeta} \delta^i_j \bar M^{\dbeta \dalpha}
	+ 4 G^{\dalpha \beta} \delta^i_j M_{\beta \alpha}
	+ 8 G_\alpha{}^\dalpha J^i{}_j~,
\end{align}
without any $\rm U(1)_R$ curvature \cite{KLRT-M08}.

Formally, this geometry is not \emph{precisely} that of \cite{Grimm, KLRT-M08} since
there remains a $\rm U(1)_R$ gauge symmetry. However, because there are no
$\rm U(1)_R$ curvatures, it is possible to eliminate it completely
as discussed in \cite{KLRT-M09}. The way to accomplish that in our
language is to use the ratio $\Phi / \Phi^\dag$ as a $\rm U(1)_R$ compensator.
Applying a similarity transformation to all fields $\Psi$ and operators $\cO$,
\begin{align}\label{GrimmRedef}
\Psi \rightarrow \left(\frac{\Phi}{\Phi^\dag}\right)^{Y/4} \Psi~,\qquad
\cO \rightarrow \left(\frac{\Phi}{\Phi^\dag}\right)^{Y/4} \cO \left(\frac{\Phi}{\Phi^\dag}\right)^{-Y/4} 
\end{align}
gives a $\rm U(1)_R$ invariant theory. The new derivatives $\cD_A$ are given by
\begin{subequations}
\begin{align}
\cD_\alpha^i &= (\Phi^\dag)^{-1/2}
	\left(\nabla_\alpha^i
	- (\nabla^{\beta i} \log\Phi) M_{\beta \alpha}
	- (\nabla_\alpha^j \log\Phi) J_j{}^i\right)~, \\
\bar\cD^\dalpha_i &= (\Phi)^{-1/2}
	\left(\bar\nabla^\dalpha_i
	+ (\bar\nabla_{\dbeta i} \log\Phi^\dag) \bar M^{\dbeta \dalpha}
	+ (\bar\nabla^\dalpha_j \log\Phi^\dag) J^j{}_i \right)~, \\
\cD_\alpha{}^\dalpha &:=
	X^{-1/2} \nabla_\alpha{}^\dalpha
	- \frac{\ri}{2} (\Phi^\dag)^{-1/2} (\nabla_\alpha^k \log \Phi) \bar\cD^\dalpha_k
	- \frac{\ri}{2} \Phi^{-1/2} (\bar \nabla^\dalpha_k \log \Phi^\dag) \cD_\alpha^k
 	\eol & \quad
	- \left(X^{-1/2} \nabla^{\ad \b} \log \Phi + 2 \ri G^{\dalpha \beta} \right)
		M_{\beta \alpha}
	+ \left(X^{-1/2} \nabla_{\a \bd} \log \Phi^\dag - 2 \ri G_{\alpha \dbeta} \right)
		\bar M^{\dbeta \dalpha}
	\eol & \quad
	+ \frac{\ri}{2} X^{-1/2} (\nabla_\alpha^k \log \Phi) (\bar\nabla^\dalpha_{j} \log \Phi^\dag)  J^j{}_k~,
\end{align}
\end{subequations}
while the torsion tensors are
\begin{subequations}
\begin{alignat}{2}
S^{ij} &:= \frac{1}{4} \frac{1}{\Phi \Phi^\dag} \nabla^{ij} \Phi~, &\qquad
\bar S_{ij} &:= \frac{1}{4} \frac{1}{\Phi \Phi^\dag} \bar \nabla_{ij} \Phi^\dag~, \\
Y_{\alpha \beta} &:= -\frac{1}{4} \frac{\Phi}{\Phi^\dag} \nabla_{\alpha \beta} \Phi^{-1}~, &
\bar Y_{\dalpha \dbeta} &:= -\frac{1}{4} \frac{\Phi^\dag}{\Phi} \bar\nabla_{\dalpha \dbeta} (\Phi^\dag)^{-1}~, \\
W'_{\a\b} &:= \Phi^{-1} W_{\a\b}~,&
\bar W'_{\ad\bd} &:= (\Phi^\dag)^{-1}  \bar W_{\ad\bd}~, \\
G_{\alpha \dalpha} &:= -\frac{1}{16} X^{1/2} [\nabla_\alpha^k, \bar\nabla_{\dalpha k}] X^{-1}~.
\end{alignat}
\end{subequations}
These obey a number of Bianchi identities \cite{KLRT-M08}:
\begin{alignat}{3}
\cD_\a^{(i} S^{jk)} &= \bar\cD_\ad^{(i} S^{jk)} = 0~,&\qquad
\cD_i^{(\a} Y^{\b \g)} &= 0~,\qquad
\bar\cD^\ad_i W_{\b\g} = 0~, \eol
\cD_\a^i S_{ij} &= -\cD^\b_j Y_{\b \a}~, &\qquad
\cD_\a^i G_{\b\bd} &= -\frac{1}{4} \bar\cD_\bd^i Y_{\a\b}
	+ \frac{1}{12} \veps_{\a\b} \bar \cD_{\bd j} S^{ji}
	- \frac{1}{4} \veps_{\a\b} \bar \cD^{\gd i} \bar W_{\bd \gd}~.
\end{alignat}
A restricted version of $\rm SU(2)$ superspace is found if $\Phi$ is
further assumed to be a reduced chiral superfield $\cW$ ({\it i.e.} an abelian vector
multiplet). In this case $S^{ij}$ is also real  \cite{KLRT-M09}.

As was shown in \cite{KLRT-M08}, $\rm SU(2)$ superspace geometry
admits a certain super-Weyl transformation, which means that
it may also describe conformal supergravity. In the formulation
used here, this super-Weyl transformation corresponds to
a reparametrization of the chiral compensator,
$\Phi \rightarrow \Phi \,e^{-\s}$.
If $\Psi$ is a conformally primary superfield with dilatation
weight $\Delta$ and $\rm U(1)_R$ weight $w$ in conformal superspace,
then it may be associated with a superfield
$\Psi'$ in $\rm SU(2)$ superspace given by
\begin{align}
\Psi' = \Phi^{-\Delta/2 + w/4} (\Phi^\dag)^{-\Delta/2-w/4} \Psi~.
\end{align}
$\Psi'$ is inert under the dilatation and $\rm U(1)_R$ symmetries of
conformal superspace; in their place we have the super-Weyl
transformations
\begin{align}\label{SW_transField}
\Psi' \rightarrow \exp\left(\frac{\Delta}{2}(\s+\bar\s) - \frac{w}{4} (\s-\bar\s)\right)\, \Psi'~.
\end{align}
Any action which is inert under super-Weyl transformations is
necessarily superconformal. At the same time, $\rm SU(2)$ superspace
may be used to describe actions which are \emph{not} superconformal.

To avoid unnecessarily long constructions ({\it e.g.} ``the
conformal supergravity introduced in \cite{KLRT-M08} based
on Grimm's geometry'') we will refer to any supergeometry
with the algebra described in this section as $\rm SU(2)$ superspace,
even though, as in the actions considered in \cite{KLRT-M08},
the super-Weyl transformation is a symmetry of the action.

%%%%%%%%%%%%%%%%%%%%%%%%%%%%%%%%%%%%%%%%%%%%%%%%%%%%%%
%%%%%%%%%%%%%%%%%%%%%%%%%%%%%%%%%%%%%%%%%%%%%%%%%%%%%%

\section{Component actions from superspace}\label{ComponentActions}

In the recent paper \cite{Butter4D}, it was demonstrated that
the superconformal geometry in section
\ref{SuperspaceGeometry} reduces in components to $\cN = 2$ superconformal tensor 
calculus. This turns out to be one of its main advantages, allowing one
to efficiently reproduce the results of tensor calculus computations
directly from superspace.
Our main goal in this section is to apply these new techniques
directly to the computation of a few simple superspace actions
involving vector and tensor multiplets.

The component field structure of \cite{Butter4D} corresponds to
the Weyl multiplet \cite{sct_rules, sct_structure, BdeRdeW}.
It involves a set of one-forms: the vierbein $e_m{}^a$, 
the gravitino $\psi_m{}^\a_i$, the $\rm{U}(1)_{R}$ and $\rm{SU}(2)_{R}$
gauge fields $A_m$ and $\phi_m{}^{ij}$ and the dilatation gauge field $b_m$,
which are each the lowest component of their corresponding superforms:
\begin{align}
e_m{}^a &:= E_m{}^a| \ , \quad \psi_m{}^\a_i := 2 E_m{}^\a_i| \ , \quad \psib_m{}^i_\ad := 2 E_m{}^i_\ad| \ , \non\\
A_m &:= \Phi_m| \ , \quad \phi_m{}^{i j} := \Phi_m{}^{ij}| \ ,\quad b_m := B_m| ~.
\end{align}
We denote the component projection of a superfield, $V(z)$ by
$V(z)| := V(z)|_{\theta = \bar{\theta} = 0}$. There are additional
\emph{composite} gauge connections: the spin connection  $\omega_m{}^{ab}$ and the special
conformal and $S$-supersymmetry connections $\frak{f}_m{}^a$ and $\phi_m{}_\a^i$,
\begin{align}
\omega_m{}^{ab} := \Omega_m{}^{ab}| \ , \quad  
\frak{f}_m{}^a := \frak{F}_m{}^a| \ , \quad
\phi_m{}^i_\a := 2 \frak F_m{}^i_\a| \ ,
\quad \bar{\phi}_m{}_i^\ad := 2 \frak F_m{}_i^\ad |~,
\end{align}
which are defined in terms of the other fields. (See appendix \ref{CompIdentities} for
a summary of these relations.)
Finally, in order to have the same number of bosonic and fermionic degrees of freedom
it is necessary to have additional non-gauge fields. These are encoded in
the components of the superfield $W_{\a\b}$:
\begin{subequations}
\begin{gather}
W_{ab} = W_{ab}^{+} + W_{ab}^{-} \ , \quad W_{ab}^{+} := (\s_{ab})^{\a\b} W_{\a\b}| \ , \quad W_{ab}^{-} := - (\tilde{\s}_{ab})_{\ad\bd} \bar{W}^{\ad\bd}| \ ,\\
\S^{\a i} := \frac{1}{3} \nabla_\b^i W^{\a\b}| \ , \quad \bar{\S}_{\ad i} := - \frac{1}{3} \bar{\nabla}^\bd_i \bar{W}_{\ad\bd} | \ ,\\
D := \frac{1}{12} \nabla^{\a\b} W_{\a\b}| = \frac{1}{12} \bar{\nabla}_{\ad\bd} \bar{W}^{\ad\bd}| \ ,
\end{gather}
\end{subequations}
where $W_{ab}^{\pm}$ satisfies the self-duality relation
$\frac{\ri}{2} \eps_{ab}{}^{cd} W_{cd}^{\pm} = \pm W_{ab}^{\pm} $.
All other components of $W_{\a\b}$ can be expressed in terms of the
previously defined fields \cite{Butter4D}. We will frequently abuse notation
and use $W_{\a\b}$ both to denote the superfield and its lowest
component. It should be clear from context to which object we are
referring.

The situation differs in $\rm SU(2)$ superspace. After having absorbed the
compensator, the structure group is effectively reduced to
$\rm SL(2,\mathbb C) \times SU(2)_R$ with the covariant derivative
\begin{align}
\cD_A = E_A{}^M \partial_M + \hf \Omega_A{}^{bc} M_{bc} + \Phi_A{}^{ij}J_{ij}
\end{align}
and torsion superfields
\begin{align}
S^{ij}~, \quad Y_{\a\b}~, \quad G_{\a\ad}~, \quad W_{\a\b} \ ,
\end{align}
together with their complex conjugates.
Grimm chose the independent component fields to be \cite{Grimm}
\begin{align}
&e_m{}^a := E_m{}^a| \ , \quad \psi_m{}^\a_i := 2 E_m{}^\a_i| \ , \quad \psib_m{}^i_\ad := 2 E_m{}^i_\ad| \ ,
	\quad \phi_m{}^{i j} := \Phi_m{}^{ij}| \ , \non\\
&W_{\a\b}| \ , \quad \bar{W}_{\ad\bd}| \ , \quad S^{ij}| \ , \quad \bar{S}^{ij}| \ , \quad Y_{\a\b}| \ , \quad \bar{Y}_{\ad\bd}| \ , \quad G_{a}| \ , \non\\
&\r_{\a i} := \frac{1}{3} \cD_\a^j S_{ij}| \ , \quad \bar{\r}^{\ad i} := \frac{1}{3} \bar{\cD}^\ad_j \bar{S}^{ij}| \ , \quad \k_{\a i} := \frac{1}{3} \cD_\a^j \bar{S}_{ij}| \ , \quad \bar{\k}^{\ad i} := \frac{1}{3} \bar{\cD}^\ad_j S^{ij}| \ , \non\\
& S := \cD^{ij} \bar{S}_{ij}| + \bar{\cD}^{ij} S_{ij}| \ , \quad C = \frac{1}{12} \cD^{ij} S_{ij}| \ , \quad \bar{C} = \frac{1}{12} \bar{\cD}^{ij} \bar{S}_{ij}| \ .
\end{align}
From now on, we will use $W^{\a\b}$, $S^{ij}$, $Y^{\a\b}$, and $G_a$
to refer to both the superfields and their component projections.

In transforming from the superconformal components to Grimm's components,
the $\rm U(1)_R$ connection $A_m$ is replaced by $e_m{}^a G_{a}$, the fields
$\S_{\a}^i$ and $D$ are replaced by $\k_{\a}^i$ and $S$, and
the dilatation connection $b_m$ is eliminated by a choice of $K_a$-gauge.
The lowest bosonic and fermionic components of $\Phi$
compensate for dilatations, $\rm U(1)_R$ gauge transformations, and $S$-supersymmetry.
The remaining components of this multiplet correspond to
$S^{ij}$, $Y_{\a\b}$, $\rho_{\a i}$ and $C$.
If the action under consideration is
invariant under super-Weyl transformations, one can show that these
component fields must always drop out.

It should be emphasized that Grimm's choice of $\k_{\a i}$
and $S$ as component fields is purely conventional; it is perhaps
more natural to trade these for $\S_{\a}^i$ and $D$. The precise
translation is given by
\begin{align}
\k^{\a i} &= - \S^{\a i} - \frac{1}{3} (\s^{ab})^{\a\b} \Psi_{ab}{}_{\b}^i
	+ \ri \psi_a{}^{\a i} G^a
	- \frac{4 \ri}{3} (\s^{ab})_\b{}^\a \psi_a{}^{\b i} G_b
	\non\\& \quad
	+ \frac{\ri}{2} (\s^a)^\a{}_\ad \psib_a{}^{\ad}_j \bar{S}^{ij}
	+ \frac{\ri}{6} (\s^a)^{\b \ad} \psib_a{}_{\ad}^i W^\a{}_\b
	+ \frac{\ri}{2} (\s^a)^\a{}_\bd \psib_a{}_{\ad}^i \bar{Y}^{\ad \bd}~, \non\\
S &= 12 D + 12 Y^{\a\b} W_{\a\b} - 12 S^{ij} \bar{S}_{ij} 
	- 4 \cR - 48 G^a G_a \non\\
	& \quad
	- 4 \ri (\psi^b{}_k \s^a \bar\Psi_{ab}{}^k) + 8 (\psi^b{}_k \s^a \psib_b{}^k) G_a
	+ 4 (\psi_a{}_k \s^a \psib_b{}^k) G^b \non\\
	& \quad
	- 6 \psi_a{}_i \psi^a{}_j S^{ij} - 2 (\s^a)_{(\a}{}^\ad (\s^b)_{\b)}{}^\bd \psi_a{}^\a_k \psi_b{}^{\b k} \bar{W}_{\ad\bd} + 6 \psi_a{}^\g_k \psi^a{}^{\d k} Y_{\g\d}
	+ {\rm c.c.} \ ,
\end{align}
where $\cR = \cR_{ab}{}^{ab}$ is the Ricci scalar and $\Psi_{ab}{}^\g_k$ is
the gravitino field strength, defined in appendix \ref{CompIdentities}.

%%%%%%%%%%%%%%%%%%%%%%%%%%%%%%%%%%%%%%%%%%%%%%%%%%%%%%

\subsection{General component chiral actions from superspace}

The conformal supergravity formulations presented in section \ref{SuperspaceGeometry}
allow the construction of superconformally invariant actions \cite{Butter4D}.
The simplest invariants are integrals over the full superspace
\be
S = \int \rd^{12}z \,E \,\cL \ , \quad \rd^{12}z \equiv \rd^4 x \,\rd^4 \q \,\rd^4\bar{\q} \ ,
\ee
where $E := {\rm sdet}(E_M{}^A)$ and $\cL$ is a real conformally primary
$\rm U(2)_{R}$ scalar with vanishing conformal dimension,
\be \mathbb{D} \cL = Y \cL = J^{ij} \cL = M_{ab} \cL = K_a \cL = S_\a^i \cL = \bar{S}^\ad_i \cL = 0 \ .
\ee
However, these tend to yield higher derivative actions and play
less of a role in $\cN=2$ superspace than in $\cN=1$.
Of more importance is the chiral action, which involves an integral
over the chiral subspace
\be S = S_c + {\rm c.c.} \ , \quad S_c = \int \rd^8 z \,\cE \, \cL_c \ ,
\quad \rd^{8}z := \rd^4 x \, \rd^4 \q \ ,
\ee
where $\cL_c$ is chiral, $\bar\nabla^\ad_i \cL_c = 0$ and $\cE$ is a
suitably chosen chiral measure \cite{Butter4D}.
The Lagrangian $\cL_c$ must be a conformally primary Lorentz and
$\rm SU(2)_R$ chiral scalar with conformal dimension two and $\rm U(1)_R$ weight $-4$:
\begin{align}
\mathbb{D} \cL_c = 2 \cL_c \ , \quad Y \cL_c = -4 \cL_c \ , \quad J^{ij} \cL_c = M_{ab} \cL_c = K_a \cL_c = S_\a^i \cL_c = \bar{S}^\ad_i \cL_c = 0 \ .
\end{align}
Any action involving an integral over the full superspace may be converted
to one over the chiral subspace by the rule \cite{Butter4D}
\be \int \rd^{12}z \,E \,\cL = \int \rd^8 z \,\cE \,\bar{\nabla}^4 \cL \ , \quad
\bar{\nabla}^4 := \frac{1}{48} \bar{\nabla}^{ij} \bar{\nabla}_{ij} \ .
\ee
In the rest of this paper we will consider only chiral actions when
dealing with conventional $\cN=2$ superspace.

It was shown in \cite{Butter4D} that the component action for a chiral
action is
\begin{align}
S_c =& \int \rd^4x \,e \Bigg( \frac{1}{48} \nabla^{ij} \nabla_{ij} -  \frac{\ri}{12} \psib_d{}^l_\dd (\tilde{\s}^d)^{\dd \a} \nabla_\a^q \nabla_{lq} 
+ \frac{\ri}{2} \psib_d{}_\dd^l (\s^d)_{\a \ad} \bar{W}^{\ad \dd} \nabla^\a_l + \bar{W}^{\ad\bd} \bar{W}_{\ad\bd} \non\\
&+\frac{1}{4} \psib_c{}_\gd^k \psib_d{}^l_\dd \Big( (\tilde{\s}^{cd})^{\gd \dd} \nabla_{kl} - \frac{1}{2} \ve^{\gd \dd} \ve_{kl} (\s^{cd})_{\b \g} \nabla^{\b \g} 
- 4 \ve^{\gd\dd} \ve_{kl} (\tilde{\s}^{cd})_{\ad\bd} \bar{W}^{\ad\bd} \Big) \non\\
&- \frac{1}{4} \ve^{abcd} (\tilde{\s}_a)^{\bd \a} \psib_b{}_\bd^j \psib_c{}_\gd^k \psib_d{}^\gd_k \nabla_{\a j} 
- \frac{\ri}{4} \ve^{abcd} \psib_a{}_{\ad}^i \psib_b{}^{\ad}_i \psib_c{}_{\bd}^j \psib_d{}^{\bd}_j \Bigg) \cL_c |~. \label{CA}
\end{align}
This corresponds to the action of a chiral multiplet coupled to
$\cN=2$ conformal supergravity constructed in \cite{deRoo}.

When dealing with actions which are not superconformal, one must resort to
a different superspace geometry, or equivalently, introduce compensators.
For example, in $\rm SU(2)$ superspace, one may introduce the chiral
action\footnote{It should be mentioned that $\cE$ here is defined using the
vierbein associated with $\cD_A$.}
\begin{align}
S_c &= \int \rd^4x\, \rd^4\q\, \cE\, \cL_c~.
\end{align}
Its component form is given by
\begin{align}
S_c =& \int \rd^4x \,e \Bigg[ \frac{1}{48} \cD^{ij} \cD_{ij} + \frac{7}{12} S^{ij} \cD_{ij} - \frac{1}{4} Y^{\a \b} \cD_{\a \b} 
+ \frac{1}{2} (\cD_i^\a S^{ij}) \cD_{\a j} + \frac{1}{6}(\cD^{ij} S_{ij}) \non\\
&+ 3 S^{ij} S_{ij} - Y^{\a \b} Y_{\a \b} + \bar{W}^{\ad \bd} \bar{W}_{\ad \bd} \non\\
&- \frac{\ri}{6} \psib_d{}^l_\dd (\tilde{\s}^d)^{\dd \a} \Big( \frac{1}{2} \cD_\a^q \cD_{lq} + 5 S_{lq} \cD_\a^q - 3 Y_{\a \b} \cD^\b_l 
+ 4 (\cD_\a^q S_{lq}) \Big) + \frac{\ri}{2} \psib_d{}_\dd^l (\s^d)_{\a \ad} \bar{W}^{\ad \dd} \cD^\a_l \non\\
& + \frac{1}{4} \psib_c{}_\gd^k \psib_d{}^l_\dd \Bigg( (\tilde{\s}^{cd})^{\gd \dd} \cD_{kl} + 8 (\tilde{\s}^{cd})^{\gd \dd} S_{kl} 
- 4 \ve^{\gd \dd} \ve_{kl} (\tilde{\s}^{cd})_{\ad \bd} \bar{W}^{\ad \bd} \non\\
&\qquad- \frac{1}{2} \ve^{\gd \dd} \ve_{kl} (\s^{cd})_{\b \g} \cD^{\b \g} - 4 \ve^{\gd \dd} \ve_{kl} (\s^{cd})_{\a \b} Y^{\a \b} \Bigg) \eol
& - \frac{1}{4} \ve^{abcd} (\tilde{\s}_a)^{\bd \a} \psib_b{}_\bd^j \psib_c{}_\gd^k \psib_d{}^\gd_k \cD_{\a j}
- \frac{\ri}{4} \ve^{abcd} \psib_a{}_{\ad}^i \psib_b{}^{\ad}_i \psib_c{}_{\bd}^j \psib_d{}^{\bd}_j \Bigg] \cL_c | \label{CAG}~.
\end{align}
If $\cL_c$ transforms under super-Weyl transformations as
\eqref{SW_transField} with $\Delta=2$ and $w=-4$, then one
can show that the action is super-Weyl invariant. For this case,
this action and \eqref{CA} can be shown to coincide.

It is worth noting that the component action \eqref{CAG} has been obtained via
the superform method \cite{GKT-M}; one expects the same should be true of the
superconformal component action \eqref{CA}.

In the rest of this section we will present component results and actions
for the $\cN = 2$ abelian vector and tensor multiplets.

%%%%%%%%%%%%%%%%%%%%%%%%%%%%%%%%%%%%%%%%%%%%%%%%%%%%%%

\subsection{The abelian vector multiplet in components}

As discussed in subsection \ref{VectorMultipletGeometry}, the abelian
vector multiplet is described by a primary reduced chiral superfield $\cW$ of
dimension 1,
\be \bar{\nabla}^\ad_i \cW = 0 \ , \quad
\nabla^{ij} \cW = \bar{\nabla}^{ij} \bar{\cW} \ ,
\quad \mathbb{D} \cW = \cW \ , \quad Y \cW = - 2 \cW \ , \quad K_A \cW = 0~.
\ee
The vector field strength $F$ in superspace is given by
\eqref{2formConstraint} and \eqref{2formCurvs}.
Within the superfield $\cW$ are found the matter components
of the abelian vector multiplet: a complex scalar field $\phi$,
a gaugino $\l_\a^i$ and a real $\rm{SU}(2)$ isotriplet $X^{ij}$,
\begin{align}
\phi = \cW| \ , \quad \l{}_\a^i =\nabla_\a^i \cW| \ , \quad X^{ij} = \nabla^{ij} \cW | \ .
\end{align}
The reality of $X^{ij}$ follows from the Bianchi identity.
The remaining component field, the gauge connection $v_m$ is
given by the lowest component of the corresponding superspace
connection, $v_m = V_m\vert$.

The component two-form field strength is constructed from a
projection of the superspace two-form,
\begin{align}
f_{mn} = F_{mn} \lc = 2 \partial_{[m} V_{n]}\lc = 2 \partial_{[m} v_{n]}~.
\end{align}
Making use of the identity
\begin{align}\label{FmnFAB}
F_{mn} = E_m{}^A E_n{}^B F_{AB} (-)^{ab}
\end{align}
and projecting to lowest component, we may solve for $F_{ab}\vert$ to give
\begin{align}\label{covF}
\hat F_{ab} := F_{ab}\vert =& e_a{}^m e_b{}^n {f}_{mn}
	- \frac{\ri}{2} (\s_{[a})_\a{}^\ad \psi_{b]}{}^\a_k \bar{\l}^k_\ad
	+ \frac{\ri}{2} (\tilde{\s}_{[a})_\ad{}^\a \l^k_\a \psib_{b]}{}^\ad_k \non\\
	&- \hf \psi_{a}{}^\g_k \psi_{b}{}^k_\g \bar{\phi}
	+ \hf \psib_{a}{}^\gd_k \psib_{b}{}^k_\gd \phi \ .
\end{align}
In the language of the superconformal tensor calculus, $\hat F_{ab}$ is
referred to as the supercovariant field strength whereas
$f_{mn}$ is the conventional field strength. Recall that $F_{ab}$
is given by equation \eqref{2formCurvsb}.

The supersymmetry transformations for this (or any) multiplet may
be derived via a covariant Lie derivative in superspace. For any
covariant superfield $\Psi$, this reduces to the usual covariant derivative:
\begin{align}
\delta \Psi = \xi^\a_i \nabla_\a^i \Psi + \bar\xi_\ad^i \bar\nabla^\ad_i \Psi~.
\end{align}
Projecting to lowest components yields the transformation of
the field $\Psi\vert$. Applying this rule to $\Psi = \cW$ and
$\Psi = \nabla_\a^i \cW$ leads to
\begin{align}
\d \phi &= \x^\a_i \l_\a^i
\end{align}
and
\begin{align}
\d \l_\a^i &= - \hf \x_{\a j} X^{ij} - 2 (\s^{ab})_{\a\b} \x^{\b i} \hat{F}_{ab}
	- 2 \x^{\b i} W_{\a\b} \bar{\phi} - 2 \ri \bar{\x}_\bd^i {\nabla}'_{\a}{}^\bd \phi
	+ \ri \bar{\x}_\bd^i \psi_\a{}^\bd{}^\g_k \l_\g^k ~,
\end{align}
where $\nabla'_a$ is defined in \eqref{nabla'}.
The other supersymmetry transformations may be derived similarly.

The vector multiplet in $\rm SU(2)$ superspace can be easily worked out.
We find the same result as in \cite{KLRT-M08}. The constraints on $\cW$ are
\be
\bar{\cD}^\ad_i \cW = 0 \ , \quad (\cD^{ij} + 4 S^{ij}) \cW = ( \bar{\cD}^{ij} + 4 S^{ij} )\bar{\cW}~,
\ee
and the matter fields are
\be
\phi = \cW| \ , \quad \l{}_\a^i =\cD_\a^i \cW| \ , \quad X^{ij} = (\cD^{ij} + 4 S^{ij}) \cW| \ .
\ee
One defines the superspace two-form using the derivatives $\cD_A$.
The result is formally identical to \eqref{2formConstraint} and
\eqref{2formCurvs}, but with the replacements
\begin{align}
\nabla_\a^i \cW \rightarrow \cD_\a^i \cW~,\quad
\nabla_{\a\b} \cW \rightarrow \cD_{\a\b} \cW + 4 Y_{\a\b}\cW~.
\end{align}
The component two-form is still given by \eqref{covF}.

%%%%%%%%%%%%%%%%%%%%%%%%%%%%%%%%%%%%%%%%%%%%%%%%%%%%%%

\subsection{Abelian vector multiplet action}

Let us now consider the general chiral action $\cL_c = \cF(\cW^I)$
of $n$ abelian vector multiplets\footnote{The non-abelian case is a straightforward generalization.}
$\cW^I$. We will consider its evaluation in two distinct cases,
first for the superconformal case and then for the general case.

\subsubsection{The superconformal case}

In the superconformal case, $\cF$ must be homogeneous of degree two in $\cW^I$,
\be
\cW^I \cF_I \equiv \cW^I \frac{\partial}{\partial \cW^I} \cF = 2 \cF \ .
\ee
The component action coupled to conformal supergravity was given first
in \cite{sct_lagrangians}. Here we describe how to reconstruct that result
directly from superspace.

Using the component reduction formula \eqref{CA}, it is straightforward
to project to components. The following two identities prove useful:
\begin{subequations}
\begin{align}
\nabla_\a^i \nabla_{ij} \cW^I &= -6\ri \nabla_{\alpha \dbeta} \bar\nabla^{\dbeta}_j \cW^I
	= -6\ri \bar\nabla^{\dbeta}_j \nabla_{\alpha \dbeta}  \cW^I~, \\
\nabla^{ij} \nabla_{ij} \cW^I
     &= 48 \Box \bar \cW^I + 12 \bar W_{\dalpha \dbeta} \bar \nabla^{\dbeta \dalpha} \bar \cW^I
     + 12 \bar \nabla_{\dalpha \dbeta} \bar W^{\dalpha \dbeta} \bar \cW^I
     + 24 \bar \nabla_{\dbeta j}  \bar W^{\dbeta \dalpha} \bar \nabla_{\dalpha}^j \bar \cW^I \ ,
\end{align}
\end{subequations}
where $\Box = \nabla^a \nabla_a$ is the supercovariant d'Alembertian.
When taken to lowest components they yield
\begin{subequations}
\begin{align}
\nabla_\a^i \nabla_{ij} \cW^I \lc &= -6\ri \nabla_{\alpha \dbeta} \bar\lambda^{I \dbeta}_j~,\\
\frac{1}{48} \nabla^{ij} \nabla_{ij} \cW^I \lc
     &= \Box \bar \phi^I + 2 \bar W_{\ad \bd} \hat F^{\ad \bd}
	- \bar W_{\ad \bd} \bar W^{\ad \bd} \phi^I
	+ 3 D \bar \phi^I
	+ \frac{3}{2} \bar\S^\ad_j \bar\lambda_\ad^j~.
\end{align}
\end{subequations}
The full component action is:
\begin{align}
S_c &=
     \int \rd^4x \, e \,\Big(\cF_I \Box \bar \phi^I
     - \frac{\ri}{2} \cF_{IJ} \lambda^{I\alpha j} \nabla_{\alpha \dalpha} \bar \lambda^J{}^\dalpha_j
     + \frac{1}{32} \cF_{IJ} X^{I ij} X^J_{ij}
     - 2 \cF_{IJ} \hat{F}^{I \alpha \beta} \hat{F}^J_{\alpha \beta}
	\eol & \quad
	- \cF \, \bar W_{\dalpha \dbeta} \bar W^{\dalpha \dbeta}
     - 2 \cF_I \bar W_{\dalpha \dbeta} \hat F^{I \dalpha \dbeta}
     + 3 \cF_I \bar \phi^I\, D 
     + \frac{3}{2} \cF_I \bar \S^k \bar\lambda_k^I
     - 2 \cF_{IJ} \bar \phi^I \, W^{\alpha \beta} \hat{F}^J_{\alpha \beta}
     \eol & \quad
     - \frac{1}{2} \cF_{IJ}  \bar \phi^I \bar \phi^J \, W^{\alpha \beta} W_{\alpha \beta}
     + \frac{1}{16} \cF_{IJK} (\lambda^{I i} \lambda^{Jj}) X^K_{ij}
     + \hf \cF_{IJK} \lambda^{I\alpha k} \lambda^J{}^\beta_k \hat{F}^K_{\alpha \beta}
     \eol & \quad
     + \frac{1}{4} \cF_{IJK} \bar \phi^I\, \lambda^{J\alpha k} \lambda^K{}^\beta_k W_{\alpha \beta} 
	+ \frac{1}{48} \cF_{IJKL} (\lambda^{Ii} \lambda^{Jj}) (\lambda^K_i\lambda^L_j)
	- \hf \cF_I (\bar\psi_m{}^j {\tilde{\sigma}}^m \sigma^b)_\dalpha \nabla_b \bar \lambda^I{}^\dalpha_j
     \eol & \quad
     + \frac{\ri}{2} \cF_I\bar\psi_b{}_\dgamma^j \,\bar W^\dgamma{}_\dbeta
          \,{\tilde{\sigma}}_b^{\dbeta \alpha} \lambda^I_{\alpha j}
     - \frac{\ri}{8} \cF_{IJ} (\bar{\psi}_m{}^i {\tilde{\sigma}}^m \lambda^{Ij}) X^J_{ij}
     - \ri \cF_{IJ} (\bar{\psi}_m{}^k {\tilde{\sigma}}^m)^\alpha \lambda^I{}^\beta_k \hat{F}^J_{\alpha \beta}
     \eol & \quad
     - \frac{\ri}{2} \cF_{IJ}\bar\phi^I (\bar{\psi}_m{}^k {\tilde{\sigma}}^m)^\alpha \lambda^J{}^\beta_k W_{\alpha \beta}
     - \frac{\ri}{12} \cF_{IJK} (\bar{\psi}_m{}^i {\tilde{\sigma}}^m \lambda^{Ij}) (\lambda_i^J \lambda_j^K)
     - \frac{1}{4} \cF_I (\bar\psi_m{}^i {\tilde{\sigma}}^{mn} \bar\psi_n{}^j) X^I_{ij}
     \eol & \quad
     + \cF_I (\bar\psi_m \bar\psi_{n})( \sigma^{mn})^{\alpha \beta} \hat{F}^I_{\alpha \beta}
     + \frac{1}{2} \cF_I \bar\phi^I (\bar\psi_m \bar\psi_{n})( \sigma^{mn})^{\alpha \beta} W_{\alpha \beta} 
     - \cF (\bar\psi_m \bar\psi_n) ({\tilde{\sigma}}^{mn})^{\dalpha \dbeta} \bar W_{\dalpha \dbeta}
     \eol & \quad
     - \frac{1}{4} \cF_{IJ} (\bar\psi_m{}^i {\tilde{\sigma}}^{mn} \bar\psi_n{}^j) (\lambda^I_i \lambda^J_j)
     + \frac{1}{8} \cF_{IJ} (\bar{\psi}_m \bar{\psi}_n) (\lambda^{Ik} \sigma^{mn} \lambda^J_k)
     \eol & \quad
     + \frac{1}{4} \cF_I \eps^{mnpq} (\bar\psi_{m} \bar\psi_{n}) (\bar\psi_p{}^i {\tilde{\sigma}}_q \lambda_i^I)
     - \frac{\ri}{4} \cF \eps^{mnpq} (\bar\psi_m \bar\psi_n) (\bar\psi_p \bar\psi_q) \Big) \ ,
\end{align}
where
\begin{subequations}
\begin{align}
\nabla_a \phi^I &\equiv \nabla'_a \phi^I - \hf (\psi_a{}_j \lambda^{Ij})~, \\
\nabla_b \bar\lambda^I{}^{\dalpha}_j &\equiv
     \nabla'_b \bar\lambda^I{}^\dalpha_j
     + 2 \bar \phi^I \bphi_b{}^\dalpha_j 
     - \ri \psi_{b \beta j} \nabla^{\dalpha \beta} \bar \phi^I
     + \frac{1}{4} \bar{\psi}_b{}^{\dalpha k} X_{kj}^I
     + 2 \bar{\psi}_{b \dbeta j} \hat{F}^{I \dbeta \dalpha}
     + \bar{\psi}_{b \dbeta j} \bar W^{\dbeta \dalpha} \phi^I~, \\
\Box \bar\phi^I &\equiv
     \nabla'^a \nabla_a \bar\phi^I
	+ 2 \frak{f}_a{}^a \bar \phi^I
	- \frac{\ri}{2} (\phi_m{}^j \sigma^m \bar\lambda_j^I)
     \eol &\quad
     + \frac{\ri}{4} (\psi_{m j} \sigma^m)_\dalpha \bar W^{\dalpha \dbeta} \bar\lambda^I{}_{\dbeta}^j
     - \frac{3\ri}{4} (\psi_{m j} \sigma^m \bar\S^j) \bar\phi^I
     - \hf \bar{\psi}_{a \dbeta}^j \nabla^a \bar\lambda^I{}^\dbeta_j \ .
\end{align}
\end{subequations}
The derivative $\nabla_a'$ which we have introduced above is given by
\begin{align}\label{nabla'}
\nabla_a' := e_a{}^m \left(\partial_m + \frac{1}{2} \omega_m{}^{ab} M_{ab}
		+ \phi_m{}^{ij} J_{ij}
		+ \ri A_m Y
		+ b_m \mathbb D\right)~.
\end{align}
This result may be compared with that given in \cite{sct_lagrangians}.

\subsubsection{The general case}

Next, we consider the situation in $\rm SU(2)$ superspace for the action
\begin{align}\label{GrimmF}
S = \int \rd^4x\, \rd^4\q\, \cE\, \cF(\cW^I)~.
\end{align}
Here $\cF$ may be an \emph{arbitrary} function of $\cW^I$, not necessarily
of degree 2. The component reduction is straightforward by making use of
\eqref{CAG} and the results in appendix \ref{CompIdentities}.
A number of identities for derivatives of $\cW$ are useful:
\begin{subequations}
\begin{align}
\cD^{ij} \cW &= X^{ij} - 4 S^{ij} \cW ~,\\
\cD^{\a j} \cD_{ij} \cW &= 4 (\cD^{\a j} \bar{S}_{ij}) \bar{\cW} - 4 (\cD^{\a j} S_{ij}) \cW - 6 G^{\a \ad} \bar{\cD}_{\ad i} \bar{\cW} \non\\
&- 4 S_{ij} \cD^{\a j} \cW + 6 \ri \cD^{\a \ad} \bar{\cD}_{\ad i} \bar{\cW} ~,\\
\cD^{ij} \cD_{ij} \cW &= 48 \cD^a \cD_a \bar{\cW} + 192 \ri G_a \cD^a \bar{\cW} - 16 S^{ij} \bar{\cD}_{ij} \bar{\cW} + 12 \bar{W}_{\ad \bd} \bar{\cD}^{\ad \bd} \bar{\cW} \non\\
&+ 8 (\bar{\cD}^{\ad i} S_{ij}) \bar{\cD}^{j}_\ad \bar{\cW} - 8 \cD^\a_{i} S^{ij} \cD_{\a j} \cW - 24 (\bar{\cD}^{\ad i} \bar{W}_{\ad \bd})\bar{\cD}^{\bd}_i \bar{\cW} \non\\
&+ 4 \cD^{ij} \bar{S}_{ij} \bar{\cW} - 4 \cD^{ij} S_{ij} \cW  - 16 \bar{S}^{ij} S_{ij} \bar{\cW} +16 S^{ij} S_{ij} \cW~.
\end{align}
\end{subequations}
In writing the action, and in particular showing that it must be
real (up to a total derivative) for the case $\cF(\cW) = \cW^2$,
it is helpful to introduce vector derivatives which do not contain the
gravitino. As a first step, we introduce
\begin{align}
\cD_a' = e_a{}^m \left(\partial_m + \frac{1}{2} \omega_m{}^{bc} M_{bc}
		+ \phi_m{}^{ij} J_{ij}\right)~.
\end{align}
Note that $\cD_a' = e_a{}^m \cD_m\lc$, where $\cD_m = E_m{}^A \cD_A$
is defined in superspace. We may calculate
\be
[{\cD}_a' , {\cD}_b'] = \cT_{ab}{}^c(x) \cD'_c + \hf \cR_{ab}{}^{cd}(x) M_{cd}
	+ \cR_{ab}{}^{ij} J_{ij}~.
\ee
The torsion tensor $\cT_{ab}{}^c$ and the curvature tensors $\cR_{ab}{}^{cd}$
and $\cR_{ab}{}^{ij}$ coincide with certain projections of superspace torsions and
curvatures,
\begin{align}
\cT_{ab}{}^c = e_a{}^m e_b{}^n T_{mn}{}^c\lc~,\qquad
\cR_{ab}{}^{cd} = e_a{}^m e_b{}^n R_{mn}{}^{cd}\lc~,\qquad
\cR_{ab}{}^{ij} = e_a{}^m e_b{}^n R_{mn}{}^{ij}|~.
\end{align}
An important property of these derivatives is that the torsion tensor
does not vanish, but is given by a gravitino bilinear, due to the presence
of gravitinos in the spin connection. To extract this
bilinear, we can introduce the torsion-free covariant derivatives
(see {\it e.g.} \cite{Ideas})
\begin{align}
\tilde{\cD}_a &:= e_a{}^m \left(\partial_m + \frac{1}{2} \tilde \omega_m{}^{bc} M_{bc}
		+ \phi_m{}^{ij} J_{ij}\right)~,\\
\tilde\omega_{mbc} &:= \omega_{mbc} + \hf e_m{}^a(\cT_{acb} + \cT_{bcd} - \cT_{abc})~.
\end{align}
The new spin connection $\tilde \omega_{mbc}$ depends only on the
vierbein, $\tilde \omega_{mbc} = \tilde\omega_{mbc}(e)$,
so the algebra of these covariant derivatives is torsion-free
\begin{align}
[\tilde{\cD}_a, \tilde{\cD}_b] &= \hf \tilde{R}_{ab}{}^{cd} M_{cd} + {\cR}_{ab}{}^{ij} J_{ij} \ ,
\end{align}
where $\tilde R_{ab}{}^{cd}$ depends only on the vierbein,
$\tilde R_{ab}{}^{cd} = \tilde R_{ab}{}^{cd}(e)$.

Now we present the action, separating it in terms of the number of gravitinos appearing explicitly within:
\begin{align}
 S_c &= \int \rd^4x \,e \,\big( \cL_0 + \cL_1 + \cL_2 + \cL_3 + \cL_4 \big) \ ,
\end{align}
where $\cL_n$ contains $n$ explicit gravitinos.
The terms are
\begin{align}
\cL_0 &= \cF \Big( 2 C + 3 S^{ij} S_{ij} - Y^{\a \b} Y_{\a \b} + \bar{W}^{\ad \bd} \bar{W}_{\ad \bd} \Big)\non\\
	&\quad
	+ \cF_I \Big(\frac{1}{24} S \bar{\phi}^I + 2 Y^{\a \b} \hat{F}^I_{\a \b}
	- 2 \bar{W}^{\ad \bd} \hat{\bar{F}}^I_{\ad\bd} + \frac{1}{4} S^{ij} X^I_{ij}
	+ \tilde{\cD}^a \tilde{\cD}_a \bar{\phi}^I
	\eol &\qquad\qquad
	+ 2 \ri \tilde{\cD}_a (G^a \bar{\phi}^I)+ 2 \ri G^a \tilde{\cD}_a \bar{\phi}^I
	- C \phi^I - 2 S^{ij} S_{ij} \phi^I - \bar{W}^{\ad\bd} \bar{W}_{\ad\bd} \phi^I
	\eol &\qquad\qquad
	+ Y^{\a \b} Y_{\a\b} \phi^I + \hf Y^{\a \b} W_{\a\b} \bar{\phi}^I
	- \hf \bar{Y}^{\ad\bd} \bar{W}_{\ad\bd}  \bar{\phi}^I
	+ \bar{S}^{ij} S_{ij} \bar{\phi}^I + \r^i \l_i^I + \bar{\k}^i \bar{\l}_i^I \Big) \non\\
	&\quad
	+  \cF_{IJ} \Big( \frac{1}{32} X^{Iij} X^J_{ij} - \hf \hat{F}^{Iab} \hat{F}^J_{ab}
	- \frac{\ri}{4} \ve^{mnpq} f_{mn} f_{pq}
	- \frac{1}{4} X^{Iij} S_{ij} \phi^J - 2 Y_{\a\b} \hat{F}^{I\a\b} \phi^J
	\non\\&\qquad\qquad
	- 2 W_{\a\b} \hat{F}^{I\a\b} \bar{\phi^J}
	+ \hf S^{ij} S_{ij} \phi^I \phi^J - \hf Y^{\a\b} Y_{\a\b} \phi^I \phi^J
	-Y^{\a\b} W_{\a\b} \phi^I \bar{\phi}^J
	\non\\&\qquad\qquad
	- \hf W^{\a\b} W_{\a\b} \bar{\phi}^I \bar\phi^J
	+ \k^i \l_i^I \bar{\phi}^J - \r^i \l_i^I \phi^J
	- \frac{\ri}{2} \l^{I\a i} \tilde{\cD}_{\a\ad} \bar{\l}^J{}^{\ad}_i
	+ \hf G^a (\l^{Ii} \s_a \bar{\l}^J_i) \Big) \non\\
	&\quad
	+  \cF_{IJK} \Big( \frac{1}{16} X^I_{ij} \l^{Ji} \l^{Kj} + \hf \hat{F}^{I\a\b} \l^J{}_{\a}^i \l^K{}_{\b i}
	- \frac{1}{4} S_{ij} \phi^I \l^{Ji} \l^{Kj} \non\\
	&\qquad\qquad
	+ \frac{1}{4} Y^{\a\b} \phi^I \l^J{}_{\a}^i \l^K{}_{\b i}
	+ \frac{1}{4} W^{\a\b} \bar{\phi}^I \l^J{}_{\a}^i \l^K{}_{\b i} \Big)\non\\
	&\quad
	+ \frac{1}{48} \cF_{IJKL} (\l^{Ii} \l^{Jj}) (\l^K_i \l^L_j)  \ ,
\end{align}

\begin{align}
\cL_1 &=
	- 2\ri \cF \,\psib_a{}^i \tilde{\s}^a \r_i
	\non\\&\quad
	+  \cF_I \, \Big(
	(\tilde{\s}^{ab})^{\ad\bd} \tilde{\cD}_a (\psib_b{}^{\ad i} \bar{\l}^I{}^\bd_i)
	-\hf \tilde{\cD}^a (\psib_a{}^k \bar{\l}^I_k)
	\non\\&\qquad\qquad
	-\hf \psi_{\a \ad}{}_{\g k} ( \ri \ve^{\a \g} \bar{\k}^{\ad k} \bar{\phi}^I
		+ \ri \ve^{\a \g} S^{kl} \bar{\l}^I{}^\ad_l
		+ \ri \ve^{\a\g} \bar{W}^{\ad\bd} \bar{\l}^I{}^k_\bd
		- \ri Y^{\a \g} \bar{\l}^I{}^{\ad k} )
	\non\\&\qquad\qquad
	- \hf \psib_{\a \bd} {}^j_\gd  ( \ri \ve^{\bd \gd} \k^\a_j \bar{\phi}^I
		- 2 \ri \ve^{\bd \gd} \r^\a_j \phi^I  + i \ve^{\bd \gd} S_{jk} \l^{I\a k}
		- \ri \bar{W}^{\bd \gd} \l^I{}^\a_j + \ri \ve^{\bd \gd} Y^{\a \b} \l^I{}_{\b j} ) \Big)
	 \non\\&\quad
	+ \cF_{IJ}\, \Big( \frac{1}{4} \ve_{abcd} (\s^a)_\a{}^\ad f^{Ibc} \psi^d{}^\a_k \bar{\l}^J{}^k_\ad
	- \frac{1}{4} \ve_{abcd} (\tilde{\s}^a)_\ad{}^\a f^{Ibc} \psib^d{}^\ad_k \l^J{}^k_\a
	+ \hf \psi_a{}^j_\a \l^I{}^\a_j \tilde{\cD}^a \bar{\phi}^J
	\non\\&\qquad\qquad
	+ (\s^{ab})_\a{}^\b \psi_a{}^j_\b \l^I{}^\a_j \tilde{\cD}_b \bar{\phi}^J
	-\hf \psib_{\a\ad}{}_\bd^j ( - \ri \ve^{\ad\bd} S_{jk} \l^{I\a k} \phi^J
	+ \ri  \bar{W}^{\ad\bd} \l^I{}^\a_j \phi^J
	\non\\&\qquad\qquad
	- \ri \ve^{\ad\bd} Y^{\a\b} \l^I_{\b j} \phi^J
	+ \ri \ve^{\ad\bd} \bar{S}_{jk} \l^{I\a k} \bar{\phi}^J
	- \ri \ve^{\ad\bd} W^{\a\b} \l^I_{\b j} \bar{\phi}^J
	+ \ri \bar{Y}^{\ad\bd} \l^I{}^\a_j \bar{\phi}^J) \Big)
	\non\\&\quad
	- \frac{\ri}{12}  \cF_{IJK} (\psib_a{}^k  \tilde{\s}^a \l^{Ij}) (\l^J_j \l^K_k)\ ,
\end{align}

\begin{align}
\cL_2 &=
	 \cF\, \psib_c{}_\gd^k \psib_d{}^l_\dd (2 (\tilde{\s}^{cd})^{\gd \dd} S_{kl}
		- \ve^{\gd \dd} \ve_{kl} (\tilde{\s}^{cd})_{\ad \bd} \bar{W}^{\ad \bd}
		- \ve^{\gd \dd} \ve_{kl} (\s^{cd})_{\a \b} Y^{\a \b})
	\eol & \quad
	+ \cF_I \,\Big(\frac{\ri}{2} (\s^{[a})_\a{}^\ad \psi^{b]}{}^\a_k \tilde{\cD}_a \psib_b{}^k_\ad \bar{\phi}^I
		+ \frac{\ri}{2} (\s^{[a})_\a{}^\ad \tilde{\cD}_a \psi_b{}^\a_k \psib^{b]}{}^k_\ad \bar{\phi}^I
		+ \frac{1}{4} \ve^{abcd} \tilde{\cD}_b (\psi_c{}_k \s_a \psib_d{}^k) \bar{\phi}^I
	\eol & \qquad\qquad
	- \ri (\psi_a{}_k \s^b \psib^a{}^k) \tilde{\cD}_b \bar{\phi}^I
	- \ri (\psi_a{}_k \s^b \psib_b{}^k) \tilde{\cD}^a \bar{\phi}^I
	+ \ri (\psi_a{}_k \s^a \psib_b{}^{k}) \tilde{\cD}^b \bar{\phi}^I
	\eol & \qquad\qquad
	- \hf \ve^{abcd} (\psi_a{}_k \s_b  \psib_c{}^k) \tilde{\cD}_d \bar{\phi}^I
	+ \frac{1}{4} \psi_c{}^\g_k \psi_d{}^\d_l (2 (\s^{cd})_{\g\d} S^{kl} \bar{\phi}^I
	+ \ve_{\g\d} \ve^{kl} (\tilde{\s}^{cd})_{\ad\bd} \bar{W}^{\ad\bd} \bar{\phi}^I
	\eol & \qquad\qquad
	+ \ve_{\g\d} \ve^{kl} (\s^{cd})_{\a\b} Y^{\a\b} \bar{\phi}^I)
	+ \frac{1}{4} \psib_c{}_\gd^k \psib_d{}^l_\dd (2 (\tilde{\s}^{cd})^{\gd\dd} \bar{S}_{kl} \bar{\phi}^I
	+ \ve^{\gd \dd} \ve_{kl} (\s^{cd})_{\b \g} W^{\b\g} \bar{\phi}^I
	\eol & \qquad\qquad
	+ \ve^{\gd \dd} \ve_{kl} (\tilde{\s}^{cd})_{\ad \bd} \bar{Y}^{\ad \bd} \bar{\phi}^I
	- 4 (\tilde{\s}^{cd})^{\gd\dd} S_{kl} \phi^I
	+ 2 \ve^{\gd \dd} \ve_{kl} (\tilde{\s}^{cd})^{\ad \bd} \bar{W}_{\ad\bd} \phi^I
	\eol & \qquad\qquad
	+ 2 \ve^{\gd \dd} \ve_{kl} (\s^{cd})_{\b \g} Y^{\b\g} \phi^I )
	+ \frac{\ri}{2} \ve^{abcd} \tilde{F}_{ab}^I \psib_c \psib_d \Big)
	\eol & \quad
	+ \frac{1}{4} \cF_{IJ} \Big(\psi_a{}^j_\a \psib^a{}^k_\gd \l^I{}^\a_j \bar{\l}^J{}^\gd_k
	+ (\tilde{\s}^{ab})_{\ad\bd} \psi_a{}^\a_{(k} \psib_b{}^\ad_{l)} \l^I{}^k_\a \bar{\l}^{J\bd l}
	+ (\s^{ab})_{\a\b} \psi_a{}^\a_{(k} \psib_b{}^\ad_{l)} \l{}^{I\b k} \bar{\l}^J{}^l_\ad
	\eol & \qquad\qquad
	+ \ri \ve_{abcd} \tilde{F}^{Iab} \psi^c \psi^d \bar{\phi}^J
	- \psi_a{}^\a_k \psi_b{}^\b_l \big(\hf (\s^{ab})_{\a\b} \bar{\l}{}^{Ik} \bar{\l}^{Jl}
	+ \frac{1}{4} \ve_{\a\b} \ve^{kl} (\tilde{\s}^{ab})^{\ad\bd} \bar{\l}^I{}^j_\ad \bar{\l}^J_{\bd j}\big)
	\eol & \qquad\qquad
	- \ri \ve_{abcd} \tilde{F}^{Iab} \psib^c \psib^d \phi^J
	- \psib_c{}_\gd^k \psib_d{}^l_\dd \big(\hf (\tilde{\s}^{cd})^{\gd \dd} \l^I_k \l^J_l
	+ \frac{1}{4} \ve^{\gd \dd} \ve_{kl} (\s^{cd})_{\b \g} \l^{I\b i} \l^J{}_i^\g\big) \Big)\ ,
\end{align}

\begin{align}
\cL_3 &= \frac{1}{8} \cF_{IJ} \Big( \ve^{abcd} (\bar{\l}^{Ik} \ts_a  \psi_b{}_k) (\psi_c \psi_d) \bar{\phi}^J
	+ \ve^{abcd} (\l^{Ik} \s_a \psib_b{}_k) (\psib_c \psib_d) \phi^J
	\eol & \qquad\qquad
	- \ve^{abcd} (\l^{Ik} \s_a \psib_b{}_k) (\psi_c \psi_d) \bar{\phi}^J
	- \ve^{abcd} (\bar{\l}^{Ik} \tilde{\s}_a \psi_b{}_k) (\psib_c \psib_d ) \phi^J \Big)\ ,
\end{align}

\begin{align}
 \cL_4 &= - \frac{\ri}{4} \cF\, \ve^{abcd} (\psib_a \psib_b) (\psib_c \psib_d) 
	\eol & \quad
	+ \cF_I \Big(\frac{\ri}{4} \ve^{abcd} (\psib_a \psib_b) (\psib_c \psib_d) \phi^I
	- \frac{\ri}{8} \ve^{abcd} (\psi_a \psi_b) (\psib_c \psib_d) \bar{\phi}^I \Big) 
	\eol & \quad
	- \cF_{IJ} \Big( \frac{\ri}{16} \ve^{abcd} (\psi_a \psi_b) (\psi_c \psi_d) \bar{\phi}^I \bar\phi^J
	+ \frac{\ri}{16} \ve^{abcd} (\psib_a{} \psib_b) (\psib_c \psib_d) \phi^I \phi^J
	\eol & \qquad\qquad
	- \frac{\ri}{8} \ve^{abcd} (\psi_a \psi_b) (\psib_c \psib_d) \phi^I \bar{\phi}^J \Big)  \ .
\end{align}

The use of torsion-free covariant derivatives $\tilde{\cD}_a$ and the way
we have grouped terms in the Lagrangian makes it easy
to verify that in the case where $\cF(\cW) = \cW^2$, the action is real.

One can check that the components $S^{ij}$, $Y_{\a\b}$, $\rho_{\a i}$ and $C$ intrinsic
to $\rm SU(2)$ superspace drop out of the action if we restrict to the superconformal
case ({\it i.e.} $\cF$ of degree two). This is easy to understand.
If we transform the action \eqref{GrimmF} to the manifestly
superconformal framework, it becomes
\begin{align}
S_c = \int \rd^4x\, \rd^4\q\, \cE\, \Phi^2 \cF (\cW^I / \Phi)~,
\end{align}
where $\Phi$ is the chiral compensator intrinsic to $\rm SU(2)$ superspace.
In the case where $\cF$ is of degree two, the dependence on the compensator
vanishes and so the additional component fields associated with it
must certainly vanish from the component action.

%%%%%%%%%%%%%%%%%%%%%%%%%%%%%%%%%%%%%%%%%%%%%%%%%%%%%%

\subsection{The tensor multiplet in components}\label{TensorMultipletComponents}

As discussed in subsection \ref{TensorMultipletGeometry}, the
tensor multiplet is described by a real primary superfield $\cG^{ij}$ of
dimension 2 \eqref{Gweight} satisfying the constraint \eqref{Gconstraint}.
The corresponding 3-form field strength $H$ in superspace is given by
\eqref{3formConstraint} and \eqref{3formCurvs}.
Within the superfield $\cG^{ij}$ are found the matter components
of the tensor multiplet: a real isotriplet field $G^{ij}$,
a fermion $\chi_{\a i}$ and a complex scalar $F$,
\begin{subequations}
\begin{alignat}{2}
G^{ij} &:= \cG^{ij}|~, \\
\c_{\a i} &:= \frac{1}{3} \nabla_\a^j \cG_{i j}| ~, &\quad 
	\bar{\c}^{\ad i} &:= \frac{1}{3} \bar{\nabla}^\ad_j \cG^{ij}|~, \\
F &:=  \frac{1}{12} \nabla^{ij} \cG_{ij}|~, &\quad
\bar{F} &:=  \frac{1}{12} \bar{\nabla}^{ij} \cG_{ij}| \ .
\end{alignat}
\end{subequations}
The remaining component field, the two-form, is given by $b_{mn} := B_{mn}\lc$.
Owing to the superspace identity
\begin{align}
\nabla_\a^i \nabla_\b^j \cG^{kl} &= - \frac{1}{6} \ve_{\a\b} \ve^{i (k } \ve^{l ) j} \nabla^{pq} \cG_{pq} \ ,
\end{align}
there are no other independent component fields.

In constructing the three-form $H_{mnp} = 3 \partial_{[m} B_{np]}$
we can make use of the superspace identity
\begin{align}
H_{mnp} = E_m{}^A E_n{}^B E_p{}^C H_{ABC} (-)^{ab+ac+bc}
\end{align}
projected to lowest component. Defining
\begin{align}
h_{mnp} := H_{mnp}\lc = 3 \partial_{[m} b_{np]}
\end{align}
it is easy to show that
\begin{align}\label{h_eqn}
h_{mnp} = \eps_{mnpq} \tilde H^a\lc e_a{}^q
	+ 3 \ri (\sigma_{[mn})_\a{}^\b \psi_{p]}{}_j^\a \chi^j_\b
	+ 3 \ri (\ts_{[mn})^\ad{}_\bd \bar\psi_{p]}{}^j_\ad \bar\chi^\bd_j
	- 3 (\sigma_{[m})_{\a\ad} \psi_{n}{}^{\a i} \bar\psi_{p]}{}^{\ad j} G_{ij}
\end{align}
or, equivalently,
\begin{align}
\tilde{h}^a := \tilde{H}^a| &=  \frac{1}{6} \veps^{abcd} H_{bcd}\lc
	= \hf \ve^{abcd} (\frac{1}{3} h_{bcd}
	- \ri (\s_{cd})_\a{}^\b \psi_b{}^\a_k \chi_\b^k \non\\
	&- \ri (\tilde{\s}_{cd})^\ad{}_\bd \psib_b{}^k_\ad \bar{\chi}^\bd_k
	+  (\s_b)_\a{}^\bd \psi_c{}^\a_k \psib_d{}^l_\bd G^k{}_l ) \ .
\end{align}

We have emphasized the construction of the two-form multiplet
completely geometrically, but it is worth noting that, as
discussed in subsection \ref{TensorMultipletGeometry}, the two-form multiplet can be
encoded in a chiral superfield $\Psi$. As shown in \cite{Muller:ChiralActions},
by making use of the gauge transformations \eqref{gaugeT},
one can choose all components of $B_{AB}$ to vanish except for
$B_{ab}|$ by imposing the component constraints\footnote{The
third constraint is not actually necessary to eliminate the other
components of the two-form, but it does substantially simplify
the component evaluation later.}
\begin{align}\label{eq_PsiConstraints}
\Psi \lc = 0~, \quad
\nabla_\a^i \Psi \lc = 0~, \quad
\nabla^{ij} \Psi \lc = \bar\nabla^{ij} \bar\Psi \lc~.
\end{align}
One may easily construct $b_{mn}$ using
$b_{mn} = e_m{}^a e_n{}^b B_{ab}\lc$.
As usual, the supersymmetry transformation laws of the component
fields may be derived by using the constraints. 

%%%%%%%%%%%%%%%%%%%%%%%%%%%%%%%%%%%%%%%%%%%%%%%%%%%%%%

\subsection{Tensor multiplet action}\label{TensorMultAction}

The most general actions involving the self-couplings of tensor
multiplets are naturally constructed in projective superspace.
However, as was shown in \cite{Siegel85} (see \cite{Butter:2010jm}
for the curved superspace generalization), such actions
can always be constructed from the chiral action\footnote{This action in superspace appeared
in \cite{Siegel83} for the flat case, and was generalized to
curved superspace in \cite{Muller:ChiralActions}. Its component
form, coupled to conformal supergravity, was given in
\cite{deWvHVP2, deWPV}. A gauge fixing relates this to
the component linear multiplet Lagrangian constructed in
\cite{BS}.}
\begin{align}\label{PsiW}
S = \int \rd^4x\, \rd^4\q\, \cE\, \Psi \cW + \HC
\end{align}
where $\Psi$ is the prepotential for the tensor multiplet
and $\cW$ is the chiral field strength for some (possibly
composite) vector multiplet.
This action is invariant under the transformation \eqref{gaugeT}
which ensures that only the physical components of the tensor
multiplet appear. At the component level, it corresponds
to a supersymmetric generalization of the $b\wedge f$ topological
action, where $f_{mn}$ is the two-form
field strength of the vector multiplet $\cW$. This
model can describe self-couplings of tensor multiplets if
a suitable composite vector multiplet can be constructed
out of the tensor multiplets.

Using the gauge conditions on $\Psi$ \eqref{eq_PsiConstraints},
the component action \eqref{CA} is
\begin{align}
S &= \int \rd^4x \,e \,\Bigg( F \phi + \chi^\a_i \l_\a^i + \frac{1}{8} G^{ij} X_{ij}
	-\frac{1}{4} \eps^{mnpq} b_{mn} f_{pq} \non\\
&- \frac{\ri}{2} \psib^{\ad \a}{}^i_\ad ( 2 \chi_{\a i} \phi + G_{ij} \l_\a^j ) + (\tilde{\s}^{cd})^{\gd \dd} \psib_c{}_\gd^k \psib_d{}^l_\dd G_{kl} \phi \Bigg) + {\rm c.c.}  \label{PsiWAction}
\end{align}
This can be compared with the results in \cite{deWvHVP2, deWPV, Muller:ChiralActions}.

In the case of the improved tensor multiplet \cite{deWPV, LR:ScalarTensor},
$\cW$ is a composite field, which we denote by $\mathbb W$:
\be
\mathbb W = - \frac{1}{24 \cG} \bar{\nabla}_{ij} \cG^{ij} + \frac{1}{36 \cG^3} \bar{\nabla}_{\ad k} \cG^{ki} \bar{\nabla}^\ad_l \cG^{lj} \cG_{ij} \ ,
\ee
where $\cG^2 = \hf \cG^{ij} \cG_{ij}$. This multiplet was first reconstructed in curved
superspace in \cite{Muller:ChiralActions}, based on the results of
\cite{Siegel85, deWPV}.
The components $\phi$, $\l_\a^i$, and $X_{ij}$ of this composite
vector multiplet are given by
\begin{subequations}
\begin{align}
\phi = \cW| &= - \frac{1}{2 G} \bar{F}+ \frac{1}{4 G^3} \bar{\chi}^i \bar{\chi}^j G_{ij}~,\\
\l_\a^i = \nabla_\a^i \cW| &= \frac{1}{G} (\ri \nabla_{\a\ad} \bar{\chi}^{\ad i}
	- W_{\a\b} \chi^{\b i}  - 3 \S_{\a j} G^{ij}) \non\\
	&\quad
	+ \frac{1}{2 G^3} (\bar{F} \chi_{\a j} G^{ij} - \ri \tilde{h}_{\a\ad} \bar{\chi}^\ad_j G^{ij} - \ri \nabla_{\a\ad} G^{ij} \bar{\chi}^{\ad k} G_{jk} + \chi_{\a j} \bar{\chi}^i_\ad \bar{\chi}^{\ad j}) \non\\
	&\quad
	- \frac{3}{4 G^5} \chi_{\a j} \bar{\chi}^k_\ad \bar{\chi}^{\ad l} G_{kl} G^{ij} ~,\\
X^{ij} = \nabla^{ij} \cW| &= \frac{1}{G} \Big(- 2 \Box G^{ij} - 6 G^{ij} D + 6 \bar{\S}_\bd^{(i} \bar{\chi}^{\bd j)} + 6 \S^{\b (i} \chi^{j)}_\b\Big) \non\\
	&\quad + \frac{1}{G^3} \Big( -6 \S^{\b k} \chi_\b^l G_{kl} G^{ij} - 6 \bar{\S}_\bd^k \bar{\chi}^{\bd l} G_{kl} G^{ij} - W^{\a\b} \chi_\a^k \chi_{\b k} G^{ij} + \bar{W}_{\ad\bd} \bar{\chi}^{\ad k} \bar{\chi}^\bd_k G^{ij}
	\non\\&\qquad \qquad
	- 2 \ri \bar{\chi}^{\ad k} G_k{}^{(i} \nabla_{\a \ad} \chi^{\a j)} - 2 \ri \chi^\a_k G^{k (i} \nabla_{\a\ad} \bar{\chi}^{\ad j)}
	+ \nabla^a G^{ik} \nabla_a G^{jl} G_{kl}
	\non\\&\qquad \qquad
	+ \tilde{h}^a \tilde{h}_a G^{ij} + G^{ij} F \bar{F} + \bar{\chi}^i_\ad \bar{\chi}^{\ad j} F + \chi^{\a i} \chi^j_\a \bar{F}
	- 2 \tilde{h}^a \nabla_a G^{k (i} G^{j)}{}_k
	\non\\&\qquad \qquad
	+ \ri \tilde{h}^{\a\ad}  \chi_\a^{(i} \bar{\chi}_\ad^{j)} - \frac{\ri}{2} \nabla^{\a\ad} G^{k (i} \chi_{\a k} \bar{\chi}_\ad^{j)} - \frac{\ri}{2} \nabla^{\a \ad} G^{k (i} \chi_\a^{j)} \bar{\chi}_{\ad k} \Big) \non\\
	&\quad  + \frac{1}{G^5} \Big( - \frac{3}{2} F G^{ij} \bar{\chi}_\ad^k \bar{\chi}^{\ad l} G_{kl} - \frac{3}{2} \bar{F} G^{ij} \chi^{\a k} \chi_\a^l G_{kl} + 3 \ri \tilde{h}_{\a\ad} \chi^\a_k \bar{\chi}^{\ad}_l G^{kl}  G^{ij}
	\non\\&\qquad \qquad
	- 3 \ri \nabla_{\a\ad} G^{k (i}  G^{j) l} \chi^\a_{(l} \bar{\chi}^{\ad}_{q)} G_k{}^q - 3 \chi^{\a i} \chi^j_\a \bar{\chi}_\ad^k \bar{\chi}^{\ad l} G_{kl}
	\non\\&\qquad \qquad
	 - 3 \chi^{\a k} \chi_\a^l \bar{\chi}^i_\ad \bar{\chi}^{\ad j} G_{kl}
	+ 3 \chi^\a_k \chi_\a^{(i} \bar{\chi}^{j)}_\ad \bar{\chi}_l^\ad G^{kl} \Big)\non\\
	&\quad  + \frac{15}{4 G^7} G^{ij} \chi^{\a k} \chi_\a^l G_{kl} \bar{\chi}_\ad^p \bar{\chi}^{\ad q} G_{pq} ~,
\end{align}
\end{subequations}
where
\begin{subequations}
\begin{align}
\nabla_a G_{ij} &\equiv \nabla'_a G^{ij} - \psi_a{}^{\g(i} \chi_\g^{j)} + \psi_a{}^{\gd (i} \bar{\chi}_{\gd}^{j)} \ , \\
\nabla_a \bar{\chi}^{\ad i} &\equiv \nabla'_a \bar{\chi}^{\ad i} + \frac{\ri}{2} \psi_a{}^{\g i} \tilde{h}_\g{}^\ad
	+ \frac{\ri}{2} \psi_a{}^\g_k \nabla_\g{}^\ad G^{ik}
	+ \hf \psib_a{}^{\ad i} \bar{F} + 2 \phi_a{}^\ad_j G^{ij} \ , \\
\Box G^{ij} &\equiv \hf \nabla'_a \nabla^a G^{ij} - \psi_a{}^{\a(i} \nabla^a \chi_\a^{j)}
	- \frac{\ri}{2} \psi^{\a\ad}{}^{(i}_\a \bar{W}_{\ad\bd} \bar{\chi}^{\bd j)} 
	- \frac{3 \ri}{4} \psi^{\a\ad}{}_{\a k} \bar{\S}_\ad^k G^{ij}
	\eol & \quad
	+ \frac{3 \ri}{4} \psi^{\a\ad}{}_\a^{(i} \bar{\S}_{\ad k} G^{j) k}
	+ \frac{3 \ri}{4} \psi^{\a\ad}{}_{\a k} \bar{\S}_\ad^{(i} G^{j) k}
	\eol & \quad
	+ 2 \mathfrak f_a{}^a G^{ij} - \ri \phi_{\a\ad}{}^{\a(i} \bar{\chi}^{\ad j)} + {\rm c.c.} 
\end{align}
\end{subequations}

There is one additional component field in the composite vector multiplet:
the gauge connection $v_m$. Its field strength $f_{mn}$ is contained within
the supercovariant field strength $\hat F_{ab}$ \eqref{covF}, which possesses
the self-dual part
\begin{align}
\hat F_{\a\b} &=  - \frac{1}{8} \nabla_{\a\b} \cW|  - \hf W_{\a\b} \bar{\cW}|   \eol
	&= - \frac{1}{4} \nabla_{(\a \ad} \Big( \frac{1}{\cG} \tilde{H}_{\b)}{}^\ad
	- \frac{\ri}{9 \cG^3} \nabla_{\b)}^k \cG_k{}^i \bar{\nabla}^\ad_l \cG^{lj} \cG_{ij} \Big)|
	+ \frac{1}{16 G^3} \nabla_{\a \ad} G_{k}{}^i \nabla_{\b}{}^\ad G^{jk} G_{ij}
	\eol & \quad
	+ \frac{1}{8 G} \chi^\g_k \nabla^k_\g W_{\a\b} \lc - \frac{1}{16 G} G_{ij} \nabla^{ij} W_{\a\b} \lc ~.
\end{align}
This is a rather complicated result. The last two terms must be evaluated
to lowest components, while the remaining terms involve a supercovariant
derivative $\nabla_a$. Evaluating this explicitly requires the identities
\begin{align}
K_a \tilde{H}_b &= 0 \ , \quad S^\g_k \tilde{H}_a| = - 3 \ri (\s_a)^{\g \ad} \bar{\chi}_{\ad k} \ , \quad \bar{S}_\gd^k \tilde{H}_a| = 3 \ri (\s_a)_{\a\gd} \chi^{\a k} \ , \eol
\nabla_\a^i \tilde{H}_{\b\bd}| &= - \nabla_{\a\bd} \chi_\b^i + \ve_{\a\b} \nabla_{\g \bd} \chi^{\g i} + \ri \ve_{\a\b} \bar{W}_{\bd\gd} \bar{\chi}^{\gd i} + 3 \ve_{\a\b} \bar{\S}_{\bd j} G^{ij} \ .
\end{align}

However, it is easiest to specify the field strength $f_{mn}$ directly
and reconstruct the supercovariant two-form from \eqref{covF} if needed.
Using the results of section \ref{ProjectiveComponents}, we will show
that\footnote{The second term in \eqref{ImpTensorF} transforms non-covariantly
under $\rm SU(2)$ transformations, but this is cancelled by the transformation
of the explicit $\rm SU(2)$ connection in $\Gamma_m$.}
\begin{subequations}
\begin{align} \label{ImpTensorF}
f_{mn} &= 2 \partial_{[m} \Gamma_{n]}
	+ \frac{1}{4 G^3} \partial_{m} G^{ik} \partial_{n} G_k{}^{j} G_{ij}~, \\
\Gamma_m &:= 
	\frac{1}{2G} \phi_{m}{}^{ij} G_{ij} + \frac{1}{2G} e_{m}{}^a \tilde h_a
	+ \frac{1}{4G} (\psi_{m}{}^j \chi_j) + \frac{1}{4G} (\bar\psi_{m}{}^j \bar\chi_j)
	+ \frac{\ri}{4 G^3} (\chi^i \sigma_{m} \bar\chi^j) G_{ij}~.
\end{align}
\end{subequations}
As observed in \cite{deWPV}, there is no $\rm SU(2)$ covariant
expression for the composite one-form $v_m$ from which we could
construct $f_{mn}$. Although the first term in $f_{mn}$ is exact,
the second is not. We will see in the next section an explanation within
projective superspace for this behavior.\footnote{Within harmonic superspace,
this feature and its origin were discussed in the rigid supersymmetric case
\cite{GIO_Tensor, GIOS}.}

\section{Projective superspace and component reductions}\label{ProjectiveComponents}

Recently, a projective superspace formulation of conformal
supergravity has been constructed \cite{KLRT-M08}.
Soon afterwards, it was shown how to perform
component reductions in this framework \cite{Kuzenko:2008ry},
but to our knowledge, this component reduction has not been applied
to any specific action yet. In this section, we will show how
this can be done explicitly. By first constructing the
projective superspace analogue of the action \eqref{PsiW},
we will show how to use the component reduction formula in projective
superspace to reproduce the component result \eqref{PsiWAction}.

The construction of \cite{KLRT-M08} was based on \cite{Grimm};
we will begin by generalizing it to the conformal superspace geometry
developed in \cite{Butter4D}.
After briefly describing the formulation of the vector and tensor
multiplets in projective superspace, we will proceed to the
component reduction of the projective action analogous
to \eqref{PsiW}.

\subsection{Formalism of projective superspace}
Following \cite{KLRT-M08} the supermanifold $\cM^{4|8}$ is augmented
with an additional $\rm \mathbb CP^1$ parametrized by an isotwistor
coordinate $v^i \in \mathbb C^2 \setminus \{0\}$.
Matter fields are constructed in terms of covariant projective multiplets
$\cQ^{(n)}(z,v)$ which are holomorphic in the isotwistor $v^i$
and of definite homogeneity, $\cQ^{(n)}(z, cv) = c^n \cQ^{(n)}(z, v)$,
on an open domain of $\mathbb C^2 \setminus \{0\}$. Such superfields
are intrinsically defined on $\rm \mathbb CP^1$.

Using the isotwistor coordinate, one may construct a subset of
anti-commuting spinor derivatives\footnote{Within \cite{KLRT-M08},
$\rm SU(2)$ superspace and the corresponding covariant derivatives $\cD_A$ were
used to construct projective multiplets, but the generalization
to the superconformal covariant derivative $\nabla_A$ is entirely straightforward.}
\begin{gather}
\nabla_\a^+ := v_i \nabla_\a^i~,\qquad
\bar\nabla_\ad^+ := v_i \bar\nabla_\ad^i~,\qquad
\{\nabla_\a^+, \bar\nabla_\ad^+\} = \{\nabla_\a^+, \bar\nabla_\b^+\} = 0~.
\end{gather}
The $+$ on these operators denotes that they are of degree $+1$ in $v^i$.
Projective superfields are required to be analytic with respect to these
derivatives,
\begin{align}\label{QProjCond}
\nabla_\a^+ \cQ^{(n)} = \bar\nabla_\ad^+ \cQ^{(n)} = 0~.
\end{align}
Since these superfields depend on only half of the Grassmann
coordinates of superspace, they play the same fundamental role in
$\cN=2$ supersymmetric theories as chiral superfields play in
$\cN=1$.  Using such fields, one may construct an action principle \cite{KLRT-M08}
\begin{align}\label{ProjAction}
S = \frac{1}{2\pi} \oint_C v^{i} \rd v_i \int \rd^4x\, \rd^4\q\, \rd^4\bar\q \, E\, 
	\frac{\cW_0 \bar \cW_0}{(\Sigma_0^{++})^2} \cL^{++}~,\qquad E = \textrm{Ber}(E_M{}^A) \ ,
\end{align}
with $\Sigma^{++}_0 := (\nabla^+)^2 \cW_0 / 4= (\bar\nabla^+)^2 \bar \cW_0 / 4$. The
Lagrangian $\cL^{++}(z,v)$ is a covariant real projective multiplet of weight-two.
The vector multiplet $\cW_0$ is used here essentially as a compensator, and the form
of the action simplifies if the gauge $\cW_0=1$ is taken (or, equivalently,
if all objects are redefined in terms of $\cW_0$ as in section \ref{GrimmGeometry}).
This action appears to depend on the choice of $\cW_0$, but one can show
that this is purely illusory. The role of the compensator
is simply to allow us to write the action over the full set of
superspace coordinates.

As discussed in \cite{KLRT-M08}, it is useful in the evaluation of this
action to introduce an additional fixed isotwistor $u_i$ which obeys
$v^i u_i \neq 0$ along the contour of integration.
In terms of $v^i$ and $u_i$, one may
introduce the objects\footnote{In \cite{KLRT-M08}, $u_i^-$ was defined as 
$u_i$ so that $u^{i+} u_i^- \neq 1$. We find it more convenient to normalize $u_i^-$.}
\begin{align}
u^{i+} = v^i~,\qquad u_i^- = \frac{u_i}{v^k u_k} \ ,
\end{align}
which obey $u^{i+} u_i^- = 1$. Associated with these are derivative operations
\begin{align}
D^{++} = (v^j u_j) v_i\frac{\partial}{\partial u_i}~,\qquad
D^0 = v^i \frac{\partial}{\partial v^i} - u_i \frac{\pa}{\pa u_i}~,\qquad
D^{--} = \frac{1}{(v^j u_j)} u^i\frac{\partial}{\partial v^i}~.
\end{align}
The charges on the derivative operators and on $u_i^\pm$
denote their homogeneity in $v^i$. Each is defined to be of
degree \emph{zero} in $u_i$. All other operators and fields which we
introduce will similarly be degree zero in $u_i$ but of some
fixed homogeneity in $v^i$.

We should note that the derivative operators may be defined formally as
\begin{align}
D^{++} = u_i^+ \frac{\partial}{\partial u_i^-}~,\qquad
D^0 = u^{i+} \frac{\partial}{\partial u^{i+}} - u_{i}^- \frac{\partial}{\partial u_{i}^-}~,\qquad
D^{--} = u^{i -} \frac{\partial}{\partial u^{i+}} \ ,
\end{align}
acting on the variables $u^{i+}$ and $u_i^-$. Fields and operators
of definite homogeneity in $v^i$ can be reinterpreted as possessing
definite $D^0$ charge. Written in this form, the operators superficially
resemble the corresponding objects defined in harmonic superspace \cite{GIOS}.
For this reason we will frequently refer to $u_i^\pm$ as ``harmonics'',
but their real origin should be kept in mind.

Just as superspace geometrizes supersymmetry, the auxiliary $\rm \mathbb CP^1$
geometrizes isospin. The components $q^i$ of an isospinor are naturally
associated with a weight-one isotwistor $q^+ = q^i u_i^+$. The 
operator $\lambda^{jk} J_{jk}$ acts on $q^i$ as
$\lambda^{jk} J_{jk} q^i = \lambda^i{}_j q^j$. In order to mimic
this action on $q^+$, we identify
\begin{align}
J_{ij} = -u_i^+ u_j^+ D^{--} + u_{(i}^+ u_{j)}^- D^0 + u_i^- u_j^- D^{++}~,
\end{align}
leading to
\begin{align}
\lambda^{ij} J_{ij} q^+ = (-\lambda^{++} D^{--} + \lambda^{+-} D^0 + \lambda^{--} D^{++}) q^+
	= -\lambda^{++} q^- + \lambda^{+-} q^+ = \lambda^+{}_j q^j~,
\end{align}
where we have denoted $\lambda^{++} = \lambda^{ij} u_i^+ u_j^+$,
$\lambda^{+-} = \lambda^{ij} u_i^+ u_j^-$ and so on. 
The above definition similarly works for $q^-$, as well as
for objects with any number of $\rm SU(2)$ indices.

This construction becomes nontrivial when we consider \emph{general}
functions $\cQ^{(n)}$ on $\rm \mathbb CP^1$ of fixed homogeneity which are defined
only in certain regions. They need not be polynomial in $v^i$
and furnish infinite dimensional representations of $\rm SU(2)$.
Such functions are called ``arctic'' if they are well-defined in the
northern chart of $\rm \mathbb CP^1$ and ``antarctic''
if they are well-defined in the southern chart. An arctic multiplet $\U^{(n)}$
is necessarily complex, and is related to an antarctic multiplet
$\breve \U^{(n)}$ by the smile conjugation operation \cite{KLRT-M08}.

Along with the operators $\nabla_\a^+$ and $\bar\nabla_\ad^+$,
we introduce $\nabla_\a^-$ and $\bar\nabla_\ad^-$ constructed
using $u_i^-$. Their algebra is easily found by contracting
the harmonics $u_i^\pm$ with eq. \eqref{CSGAlgebra}.
Of particular interest is the isotwistor reformulation
of the algebra of $S^\a_i$ with $\nabla_\b^j$:
\begin{subequations}
\begin{align}
\{S_\alpha^\pm, \nabla_\beta^\pm\} = \pm 4 \veps_{\alpha \beta} D^{\pm\pm}, \qquad
\{\bar S^{\dalpha \pm}, \bar \nabla^{\dbeta \pm}\} = \mp 4 \veps^{\dalpha \dbeta} D^{\pm\pm}~, \\
\{S_\alpha^\mp, \nabla_\beta^\pm\} = \pm (2 \veps_{\alpha \beta} \mathbb D - 4 M_{\alpha \beta}
     - \veps_{\alpha \beta} Y) - 2 \veps_{\alpha \beta} D^0~, \\
\{\bar S^{\dalpha \mp}, \bar \nabla^{\dbeta \pm}\} = \mp (2 \veps^{\dalpha \dbeta} \mathbb D + 4 \bar M^{\dalpha \dbeta}
     + \veps^{\dalpha \dbeta} Y) + 2 \veps^{\dalpha \dbeta} D^0~.
\end{align}
\end{subequations}
These are quite powerful equations. They tell us that if a superfield
$\cQ^{(n)}$ obeys the analyticity condition \eqref{QProjCond} and is also primary,
then it must obey the further conditions implied by
$\{S_\beta^+, \nabla_\a^+\} \cQ^{(n)} = \{S_\beta^-, \nabla_\a^+\} \cQ^{(n)} = 0$
and their complex conjugates.
These conditions lead to\footnote{These relations were first discussed by Kuzenko
for the case of globally superconformal multiplets in 4D \cite{Kuzenko:SPH},
extending the work in 5D \cite{K:5D_compact}.
The generalization to the local case appeared in \cite{KLRT-M08}.}
\begin{align}
Y \cQ^{(n)} = 0~,\qquad \mathbb D\cQ^{(n)} = D^0 \cQ^{(n)} = n \cQ^{(n)}~,\qquad
D^{++} \cQ^{(n)} = 0~.
\end{align}
In particular, the last condition implies that $\cQ^{(n)}$ must be
independent of $u_i$.

An important property of the contour integral is the vanishing of a
total $D^{--}$ derivative,
\begin{align}
\oint v^{i} \rd v_i D^{--} \cF = 0
\end{align}
where $\cF$ is an arbitrary weight-zero function of $v^i$ and $u_i$.
A simple proof is given in appendix D of \cite{Butter:2010jm}.

\subsection{Component reductions in projective superspace}
The reduction of the general projective action \eqref{ProjAction}
to components was carried out in \cite{Kuzenko:2008ry} using 
$\rm SU(2)$ superspace. It is straightforward
to convert the result given there to one involving the
superconformally covariant derivative $\nabla_A$. The action \eqref{ProjAction}
reduces to
\begin{align}
S = \frac{1}{2\pi} \oint u^{i+} \rd u_i^+ \int \rd^4x\, e\, \cL^{--}~,\qquad e = \mathrm{det}(e_m{}^a)~,
\end{align}
where the isotwistor-dependent component Lagrangian $\cL^{--}$
depends on the auxiliary isotwistor $u_i$ in the combination
$u_i^- = u_i / (v^k u_k)$. We have written the contour measure in
terms of $u^{i+}$ to emphasize its degree of homogeneity in $v^i$.
As discussed in \cite{Kuzenko:2008ry}, the action is independent
of the choice of the isotwistor $u_i$; it is subject only to
the requirement that $v^k u_k \neq 0$ along the integration contour.

The full expression for the component Lagrangian is
\begin{align}\label{Component_L--}
\cL^{--} &= \frac{1}{16} (\nabla^-)^2 (\bar\nabla^-)^2 \cL^{++}
	- \frac{\ri}{8} (\bar\psi_m^- \ts^m)^\alpha \nabla_\alpha^- (\bar\nabla^-)^2 \cL^{++}
	- \frac{\ri}{8} (\psi_m^- \sigma^m)_\dalpha \bar\nabla^{\dalpha -} (\nabla^-)^2 \cL^{++}
	\eol & \quad
	- \frac{1}{4} \Big(
	(\psi_m^- \sigma^{mn})^\alpha \bar\psi_n{}^{\dalpha-}
	+ \psi_m{}^{\alpha-}  (\ts^{mn}\bar\psi_n^-)^\dalpha
	- \ri \phi_m{}^{--} (\sigma^m)^{\dalpha \alpha} \Big) [\nabla_\alpha^{-}, \bar\nabla_\dalpha^{-}]  \cL^{++}
	\eol & \quad
	- \frac{1}{4} (\psi_m^- \sigma^{mn} \psi_n^-) (\nabla^-)^2 \cL^{++}
	- \frac{1}{4} (\bar\psi_m^- \ts^{mn} \bar\psi_n^-) (\bar \nabla^-)^2 \cL^{++}
	\eol & \quad
	- \Big(
	\frac{1}{2} \eps^{mnpq} (\psi_m^- \sigma_n \bar\psi_p^-) \psi_q{}^{\alpha -}
	- 2 (\psi_m^- \sigma^{mn})^\alpha \phi_n{}^{--} \Big) \nabla_\alpha^- \cL^{++}
	\eol & \quad
	+ \Big(
	\frac{1}{2} \eps^{mnpq} (\bar\psi_m^- \ts_n \psi_p^-) \bar\psi_q{}_\dalpha^-
	- 2 (\bar\psi_m^- \ts^{mn})_\dalpha \phi_n{}^{--} \Big) \bar\nabla^{\dalpha -} \cL^{++}
	\eol & \quad
	- 3 \ve^{mnpq} (\psi_m^- \sigma_n \bar\psi_p^-) \phi_q{}^{--} \cL^{++}~,
\end{align}
where projection to components is implicitly applied to all covariant
spinor derivatives of $\cL^{++}$. The auxiliary isotwistor $u_i$
appears explicitly in the expressions
\begin{gather}
\nabla_\a^- := u_i^- \nabla_\a^i~,\qquad \bar\nabla_\ad^- := u_i^- \bar\nabla_\ad^i~,\qquad \\
\psi_m{}_\a^- := u_i^- \psi_m{}_\a^i~, \qquad
\bar\psi_m{}_\ad^- := u_i^- \bar\psi_m{}_\ad^i~, \qquad
\phi_m{}^{--} := u_i^- u_j^- \phi_m{}^{ij} \ ,
\end{gather}
as well as implicitly in the definition of the $\rm SU(2)$ connection
within the covariant derivative. Note that in addition to the gravitino,
the $\rm SU(2)$ connection $\phi_m{}^{--}$ appears \emph{explicitly}
in the action, and for much the same reason: they both correspond to
connections on an auxiliary manifold which is integrated over --
a Grassmann manifold in the case of the gravitino and a $\rm \mathbb CP^1$
manifold in the case of the isospin connection.

In later sections, we will also use the following notations:
\begin{align}
\psi_m{}_\a^+ := u_i^+ \psi_m{}_\a^i~, \qquad
\bar\psi_m{}_\ad^+ := u_i^+ \bar\psi_m{}_\ad^i~.
\end{align}

\subsection{Abelian vector and tensor multiplets in projective superspace}
Projective superspace provides a natural realization for both the $\cN=2$
abelian vector multiplet and the $\cN=2$ tensor multiplet. We briefly review
their constructions here.

The tensor multiplet $\cG^{ij}$ is naturally associated with a
projective multiplet $\cG^{++}$ of weight-two, which is required to
be defined everywhere on $\rm \mathbb CP^1$. This last condition
implies that $\cG^{++} = \cG^{ij} u_i^+ u_j^+$ and then the
analyticity condition \eqref{QProjCond} becomes equivalent to
\eqref{Gconstraint}.

The abelian vector multiplet structure is a bit more intricate.
Recall that it is described by a chiral
superfield $\cW$ obeying the Bianchi identity \eqref{BianchiYM}.
The chirality condition and the Bianchi identity are naturally satisfied
by the definition
\begin{align}\label{Wproj}
\cW := \frac{1}{8 \pi} \oint u^{i+} \rd u_i^+ (\bar\nabla^-)^2 \cV~.
\end{align}
$\cV$ is a real weight-zero projective multiplet, which serves
as a prepotential for the $\cN=2$ abelian vector multiplet. The 
reduced chirality of this expression was demonstrated in
\cite{Kuzenko:Dual, KT-M-AdS} using $\rm SU(2)$ superspace,
but the extension to conformal superspace is straightforward.
The prepotential $\cV$ possesses a gauge transformation
\begin{align}
\cV \rightarrow \cV + \Lambda + \breve \Lambda \ ,
\end{align}
where $\Lambda$ is an arctic multiplet of weight zero and
$\breve \Lambda$ is its smile conjugate. One can check that
$\cW$ is invariant under this transformation \cite{Kuzenko:Dual, KT-M-AdS}.
In the Minkowski case, one can further show that $\cV$ decomposes into separate
prepotentials respectively for the $\cN=1$ abelian vector and chiral
multiplets, along with an infinite set of unconstrained
$\cN=1$ multiplets which are pure gauge degrees of freedom \cite{LR:SYM}.

\subsection{Component evaluation}
General superconformal two-derivative actions
involving several tensor multiplets $\cG_A^{++}$ are naturally constructed in
projective superspace by choosing a Lagrangian $\cL^{++}$ which
is an analytic function of $\cG_A^{++}$ of degree one. It
was shown in \cite{Siegel85, Butter:2010jm} how to relate this to the
formulation \eqref{PsiW}.
Such models of several tensor multiplets coupled to conformal
supergravity were also considered at the component level in \cite{deWS}.

For a single tensor multiplet, there is a unique action in
projective superspace known as the improved tensor multiplet action,\footnote{
The improved tensor multiplet action coupled to supergravity appeared
for the first time in \cite{Kuzenko:Dual}. The flat analogue appeared
originally in \cite{KLR}.}
\begin{align}\label{ImprovedTensorAction}
\cL^{++} = \cG^{++} \ln (\cG^{++} /\ri \U^+ \breve \U^+) \ ,
\end{align}
where $\U^+$ is a weight-one arctic multiplet.
As discussed in \cite{Kuzenko:Dual, Kuzenko:2008ry}, the appearance of the arctic multiplet
in the action has no effect on the physics.\footnote{This action
resembles the $\cN=1$ improved tensor action, $\cL = G \ln (G / \Phi \bar \Phi)$.
The multiplet $\Phi$ is chiral but has no physical effect on the action.}
Under the redefinition $\U^+ \rightarrow \U^+ e^{-\Lambda}$
where $\Lambda$ is a weight-zero arctic multiplet,
the action changes as
$\cL^{++} \rightarrow \cL^{++} + \cG^{++} (\Lambda + \breve \Lambda)$.
This additional term was shown in \cite{Kuzenko:2008ry} to vanish by
inspection of the component action.

Given this Lagrangian, we may immediately apply the component
reduction formula \eqref{Component_L--}. In principle,
this is a straightforward process and for superconformal tensor
multiplet models the contour integral can (at least in principle)
be performed since there are only a finite number of degrees
of freedom.

Here we take an approach analogous to \eqref{PsiW}.
The Lagrangian \eqref{ImprovedTensorAction} can be understood as
\begin{align}\label{Projective_GV}
\cL^{++} = \cG^{++} \cV \ ,
\end{align}
where $\cV$ is a weight-zero projective multiplet. The gauge invariance
$\cV \rightarrow \cV + \Lambda + \breve \Lambda$ allows the
identification of $\cV$ as the projective prepotential for
an abelian vector multiplet. In this case, it is a \emph{composite}
vector multiplet prepotential and completely equivalent to the action
\eqref{PsiW}.

In this section, we undertake the evaluation of the action \eqref{Projective_GV}
using the component Lagrangian \eqref{Component_L--}, so that
the comparison to \eqref{PsiWAction} is completely clear.
That is, we will organize terms in such a way that the explicit contour
integral is eliminated, with the resulting component
expression depending only on $\cG^{ij}$ and the physical
components of $\cV$. Only then will we identify $\cV$ as a composite
multiplet.
This will serve as a useful test for the component
reduction rule \eqref{Component_L--}, which can be used
in principle for cases (such as models with arctic multiplets) where other methods of
component reduction are prohibitively difficult (or non-existent).
Along the way, we will also discover a useful result about
how to construct the one-form $v_m$ from the projective
prepotential $\cV$.

This is a nontrivial calculation involving a number of integrations
by parts. It is also, to our knowledge, the first \emph{direct}
application of the component reduction rule \eqref{Component_L--}.
For this reason, we will present a detailed summary of the calculation.
First, we will present the full calculation in the \emph{globally}
superconformal case to emphasize the ideas which are important
in organizing the action. Then we will present a summary of the 
locally superconformal calculation
emphasizing the techniques which are needed to evaluate it fully.

\subsubsection{Globally superconformal action}
In evaluating the component Lagrangian \eqref{Component_L--}
corresponding to the projective Lagrangian \eqref{Projective_GV},
there is a single term in the absence of conformal supergravity:
\begin{align}
\cL^{--} = \frac{1}{16} (D^-)^2 (\bar D^-)^2 (\cG^{++} \cV)~.
\end{align}
Nevertheless, its evaluation is not entirely trivial.
One begins by constructing a number of identities for the
$\cN=2$ tensor multiplet:
\begin{subequations}
\begin{align}
(\bar D^-)^2 \cG^{++} &= (\bar D^+)^2 \cG^{--}~, \\
D_\a^- (\bar D^-)^2 \cG^{++} &= D_\a^- (\bar D^+)^2 \cG^{--}
	= 4\ri \partial_{\a \bd} \bar D^{\bd+} \cG^{--}~, \\
(D^-)^2 (\bar D^-)^2 \cG^{++} &= (D^-)^2 (\bar D^+)^2 \cG^{--}
	= 16 \Box \cG^{--}~.
\end{align}
\end{subequations}
In addition, using the components of the tensor multiplet as defined
in section \ref{TensorMultipletComponents}, one can show that
\begin{subequations}
\begin{align}
D_\a^- \cG^{++}\vert  &= -2 \chi_\a^+~,\\
(\bar D^-)^2 \cG^{++} \vert &= 4 \bar F~, \\
D_\a^- \bar D_{\bd}^- \cG^{++}\vert &= -2\ri \partial_{\a\bd} G^{+-} + 2\ri \tilde H_{\a \bd}~,
\end{align}
\end{subequations}
with the obvious definitions $\chi_\a^+ = \chi_\a^i u_i^+$
and $G^{+-} = G^{ij} u_i^+ u_j^-$.
In this section, we will abuse notation and frequently use $\chi_\a^+$, $\bar F$, 
$\tilde H_{\a\ad}$, and so on to describe both the component fields as well 
as the superfields with these corresponding lowest components.

Applying these rules, it is straightforward to show that
\begin{align}
\cL^{--} &= \Box \cG^{--} \cV
     + \ri \partial_{\a \bd} \bar\chi^{\bd -} D^{\alpha -} \cV
     + \ri \partial_{\a \bd} \chi^{\a -} D^{\bd -} \cV
	+ \frac{1}{4} F (\bar D^-)^2 \cV
	+ \frac{1}{4} \bar F (D^-)^2 \cV
     \eol & \quad
     -\frac{\ri}{2} \partial_{\alpha \dalpha} \cG^{-+} D^{\alpha -} \bar D^{\dalpha -} \cV
     + \frac{\ri}{2} \tilde H_{\alpha \dalpha} D^{\alpha -} \bar D^{\dalpha -} \cV
     - \frac{1}{4} \chi^{\alpha +} D_{\alpha}^- (\bar D^-)^2 \cV
     + \frac{1}{4} \bar\chi_{\dalpha}^+ \bar D^{\dalpha -} (D^-)^2 \cV
     \eol & \quad
     + \frac{1}{16} \cG^{++} (D^-)^2 (\bar D^-)^2 \cV~.
\end{align}
Integrating by parts and rearranging terms, we find
\begin{align}
\cL^{--} &=
	\frac{1}{4} F (\bar D^-)^2 \cV
	+ \frac{1}{4} \bar F (D^-)^2 \cV
	+ \frac{\ri}{4} \tilde H_{\alpha \dalpha} [D^{\alpha -}, \bar D^{\dalpha -}] \cV
     \eol & \quad
	+ \cG^{--} \Box \cV
     + \frac{\ri}{2} \cG^{-+} \partial_{\alpha \dalpha} D^{\alpha -} \bar D^{\dalpha -} \cV
     + \frac{1}{16} \cG^{++} (D^-)^2 (\bar D^-)^2 \cV
     \eol & \quad
     - \frac{1}{4} \chi^{\alpha +} D_{\alpha}^- (\bar D^-)^2 \cV
     + \frac{1}{4} \bar\chi_{\dalpha}^+ \bar D^{\dalpha -} (D^-)^2 \cV
	\eol & \quad
     - \ri \bar\chi^{\bd -} \partial_{\a \bd} D^{\alpha -} \cV
     - \ri \chi^{\a -} \partial_{\a \bd} D^{\bd -} \cV~.
\end{align}
To simplify further, we need the identity
\begin{align}
     \frac{1}{16} \cG^{++} (D^-)^2 (\bar D^-)^2 \cV
	&= -\frac{1}{16} \cG^{--} (D^+)^2 (\bar D^-)^2 \cV
	+ \frac{1}{16} \cG^{++} D^{\a -} D_\a^+ (\bar D^-)^2 \cV
	\eol & \quad
	+ \frac{1}{16} \cG^{++} D^{\a +} D_\a^- (\bar D^-)^2 \cV
	+ \frac{1}{16} \cG^{ij} D_{ij} (\bar D^-)^2 \cV~.
\end{align}
Then using the property that $\cV$ is annihilated by $D_\a^+$, we find
\begin{align}
     \frac{1}{16} \cG^{++} (D^-)^2 (\bar D^-)^2 \cV
	&= 
	- \cG^{--} \Box \cV
     - \frac{\ri}{2} \cG^{-+} \partial_{\alpha \dalpha} D^{\alpha -} \bar D^{\dalpha -} \cV
	+ \frac{1}{16} \cG^{ij} D_{ij} (\bar D^-)^2 \cV~.
\end{align}
Similarly, we can rewrite
\begin{align}
- \frac{1}{4} \chi^{\alpha +} D_{\alpha}^- (\bar D^-)^2 \cV
	&= - \frac{1}{4} \chi^{\alpha -} D_{\alpha}^+ (\bar D^-)^2 \cV
	- \frac{1}{4} \chi^{\alpha j} D_{\alpha j} (\bar D^-)^2 \cV \eol
	&= \ri \chi^{\a -} \partial_{\a \bd} D^{\bd -} \cV
	+ \frac{1}{4} \chi^{\alpha}_j D_{\alpha}^j (\bar D^-)^2 \cV~.
\end{align}
Together these lead to
\begin{align}
\cL^{--} &=
	\frac{1}{4} F (\bar D^-)^2 \cV
	+ \frac{1}{4} \bar F (D^-)^2 \cV
	+ \frac{\ri}{4} \tilde H_{\alpha \dalpha} [D^{\alpha -}, \bar D^{\dalpha -}] \cV
     \eol & \quad
	+ \frac{1}{16} \cG^{ij} D_{ij} (\bar D^-)^2 \cV
	+ \frac{1}{4} \chi^{\alpha}_j D_{\alpha}^j (\bar D^-)^2 \cV
	+ \frac{1}{4} \bar\chi_\ad^j \bar D^\ad_j (D^-)^2 \cV~.
\end{align}
Applying the contour integral and taking the lowest component,
$\cL = \dfrac{1}{2\pi} \oint u^{i+} \rd u_i^+ \cL^{--}\vert$, leads to
\begin{align}
\cL &= \Big[F \cW
	+ \bar F \bar \cW
	- 2 \tilde H^m v_m
	+ \frac{1}{4} G^{ij} (D_{ij} \cW)
	+ \chi^{\alpha}_j (D_{\alpha}^j \cW)
	+ \bar\chi_\ad^j (\bar D^\ad_j \bar\cW)\Big]\vert \eol
	&= F \phi
	+ \bar F \bar \phi
	- 2 \tilde h^m v_m
	+ \frac{1}{4} G^{ij} X_{ij}
	+ \chi^{\alpha}_j \lambda_\alpha^j
	+ \bar\chi_\ad^j \bar\lambda^\ad_j~,
\end{align}
where $\cW$ is given by \eqref{Wproj} and we have defined
\begin{align}\label{flat_A}
v_m := -\frac{\ri}{8} (\ts_m)^{\dalpha \alpha} \oint \frac{u^{i+} \rd u_i^+}{2\pi}
	[D_\alpha^-, \bar D_\dalpha^-] \cV\vert~.
\end{align}
It is straightforward to check that $f_{mn}$, defined by the flat
superspace limit of eq. \eqref{covF}, is
consistent with $f_{mn} = \partial_m v_n - \partial_n v_m$.
Then this action may be easily compared with the flat space limit of \eqref{PsiWAction};
these differ only by an integration by parts which converts the $h \wedge v$ term
into a $b \wedge f$ term.

Having constructed the globally superconformal component action,
our next task is to generalize it to the locally superconformal case.
In principle, this is a straightforward (if tedious) calculation
making use of \eqref{Component_L--}, but before summarizing that approach,
we will briefly discuss how to generalize the equation \eqref{flat_A}
to the locally superconformal case.

\subsubsection{Construction of the abelian vector multiplet one-form}\label{ConstructionOneForm}
At leading order, we expect $v_m$ to be given (up to an exact piece which
we cannot determine) by
\begin{align}
v_m := -\frac{\ri}{8} (\ts_m)^{\dalpha \alpha} \oint \frac{u^{i+} \rd u_i^+}{2\pi} [\nabla_\alpha^-, \bar \nabla_\dalpha^-] \cV\vert + \cdots
\end{align}
We can deduce the missing terms by imposing the requirement of
$S$-supersymmetry invariance of $v_m$. Observing that
\begin{align}
S_{\beta j} [\nabla_\alpha^-, \bar\nabla_\dalpha^-] \cV
	&= 8 \eps_{\beta \alpha} u_j^- \bar\nabla_\dalpha^- \cV \ ,
\end{align}
we may add a gravitino term to counter this. The gravitino
transforms homogeneously under $S$-supersymmetry,
\begin{align}
S_{\beta j} \psi_m{}^\alpha_i = 0~, \qquad
S_{\beta j} \bar\psi_{m}{}_\dalpha^i = -2\ri (\sigma_m)_{\beta \dalpha} \delta_j{}^i \ ,
\end{align}
so we guess
\begin{align}
v_m := -\frac{\ri}{8} \oint \frac{u^{i+} \rd u_i^+}{2\pi}
	\Big((\ts_m)^{\dalpha \alpha} [\nabla_\alpha^-, \bar \nabla_\dalpha^-] \cV
	- 4\ri \psi_m{}^{\alpha -} \nabla_\alpha^- \cV
	+ 4\ri \bar\psi_m{}_\dalpha^- \bar\nabla^{\dalpha -} \cV + \cdots
	\Big)\vert~.
\end{align}
These new terms introduce additional terms since
\begin{align}
S_{\beta k} \nabla_\alpha^- \cV = -4 u_k^+ \eps_{\beta \alpha} D^{--} \cV~, \qquad
S_{\beta k} \bar\nabla_\dalpha^- \cV = 0~.
\end{align}
These can be cancelled by observing that the $\rm SU(2)$
connection transforms homogeneously under $S$-supersymmetry,
\begin{align}
S_{\beta k} \phi_m{}^i{}_j = 2 \delta^i_k \psi_{m \beta j} - \delta^i_j \psi_{m \beta k} \ .
\end{align}
So we may choose
\begin{align}
v_m &:= -\frac{\ri}{8} \oint \frac{u^{i+} \rd u_i^+}{2\pi}
	\Big((\ts_m)^{\dalpha \alpha} [\nabla_\alpha^-, \bar \nabla_\dalpha^-] \cV
	- 4\ri \psi_m{}^{\alpha -} \nabla_\alpha^- \cV
	+ 4\ri \bar\psi_m{}_\dalpha^- \bar\nabla^{\dalpha -} \cV
	+ 8\ri \phi_m{}^{-+} D^{--} \cV
	\Big)\vert \eol
	&= \oint \frac{u^{i+} \rd u_i^+}{2\pi}
	\Big(-\frac{\ri}{8} (\ts_m)^{\dalpha \alpha} [\nabla_\alpha^-, \bar \nabla_\dalpha^-] \cV
	- \frac{1}{2} \psi_m{}^{\alpha -} \nabla_\alpha^- \cV
	+ \frac{1}{2} \bar\psi_m{}_\dalpha^- \bar\nabla^{\dalpha -} \cV
	- \phi_m{}^{--} \cV
	\Big)\vert~, \label{v_formula}
\end{align}
where we integrated $D^{--}$ by parts under the contour integral.
This expression is fully invariant under $S$-supersymmetry.

Having postulated an expression for $v_m$, we should check that it
is actually sensible. Since it should generate a field strength
$f_{mn}$ that is an $\rm SU(2)$ invariant, the expression
for $v_m$ should transform under an $\rm SU(2)$ transformation
at most by an exact form. We begin by writing
\begin{align}
v_m = \oint \frac{u^{i+} \rd u_i^+}{2\pi} \mathcal V_m^{--} \vert~.
\end{align}
To avoid the inhomogeneous term from the connection piece, let's consider
for the moment a global $\rm SU(2)$ transformation under which
\begin{align}
\delta \mathcal V_m^{--} = \lambda^{ij} J_{ij} \mathcal V_m^{--}
	= \left(-\lambda^{++} D^{--} + \lambda^{-+} D^0 + \lambda^{--} D^{++} \right)\mathcal V_m^{--}~.
\end{align}
Under the integral, one finds
\begin{align}
\delta v_m = \oint \frac{u^{i+} \rd u_i^+}{2\pi} \lambda^{--} D^{++} \mathcal V_m^{--}\vert \ ,
\end{align}
which is nonzero precisely because $\mathcal V_m^{--}$ depends on the isotwistor $u_i$.
We find
\begin{align}
D^{++} \mathcal V_m^{--}\lc &=
	e_m{}^a \nabla_a \cV\lc
	- \frac{1}{2} \psi_m{}^{\alpha +} \nabla_\alpha^- \cV\lc
	+ \frac{1}{2} \bar\psi_m{}_\dalpha^{+} \bar\nabla^{\dalpha -} \cV\lc
	+ \phi_m{}^{++} D^{--} \cV\lc ~.
\end{align}
Now observe that
\begin{align}
e_m{}^a \nabla_a \cV\lc
	&= \partial_m \cV\lc + \frac{1}{2} \psi_m{}^{\alpha +} \nabla_\alpha{}^- \cV\lc
	- \frac{1}{2} \bar\psi_m{}_{\dalpha}^+ \bar\nabla^{\dalpha -} \cV\lc
\	- \phi_m{}^{++} D^{--} \cV\lc \ ,
\end{align}
and so we see immediately that $D^{++} \mathcal V_m^{--} \lc= \partial_m \cV\lc$.
Thus,
\begin{align}\label{deltaA_global}
\delta v_m = \oint \frac{u^{i+} \rd u_i^+}{2\pi} \lambda^{--} \partial_m \cV\vert~.
\end{align}
If we allow $\lambda^i{}_j$ to be a local gauge transformation, the only missing piece
is the inhomogeneous term from the $\rm SU(2)$ connection. That extra piece is
easily found to be
\begin{align}
-\oint \frac{u^{i+} \rd u_i^+}{2\pi} \delta \phi_m{}^{--} \cV\lc
	= \oint \frac{u^{i+} \rd u_i^+}{2\pi} \partial_m \lambda^{--} \cV\lc~,
\end{align}
which gives
\begin{align}
\delta v_m = \partial_m \oint \frac{u^{i+} \rd u_i^+}{2\pi} \lambda^{--} \cV\vert~.
\end{align}
We observe that our expression for $v_m$ is actually \emph{not} $\rm SU(2)$ invariant,
but it \emph{does} generate an $\rm SU(2)$ invariant field strength $f_{mn}$.

Next, we observe that $v_m$ explicitly depends on the choice of the auxiliary
isotwistor $u_i$ (as noted above). However, $f_{mn}$ should not, so $v_m$
should transform into an exact form under small deformations of $u_i$.
Let us consider a shift
$u_i \rightarrow u_i + \delta u_i$ where $\delta u_i$ is a constant. Due to
the linear independence of $v_i$ and $u_i$, it is possible to
parametrize\footnote{Here we follow very closely the approach used in
\cite{Kuzenko:2008ry}, but with slight changes in notation.}
\begin{align}\label{ushift}
\delta u_i = v^k u_k v_i \alpha^{--} + u_i \beta ~,\qquad
\alpha^{--} := -\frac{u^i \delta u_i}{(v^k u_k)^2} ~,\qquad \beta := \frac{v^i \delta u_i}{v^k u_k}~.
\end{align}
We have chosen normalizations so that $\a^{--}$ and $\b$, which are respectively
of degree $-2$ and degree $0$ in $v^i$, are both
of degree zero in $u_i$. Now for any function $\cF(v,u)$ which is degree zero
in $u_i$,
\begin{align}
u_i \frac{\partial}{\partial u_i} \cF = 0 \ ,
\end{align}
one can easily show that
\begin{align}
\delta \cF = \a^{--} D^{++} \cF \ ,
\end{align}
under the shift \eqref{ushift}.
Applying this to the definition for $v_m$ leads immediately via arguments
we made above \eqref{deltaA_global} to
\begin{align}
\delta v_m = \oint \frac{u^{i+} \rd u_i^+}{2\pi} \a^{--} \partial_m \cV\vert
	= \partial_m \oint \frac{u^{i+} \rd u_i^+}{2\pi} \a^{--} \cV\vert \ ,
\end{align}
since $\a^{--}$ is constant in $x$. As required, the expression for $v_m$ is
independent of $u_i$ up to a gauge transformation.

Now let us verify that $v_m$ indeed generates the correct field strength $f_{mn}$.
Using the definition \eqref{FmnFAB}, it is straightforward to calculate
\begin{align}
f_{mn} = 2 \oint \frac{u^{i+} \rd u_i^+}{2\pi} &\Bigg[
	- \nabla'_{[m} \Bigg(
	\frac{\ri}{8} (\ts_{n]})^{\ad \a} [\nabla_\a^-, \bar\nabla_\ad^-] \cV\lc
	+ \frac{1}{2} \psi_{n]}{}^{\a -} \nabla_\a^- \cV\lc
	- \frac{1}{2} \bpsi_{n]}{}_\ad^{-} \bar\nabla^{\ad -} \cV\lc
	\Bigg)
	\eol & \quad
	- \partial_{[m} \phi_{n]}{}^{--} \cV \lc + \phi_{[m}{}^{- j} \phi_{n]j}{}^- \cV\lc
	\Bigg]~.
\end{align}
In the above expression, we have used
\begin{align}
\nabla'_m := \partial_m + \frac{1}{2}\omega_m{}^{ab} M_{ab} + \ri A_m Y + b_m \mathbb D + \phi_m{}^{ij} J_{ij}
\end{align}
as the derivative covariant with respect to Lorentz transformations, $\rm SU(2) \times U(1)$
transformations, and dilatations. However, only the $\rm SU(2)$ connection is
nontrivial when acting on the term in the first line.
If we extract that $\rm SU(2)$ connection and evaluate it directly, we find
\begin{align}
f_{mn} &= 2 \oint \frac{u^{i+} \rd u_i^+}{2\pi} 
		\partial_{[m} \cV_{n]}^{--} \lc = 2 \partial_{[m} v_{n]}~.
\end{align}

This explicit definition for $v_m$ is especially interesting since in the application
to models of one or more tensor multiplets, it is
natural for the composite field strength $f_{mn}$ which appears
to be closed but not exact \cite{deWPV}.
As an example, let us attempt to calculate $v_m$ for the case
$\cV = \log(\cG^{++} / \ri \U^+ \breve \U^+)$. In calculating
$v_m$, there will be terms dependent on the arctic multiplet
and its conjugate plus those dependent on just $\cG^{++}$.
We expect the dependence on the arctic multiplet to yield an
exact term, so we focus only on the terms involving $\cG^{++}$. It is
straightforward to calculate
\begin{align}
\mathcal V_m^{--}\vert &=
	\frac{1}{G^{++}}
	\Big(\partial_m G^{-+} - e_m{}^a \tilde H_a
		- \frac{1}{2} (\psi_m{}^j \chi_j)
		- \frac{1}{2} (\bar\psi_m{}^j \bar\chi_j)
		- \phi_m{}^{ij} G_{ij}\Big)
	\eol & \quad
	- \frac{\ri}{(G^{++})^2} \chi^+ \sigma_m \bar\chi^+ \ .
\end{align}
A key feature of this expression is that only the first term
depends on the auxiliary isotwistor $u_i$.

If we perform the contour integral using the techniques discussed, for example,
in \cite{Butter:2010jm}, we find the useful formulae\footnote{The contour is performed so
that it encircles one of the roots of $G^{++}$ in a counter-clockwise
fashion.} 
\begin{align}
\frac{1}{2\pi} \oint u^{i+} \rd u_i^+ \frac{1}{G^{++}} = -\frac{1}{2} \frac{1}{G}~,\qquad
\frac{1}{2\pi} \oint u^{i+} \rd u_i^+ \frac{\Omega^{++}}{(G^{++})} = -\frac{1}{4} \frac{\Omega^{ij} G_{ij}}{G^3}~,
\end{align}
for an arbitrary $\Omega^{++} = \Omega^{ij} u_i^+ u_j^+$. This leads to
\begin{align}
v_m &= \frac{1}{G} \Big(
	\frac{1}{2} \phi_m{}^{ij} G_{ij} + \frac{1}{2} e_m{}^a \tilde H_a
	+ \frac{1}{4} (\psi_m{}^j \chi_j) + \frac{1}{4} (\bar\psi_m{}^j \bar\chi_j)\Big)
	+ \frac{\ri}{4 G^3} (\chi^i \sigma_m \bar\chi^j) G_{ij}
	\eol & \quad
	+ \frac{1}{2\pi} \oint u^{i+} \rd u_i^+ \frac{\partial_m G^{-+}}{G^{++}}~.
\end{align}
As expected there is a contribution to $v_m$ which is not invariant under
shifts in the auxiliary isotwistor $u_i$.
This naturally leads to a field strength $f_{mn}$ which is closed but not exact.
However, upon taking the exterior derivative of $v_m$, this dependence vanishes,
\begin{align}
f_{mn}
	&= 2 \partial_{[m} \Bigg(
	\frac{1}{2G} \phi_{n]}{}^{ij} G_{ij} + \frac{1}{2G} e_{n]}{}^a \tilde H_a
	+ \frac{1}{4G} (\psi_{n]}{}^j \chi_j) + \frac{1}{4G} (\bar\psi_{n]}{}^j \bar\chi_j)
	+ \frac{\ri}{4 G^3} (\chi^i \sigma_{n]} \bar\chi^j) G_{ij}\Bigg)
	\eol & \qquad
	- \frac{1}{2\pi} \oint u^{i+} \rd u_i^+  \frac{\partial_{m} G^{+j} \partial_{n} G_j{}^{+}}{(G^{++})^2} \ .
\end{align}
The remaining contour integral can then be performed to yield
\begin{align}
f_{mn} &= 2 \partial_{[m} \Bigg(
	\frac{1}{2G} \phi_{n]}{}^{ij} G_{ij} + \frac{1}{2G} e_{n]}{}^a \tilde H_a
	+ \frac{1}{4G} (\psi_{n]}{}^j \chi_j) + \frac{1}{4G} (\bar\psi_{n]}{}^j \bar\chi_j)
	+ \frac{\ri}{4 G^3} (\chi^i \sigma_{n]} \bar\chi^j) G_{ij}\Bigg)
	\eol & \qquad
	+ \frac{1}{4} \frac{\partial_{m} G^{ik} \partial_{n} G_k{}^{j} G_{ij}}{G^3}~.
\end{align}
The last term is closed but cannot be written as an exact form in an
$\rm SU(2)$-covariant way.

\subsubsection{Locally superconformal action}
Armed now with a reasonable expression for $v_m$, we may calculate the
locally superconformal action. First, we present the final result
\begin{align}\label{GVprojFinalResult}
\cL &= \frac{1}{4} G^{ij} X_{ij}
	+ \Big[F \phi + \chi^\alpha_j \lambda_\alpha^j + \HC\Big]
	- 2 \tilde h^m v_m
	\eol & \quad
	+ \Big[
		\ri (\psi_{m}{}^j \sigma^m  \bar\chi_j) \bar \phi
		- \frac{\ri}{2} (\psi_m{}^j \sigma^m \bar\lambda^k) G_{jk}
		- (\psi_m{}^i \sigma^{mn} \psi_n{}^j) G_{ij} \bar \phi
		+ \HC
	\Big] \ ,
\end{align}
which exactly matches \eqref{PsiWAction}, as it must, up to a total derivative.
We now sketch the derivation in the remainder of this section.
Readers uninterested in the details are urged to continue onward to
section \ref{VectorTensorSection}.

We begin by separating the terms in the component Lagrangian \eqref{Component_L--}
into four groups, which we denote $T_n$ depending on the number $n$ of explicit gravitinos:
\begin{subequations}
\begin{align}
T_0 &= \frac{1}{16} (\nabla^-)^2 (\bar\nabla^-)^2 \cL^{++}
	+ \frac{\ri}{4} \phi^{\dalpha \alpha --} [\nabla_\alpha^-, \bar\nabla_\dalpha^-] \cL^{++}~, \\
T_1 &= - \frac{\ri}{8} (\bar\psi_m^- \ts^m)^\alpha \nabla_\alpha^- (\bar\nabla^-)^2 \cL^{++} +
	2 (\psi_m^- \sigma^{mn})^\alpha \phi_n{}^{--} \nabla_\alpha^- \cL^{++} + \HC~, \\
T_2 &= -\frac{1}{4} (\psi_m^- \sigma^{mn} \psi_n^-) \cL^{++}
	- \frac{1}{4} (\psi_m^- \sigma^{mn})^\alpha \bar\psi_m{}^{\dalpha -} [\nabla_\a^-, \bar\nabla_\ad^-] \cL^{++}
	\eol & \quad
	- \frac{3}{2} \veps^{mnpq} \phi_q{}^{--} (\psi_m{}^i \sigma_n \bar\psi_p{}^j) \cL^{++}
	+ \HC~, \\
T_3 &= -\frac{1}{2} \veps^{mnpq} (\psi_m^- \sigma_n \bar\psi_p^-) \psi_q{}^{\alpha -} \nabla_\a^- \cL^{++} + \HC
\end{align}
\end{subequations}
We begin with $T_0$ and repeat essentially the same analysis we
performed in the flat case, only this time we must take curvatures
into account and we cannot dispense so easily with total derivatives.
We find
\begin{align}
T_0 &= \frac{1}{4} F (\bar\nabla^-)^2 \cV + \frac{1}{4} \bar F (\nabla^-)^2 \cV
     + \frac{\ri}{4} \tilde H_{\alpha \dalpha} [\nabla^{\alpha -}, \bar\nabla^{\dalpha -}] \cV
     + \frac{1}{16} \cG^{ij} \nabla_{ij} (\bar\nabla^-)^2 \cV
     \eol & \quad
     + \frac{1}{4} \chi^{\alpha}_j \nabla_{\alpha}^j (\bar\nabla^-)^2 \cV
     + \frac{1}{4} \bar\chi_{\dalpha}^j \bar\nabla^{\dalpha}_j (\nabla^-)^2 \cV
     \eol & \quad
     - \frac{1}{2} \nabla^{\beta +} W_{\beta \alpha} \cG^{--}\, \nabla^{\alpha -} \cV
     + \frac{1}{2} \bar\nabla_\dbeta^+ W^{\dbeta \dalpha} \cG^{--}\, \bar\nabla_{\dalpha}^- \cV
     \eol & \quad
     + \nabla^{\beta -} W_{\beta \alpha} \cG^{-+}\, \nabla^{\alpha -} \cV
     - \bar\nabla_\dbeta^- \bar W^{\dbeta \dalpha} \cG^{-+} \bar\nabla_{\dalpha}^- \cV 
     \eol & \quad
     + \chi^{\alpha -} \nabla^{\beta -} W_{\beta \alpha} \cV
     + \bar\chi_{\dalpha}^- \bar\nabla_{\dbeta}^- \bar W^{\dbeta \dalpha} \cV
     \eol & \quad
	- \phi^{\dalpha \alpha --}
	\left(
	\tilde H_{\alpha \dalpha} \cV
	- \nabla_{\alpha \dalpha} \cG^{-+} \cV
	- \frac{\ri}{2} \cG^{++} \nabla_\alpha^- \bar\nabla_\dalpha^- \cV
	+ \ri \chi_\alpha^+ \bar\nabla_\dalpha^- \cV
	+ \ri \bar\chi_\dalpha^+ \nabla_\alpha^- \cV
	\right)
     \eol & \quad
     + \nabla^{\dalpha \alpha} \left(
          \frac{1}{2} \cG^{--} \nabla_{\alpha \dalpha} \cV - \frac{1}{2} \nabla_{\alpha \dalpha}  \cG^{--} \cV
          -\frac{\ri}{2} \cG^{-+} \nabla_{\alpha}^- \bar\nabla_{\dalpha}^- \cV
          + \ri \bar\chi_{\dalpha}^- \nabla_{\alpha}^- \cV
          + \ri \chi_{\alpha}^- \bar\nabla_{\dalpha}^- \cV
          \right)~.
\end{align}
The first two lines match the flat space result, the next four contain
conformal supergravity corrections, and the last appears to be
a total derivative.
This, of course, is not actually the case. The vector derivative
$\nabla_a$ contains a spin-connection with torsion along with several
connections -- supersymmetry, $\rm SU(2)$, special conformal and $S$-supersymmetry
-- which do not vanish when we rewrite this as a surface term.
Let us demonstrate this explicitly.

Given an action of the form
$\int \rd^4x\, e\, \nabla_a V^a$
for some vector $V^a$, one can show that the expression can be rearranged
into the form
\begin{align}
\int \rd^4x\, \left(\nabla_m (e\, V^a e_a{}^m)
		- \frac{e}{2} \psi_a{}^{\beta}_j \nabla_\beta^j V^a
		- \frac{e}{2} \bar\psi_a{}_\dbeta^j \bar\nabla^\dbeta_j V^a
		+ e T_{m n}{}^b e_b{}^n V^a e_a{}^m
	\right)~,
\end{align}
where the torsion tensor is $T_{mn}{}^a = -\ri (\psi_{[mj} \sigma^a \bar\psi_{n]}^j)$.
Moreover, the full covariant derivative $\nabla_m$ still carries
connections (the special conformal and $S$-supersymmetry) which do not
necessarily annihilate $e\, V^a e_a{}^m$.

In our case, we have a slightly more complicated situation with
an auxiliary contour integration
\begin{align}
\frac{1}{2\pi} \oint u^{i+} \rd u_i^+ \int \rd^4x\, e\, \nabla_a V^{a --}~,
\end{align}
and so we must pay attention to not only the special conformal and
$S$-supersymmetry connections but also to the $\rm SU(2)$ connection within
$\nabla_a$. Taking all of these considerations into account, we find
(after performing a contour integration by parts)
\begin{align}
&\frac{1}{2\pi} \oint u^{i+} \rd u_i^+ \int \rd^4x\, e\, \nabla_a V^{a --} \eol
	&= \frac{1}{2\pi} \oint u^{i+} \rd u_i^+ \int \rd^4x\, e\,
	\Big(
	- \frac{1}{2} \psi_a{}^{\beta}_j \nabla_\beta^j V^a
	- \frac{1}{2} \bar\psi_a{}_\dbeta^j \bar\nabla^\dbeta_j V^a
	+ T_{m n}{}^b e_b{}^n V^a e_a{}^m
	\eol & \quad
	+ {\frak f}_a{}^b K_b V^{a --}
	+ \frac{1}{2} \phi_{a}{}_\beta^j S^\beta_{j} V^{a --}
	+ \frac{1}{2} \phi_a{}^\dbeta_j \bar S_\dbeta^{j} V^{a --}
	+ \phi_a{}^{--} D^{++} V^{a --}
	\Big)~.
\end{align}
We emphasize that the operator $D^{++}$ \emph{cannot} be integrated by parts under
the contour integral, since it involves a derivative with respect to the
auxiliary isotwistor $u_i$ which is not a variable of integration.

If we now integrate by parts the vector derivative in $T_0$, we find
\begin{align}
T_0 &= \frac{1}{4} F (\bar\nabla^-)^2 \cV + \frac{1}{4} \bar F (\nabla^-)^2 \cV
     + \frac{1}{16} \cG^{ij} \nabla_{ij} (\bar\nabla^-)^2 \cV
     \eol & \quad
     + \frac{1}{4} \chi^{\alpha}_j \nabla_{\alpha}^j (\bar\nabla^-)^2 \cV
     + \frac{1}{4} \bar\chi_{\dalpha}^j \bar\nabla^{\dalpha}_j (\nabla^-)^2 \cV
%      \eol & \quad
	+ \frac{\ri}{4} \tilde H_{\alpha \dalpha} [\nabla^{\alpha -}, \bar\nabla^{\dalpha -}] \cV
	- \tilde H_{\alpha \dalpha} \phi^{\dalpha \alpha --} \cV
     \eol & \quad
	+ \Bigg(\frac{1}{2}  \Psi^{\bd \a}{}_{\beta \bd}{}^{\beta -} \cG^{-+} \nabla_\a^- \cV
	- \frac{1}{4}  \Psi^{\bd \a}{}_{\beta \dbeta}{}^{\beta +} \cG^{--} \nabla_\a^- \cV
	- \frac{1}{2}  \Psi^{\bd \a}{}_{\beta \bd}{}^{\beta -} \chi_\a^- \cV
		\eol & \qquad 
		- \frac{\ri}{4} \psi_{\alpha \dbeta}{}^{\alpha +} \bar W^{\dbeta \dalpha} \cG^{--} \bar\nabla_\dalpha^- \cV
		+ \frac{\ri}{2} \psi_{\alpha \dbeta}{}^{\alpha -} \bar W^{\dbeta \dalpha} \cG^{-+} \bar\nabla_\dalpha^- \cV
		+ \frac{\ri}{2} \psi_{\alpha \dbeta}{}^{\alpha -} \bar W^{\dbeta \dalpha} \bar\chi_\dalpha^- \cV
	+ \HC \Bigg)
	\eol & \quad
	- \left(\frac{1}{2} \psi^{\dalpha \alpha}{}^\beta_j \nabla_\beta^j
	+ \frac{1}{2} \bar\psi^{\dalpha \alpha}{}_{\dbeta}^j \bar\nabla^\dbeta_j\right) \Omega_{\alpha \dalpha}^{--}
	- 2 T_{mn}{}^b e_b{}^n e_a{}^m \Omega^{a--}~.
\end{align}
It is easy to see that this result matches \eqref{GVprojFinalResult}
to zeroth order in the gravitinos.

If we next include the contribution of $T_1$, perform another
set of integrations by parts, and rewrite all expressions
involving $[\nabla_\a^-,\bar\nabla_\ad^-]\cV$ to
instead involve $\mathcal V_{\a \ad}^{--}$, we find
\begin{align}
T_0 + T_1
	&= \frac{1}{4} \cG^{ij} \nabla_{ij} \cW^{--}
	+ \Big[F \cW^{--} + \chi^\alpha_j \nabla_\alpha^j \cW^{--} + \HC\Big]
	\eol & \quad
	+ \tilde H^{\dalpha \alpha} \cV_{\alpha \dalpha}^{--}
	- \Big[2 (\psi_m{}^j \sigma^{mn} \chi_j) \cV_n^{--} + \HC \Big]
	\eol & \quad
	+ \Big[
		\ri \psi_{\alpha \dalpha}{}^{\alpha j} \bar\chi^{\dalpha}_j \bar \cW^{--}
		- \frac{\ri}{2} \psi^{\dalpha \alpha}{}_\alpha^j \cG_{jk} \bar\nabla_\dalpha^k \bar \cW^{--}
		+ \HC
	\Big]
	\eol & \quad
	+ R_1 \ ,
\end{align}
where we have introduced
$\cW^{--} := \frac{1}{4} (\bar\nabla^-)^2 \cV$
as convenient shorthand and collected all terms involving two or more gravitinos into the
remainder term $R_1$
\begin{align}
R_1 &=
	\frac{1}{4} \psi^{\dalpha \alpha \, \beta j} \chi_{\beta j}
		(\psi_{\alpha \dalpha}{}^{\gamma -} \nabla_\gamma^- \cV
		- \bar\psi_{\alpha \dalpha \, \dgamma}{}^- \bar\nabla^{\dgamma -} \cV)
	\eol & \quad
	- \frac{1}{2} \psi^{\dalpha \alpha \, \beta j} \chi_{\alpha j}
		(\psi_{\beta \dalpha}{}^{\gamma -} \nabla_\gamma^- \cV
		- \bar\psi_{\beta \dalpha \, \dgamma}{}^- \bar\nabla^{\dgamma -} \cV)
	\eol & \quad
	+ \frac{1}{2} \Bigg[(\psi_m{}^j \sigma^{mn})^\alpha \psi_n{}^\gamma_k \nabla_\gamma^k
	+ (\psi_m{}^j \sigma^{mn})^\alpha \bar\psi_n{}_\dgamma^k \bar\nabla^\dgamma_k
	\Bigg] \Lambda_{\alpha j}^{--}
	\eol & \quad
	+ \frac{1}{2} \veps^{mnpq} (\bar \Psi_{mn}{}^j \bsigma_p \psi_q{}^i)
		\left(
		\frac{1}{2} u_i^+ u_j^- \cG^{--} + \frac{1}{2} u_i^- u_j^+ \cG^{--}
		- 2 u_i^- u_j^- \cG^{-+}
		\right) \cV
	\eol & \quad
	+ \ri \veps^{mnpq} (\psi_m^- \sigma_{na})^\alpha T_{pq}{}^a
		\left(\cG^{-+} \nabla_\alpha^- \cV -\chi_\alpha^- \cV\right)
	- \frac{\ri}{2} \veps^{mnpq} (\psi_m^+ \sigma_{na})^\alpha T_{pq}{}^a (\cG^{--} \nabla_\alpha^- \cV)
	\eol & \quad
	+ \HC - 2 T_{nm}{}^b e_b{}^m e_a{}^n \Omega^{a --}~.
\end{align}
We have defined
\begin{align}
\Lambda_{\alpha j}^{--} = \frac{1}{2} u_j^- \chi_\a^- \cV
		+ \frac{1}{4} u_j^+ \cG^{--} \nabla_\a^- \cV
		- \frac{1}{2} u_j^- \cG^{-+} \nabla_\a^- \cV~,
\end{align}
to keep the expression for $R_1$ somewhat simpler.
One can easily check that the above expression is correct to first order
in gravitinos. The term $R_1$ involves only two gravitinos or more and
should be cancelled when additional contributions are included.

Now let us include the two gravitino term $T_2$. Performing another
set of integrations by parts leads to
\begin{align}
T_0 + T_1 + T_2
	&= \frac{1}{4} \cG^{ij} \nabla_{ij} \cW^{--}
	+ \Big[F \cW^{--} + \chi^\alpha_j \nabla_\alpha^j \cW^{--} + \HC\Big]
	\eol & \quad
	- \Big[2 \tilde H^q + 2 (\psi_m{}^j \sigma^{mq} \chi_j + \HC)
		- \veps^{mnpq} (\psi_m{}^i \sigma_n \bar\psi_p{}^j) \cG_{ij} \Big] \cV_q^{--}
	\eol & \quad
	+ \Big[
		\ri (\psi_{m}{}^j \sigma^m  \bar\chi_j) \bar \cW^{--}
		- \frac{\ri}{2} (\psi_m{}^j \sigma^m)_\dalpha \cG_{jk} \bar\nabla^{\dalpha k} \bar \cW^{--}
		+ \HC
	\Big]
	\eol & \quad
	- \Big[(\psi_m{}^i \sigma^{mn} \psi_n{}^j) \cG_{ij} \bar \cW^{--} + \HC \Big]
	\eol & \quad
	+ R_2 \ ,
\end{align}
where the remainder $R_2$ is a three-gravitino term:
\begin{align}
R_2 &= 
	\frac{1}{2} \veps^{mnpq} (\psi_m{}^i \sigma_n \bar\psi_p{}^j) \cG_{ij}
		\psi_m{}^{\alpha -} \nabla_\alpha^- \cV
	- \frac{1}{2}  \veps^{mnpq} (\psi_m{}^i \sigma_n \bar\psi_p{}^j)
		\psi_q{}^\alpha_k \nabla_\alpha^k \Lambda_{ij}^{--}
	\eol & \quad
	- \ri\veps^{mnpq} (\psi_m{}^i \sigma_{na})^\alpha T_{pq}{}^a
		\left(\frac{1}{2} u_i^+ \cG^{--} \nabla_\alpha^- \cV
			- u_i^- \cG^{-+} \nabla_\alpha^- \cV + u_i^- \chi_\alpha^- \cV
	\right) + \HC
\end{align}

Finally, we expand out $T_3$ and add it to $R_2$. This is a straightforward
exercise of performing $\rm SU(2)$ algebra with the harmonics, and one finds that
$R_2 + T_3 = 0$. So the full component Lagrangian is given by
\begin{align}
\cL^{--}
	&= \frac{1}{4} \cG^{ij} \nabla_{ij} \cW^{--}
	+ \Big[F \cW^{--} + \chi^\alpha_j \nabla_\alpha^j \cW^{--} + \HC\Big]
	\eol & \quad
	- \Big[2 \tilde H^q + 2 (\psi_m{}^j \sigma^{mq} \chi_j + \HC)
		- \veps^{mnpq} (\psi_m{}^i \sigma_n \bar\psi_p{}^j) \cG_{ij} \Big] \cV_q^{--}
	\eol & \quad
	+ \Big[
		\ri (\psi_{m}{}^j \sigma^m  \bar\chi_j) \bar \cW^{--}
		- \frac{\ri}{2} (\psi_m{}^j \sigma^m)_\dalpha \cG_{jk} \bar\nabla^{\dalpha k} \bar \cW^{--}
		+ \HC
	\Big]
	\eol & \quad
	- \Big[(\psi_m{}^i \sigma^{mn} \psi_n{}^j) \cG_{ij} \bar \cW^{--} + \HC \Big]~.
\end{align}
All terms have been rearranged so that the contour integrals can be performed.
Doing so, and taking the component projection leads to
\begin{align}
\cL &= \frac{1}{4} G^{ij} X_{ij}
	+ \Big[F \phi + \chi^\alpha_j \Sigma_\alpha^j + \HC\Big]
	- 2 \tilde h^m v_m
	\eol & \quad
	+ \Big[
		\ri (\psi_{m}{}^j \sigma^m  \bar\chi_j) \bar \phi
		- \frac{\ri}{2} (\psi_m{}^j \sigma^m)_\dalpha G_{jk} \bar\Sigma_\dalpha{}^k
		- (\psi_m{}^i \sigma^{mn} \psi_n{}^j) G_{ij} \bar \phi + \HC \Big]~,	
\end{align}
where we have identified the term multiplying $v_m$ using eq. \eqref{h_eqn}

After a laborious calculation, we have indeed recovered \eqref{PsiWAction}.
Of course, we \emph{had} to, since a straightforward superspace argument
guarantees that the projective superspace action constructed from
the Lagrangian $\cL^{++} = \cG^{++} \cV$ must match the chiral superspace
action constructed from $\cL_c = \Psi \cW$. What is important about this
calculation is a demonstration that we can explicitly evaluate projective
superspace actions in conformal supergravity \emph{without} first converting
them to $\cN=1$ language or by first converting them to a chiral (or some other)
Lagrangian.

%%%%%%%%%%%%%%%%%%%%%%%%%%%%%%%%%%%%%%%%%%%%%%%%%%%%%%
%%%%%%%%%%%%%%%%%%%%%%%%%%%%%%%%%%%%%%%%%%%%%%%%%%%%%%

\section{Lifting superspace results from components}\label{VectorTensorSection}

In the previous sections, we have emphasized the use of
superspace for the construction of supersymmetric actions and
the derivation of supersymmetry transformation rules.
However, this process can also be reversed. Given a consistent
set of component supersymmetry transformation rules for
some multiplet, there should be a superfield formulation
obeying some set of constraints which corresponds to that
multiplet. In principle, the supersymmetry transformations may
be used to identify the constraints on the superfield.
In this section we will illustrate this point with the example of
the vector-tensor multiplet constructed in \cite{Claus3}.

The vector-tensor (VT) multiplet \cite{SSW:VT} (see also the review
\cite{SSW}) has received considerable attention
triggered by the discovery that it describes the dilaton-axion
complex in $\cN = 2$ supersymmetric vacua of heterotic string
theory \cite{deWKLL}. The VT multiplet can be obtained from an
abelian $\cN = 2$ vector multiplet by eliminating the auxiliary field
and dualizing one of the two physical fields in the vector
multiplet to a gauge two-form. The resulting multiplet (once an
auxiliary field is restored) is off-shell in the presence of a central charge.

A number of papers have addressed different variants of the VT
multiplet using superspace techniques.\footnote{The recent papers
\cite{ADS, ADST} use Free Differential Algebra to formulate a general
coupling of VT multiplets to $\cN = 2$ supergravity.} One approach is to
look for consistent deformations of the superfield constraints
and has been pursued by a number of authors \cite{DKT, DK, IS, DIKST, DT}
in flat superspace. A second line of attack is
to construct the superform geometry associated with the VT
multiplet by identifying its one-form and/or two-form
as component projections of superspace forms
\cite{HOW, GHH, BHO}. A third method is simply to work out
its component structure directly. This approach was
exhaustively applied using the superconformal tensor calculus
by Claus {\it et al.} \cite{Claus1, Claus2, Claus3},
where it was shown that there were essentially two
distinct types of VT multiplet in supergravity with
a gauged central charge: the linear VT multiplet and
the nonlinear VT multiplet.\footnote{The linear VT multiplet must
be coupled to (at least) one additional vector multiplet for
consistency. Additional Chern-Simons-type couplings are also
allowed in both cases.}

Recently, in  \cite{KN} a superspace formulation for these
two cases of the VT multiplet has been found in supergravity.
However, it remains unclear how to compare these results to
those given in \cite{Claus3}. Our goal in this section is to
recast the results of \cite{Claus3} in a superspace form
and to compare them to the results of \cite{KN}.

%%%%%%%%%%%%%%%%%%%%%%%%%%%%%%%%%%%%%%%%%%%%%%%%%%%%%%

\subsection{Vector-tensor multiplet in supergravity}

The vector-tensor multiplet consists of five component fields:
a real scalar $\ell$, a Weyl fermion $\lambda_\a^i$, a real scalar
$U$, a gauge one-form $v_m$, and a gauge two-form $b_{mn}$.
The real scalar $U$ can alternatively be understood as the action
of the central charge on the real scalar $\ell$. In the language of
Claus {\it et al.}, $U = \ell^{(z)}$ \cite{Claus1, Claus2, Claus3}.

The most general vector-tensor multiplet considered in \cite{Claus3}
was coupled not only to the central charge vector multiplet $\cZ$ but also
to a number of additional vector multiplets $\cW^A$ with $A = 2, \cdots n$.
These couplings were parametrized by a set of real numbers $\eta_{11}$,
$\eta_{1A}$, and $\eta_{AB}$.
The indices $1$ and $A$ label component one-forms, with the first one-form
associated with the vector-tensor multiplet $L$ and the rest associated
with $\cW^A$. The one-form associated with the central charge multiplet
was denoted $0$. Below we have attempted to match as much as possible
the notation used in these papers, except for the lift to
superspace.\footnote{Claus {\it et al.} denoted $X$ as the lowest component of the
central charge vector multiplet, which we denote by $\cZ$. In
addition, we use a different normalization convention for
the vector multiplet. However, since in the couplings to
the vector-tensor multiplet all vector multiplets appear
in ratios, this normalization will not matter.} In this way, we
will derive constraints for the vector-tensor multiplet in curved
superspace.

To lift the vector-tensor multiplet of \cite{Claus3} to superspace,
we must construct a superfield $L$
which possesses $\ell$ as its lowest component. The superfield $L$ should
share the same properties as its lowest component: this means
$L$ should be real, it should have vanishing dilatation weight,
and it should be conformally primary.

Our first step is a simple one: we analyze $\delta \ell$.
Since $\ell = L\vert$, we know that
\begin{align}
\delta \ell = (\xi^\a_i {\bm\nabla}_\a^i L + \bar\xi_\ad^i \bar{\bm\nabla}^\ad_i L)\vert
	= \xi^\a_i \lambda_\a^i + \bar\xi_\ad^i \bar\lambda^\ad_i \ ,
\end{align}
where the second equality follows from the component
approach. Hence we deduce the obvious definitions
\begin{align}
\lambda_\a^i := {\bm\nabla}_\a^i L \vert~,\qquad
\bar\lambda^\ad_i := \bar{\bm\nabla}^\ad_i L \vert~.
\end{align}
The next step is nontrivial. Beginning with the relation
\begin{align}
\delta \lambda_\a^i = (\delta {\bm\nabla}_\a^i L)\vert =
	(\xi^\b_j {\bm\nabla}_\b^j {\bm\nabla}_\a^i L + \bar\xi_\bd^j \bar{\bm\nabla}^\bd_j {\bm\nabla}_\a^i L)\vert \ ,
\end{align}
we rearrange the right-hand side into irreducible representations
of $\rm SU(2)$ and the Lorentz group. We use
\begin{align}
{\bm\nabla}_\b^j {\bm\nabla}_\a^i L &= \frac{1}{2} \{{\bm\nabla}_\b^j, {\bm\nabla}_\a^i\} L
	+ \frac{1}{2} [{\bm\nabla}_\b^j, {\bm\nabla}_\a^i] L \eol
	&= - \veps_{\b \a} \veps^{j i} \bar \cZ \Delta L
	+ \frac{1}{2} \veps_{\b \a} {\bm\nabla}^{ji} L
	- \frac{1}{2} \veps^{j i} {\bm\nabla}_{\b\a} L
\end{align}
and
\begin{align}
\bar{\bm\nabla}^\bd_j {\bm\nabla}_\a^i L &= \frac{1}{2} \{\bar{\bm\nabla}^\bd_j, {\bm\nabla}_\a^i\} L
	+ \frac{1}{2} [\bar{\bm\nabla}^\bd_j, {\bm\nabla}_\a^i] L  \eol
	&= -\ri \delta^i_j {\bm\nabla}_\a{}^\bd L - \frac{1}{2} [{\bm\nabla}_\a^i, \bar{\bm\nabla}^\bd_j] L \eol
	&= -\ri \delta^i_j {\bm\nabla}_\a{}^\bd L
	- \frac{1}{4} \delta^i_j [{\bm\nabla}_\a^k, \bar{\bm\nabla}^\bd_k] L
	- \frac{1}{2} \veps_{jk} [{\bm\nabla}_\a^{(i}, \bar{\bm\nabla}^{\bd k)}] L~.
\end{align}
This leads to
\begin{align}\label{dlambda}
\delta {\bm\nabla}_\a^i L &=
	- \xi^i_\a \bar \cZ \Delta L
	- \frac{1}{2} \xi_{j \a} {\bm\nabla}^{ji} L
	+ \frac{1}{2} \xi^{i \b} {\bm\nabla}_{\b\a} L
	\eol & \quad
	- \ri \bar \xi^i_\bd  {\bm\nabla}_\a{}^\bd L
	- \frac{1}{4} \bar \xi^i_\bd [{\bm\nabla}_\a^k, \bar{\bm\nabla}^\bd_k] L
	+ \frac{1}{2} \bar \xi_{k \bd} [{\bm\nabla}_\a^{(i}, \bar{\bm\nabla}^{\bd k)}] L~.
\end{align}
Taking the component projection and comparing with the expression
for $\delta \lambda_\a^i$ given in eq. (3.10) of \cite{Claus3}, we can immediately
conclude a number of relations by comparing the coefficients of $\xi$ and $\bar\xi$.
Only two of them will be of interest to us here, since they define
the constraints on the vector-tensor multiplet. These consist of:
\begin{align}\label{VTConstraint1}
{\bm\nabla}_\a^{(i} \bar{\bm\nabla}_\bd^{j)} L = -\bar{\bm\nabla}_\bd^{(j} {\bm\nabla}_\a^{i)} L = 0
\end{align}
and
\begin{subequations}\label{VTConstraint2}
\begin{align}
{\bm\nabla}^{ij} L &= -\frac{2}{\cZ} {\bm\nabla}^{\a(i} L {\bm\nabla}_\a^{j)}\cZ
	+ \frac{1}{\cZ} \frac{1}{2 \eta_{11} L - \textrm{Re g}} \Gamma^{ij}~, \\
\Gamma^{ij} &:= -\eta_{11} L^2 {\bm\nabla}^{ij} \cZ
	- 2\eta_{11} \cZ {\bm\nabla}^{\a(i} L {\bm\nabla}_\a^{j)} L
	+ 2 \eta_{11} \bar \cZ \bar{\bm\nabla}_\ad^{(i} L \bar{\bm\nabla}^{\ad j)} L
	\eol&\quad
% 	- \frac{1}{2} L \bar \cZ (\partial_{\bar I} \bar g) \bar{\bm\nabla}^{ij} \bar \cZ^{I}
	- \frac{1}{2} L \bar \cZ \frac{\pa\bar g}{\pa \bar\cW^A} \bar{\bm\nabla}^{ij} \bar \cW^{A}
	- \frac{1}{2} L \bar \cZ \frac{\pa \bar g}{\partial \bar \cZ} \bar{\bm\nabla}^{ij} \bar \cZ
	+ \cZ {\bm\nabla}^{\a(i} g {\bm\nabla}_\a^{j)} L
	- \bar \cZ \bar{\bm\nabla}_\ad^{(i} \bar g \bar{\bm\nabla}^{\ad j)} L
	\eol&\quad
	+ \ri {\bm\nabla}^{ij} (\cZ b)
	+ \ri \bar {\bm\nabla}^{ij} (\bar \cZ \bar b)~.
\end{align}
\end{subequations}
The following combinations of fields have been used:
\begin{align}
g := \ri \eta_{1 A} \frac{\cW^A}{\cZ}~,\qquad
b := -\frac{\ri}{4} \eta_{AB} \frac{\cW^A \cW^B}{\cZ^2}~.
\end{align}
It was observed in \cite{Claus3} that the constraints (and the action)
are unchanged by certain constant shifts in the parameters $g$ and
$b$. These shifts turn out to coincide with constant shifts in the
real part of $\cW^A / \cZ$.

The two equations \eqref{VTConstraint1} and \eqref{VTConstraint2}
constitute the lift of the VT multiplet to superspace. Just as at
the component level \cite{Claus1, Claus2, Claus3}, different
superspace formulations of the VT multiplet can be recovered by
different choices of the coefficients $\eta$: the nonlinear VT
multiplet ($\eta_{1A}=0$) and the linear
VT multiplet ($\eta_{11}=0$),
both with ($\eta_{AB}\neq 0$) and
without ($\eta_{AB} = 0$) Chern-Simons terms.

\subsection{Reinterpreting the constraints}
Due to the importance of these constraints in describing the
multiplet, we would like to have a better understanding of them.
It turns out that they have an elegant physical interpretation
from dimensional reduction of a 5D theory in the Minkowski limit \cite{KL}. Before giving
the interpretation, let us perform a number of relabellings
which will allow us to dramatically simplify \eqref{VTConstraint2}.

First, let us take the quantity $\cW^A / \cZ$ and consider its imaginary part
\begin{align}
Y^A := \frac{1}{2 \ri} \left(\frac{\cW^A}{\cZ} - \frac{\bar \cW^A}{\bar \cZ} \right)~.
\end{align}
It turns out the constraint \eqref{VTConstraint2} can be rewritten entirely
in terms of the quantity $Y^A$:
\begin{align}
{\bm\nabla}^{ij} L = -\frac{2}{\cZ} {\bm\nabla}^{\a(i} L {\bm\nabla}_\a^{j)}\cZ
	+ \frac{1}{\cZ} \frac{1}{2 \eta_{11} L + \eta_{1A} Y^A} \Gamma^{ij}
\end{align}
with
\begin{align}
\Gamma^{ij} &= -\eta_{11} L^2 {\bm\nabla}^{ij} \cZ
	- 2\eta_{11} \cZ {\bm\nabla}^{\a(i} L {\bm\nabla}_\a^{j)} L
	+ 2 \eta_{11} \bar \cZ \bar{\bm\nabla}_\ad^{(i} L \bar{\bm\nabla}^{\ad j)} L
	\eol&\quad
	- \frac{1}{2} L \bar \cZ \eta_{1A} Y^A {\bm\nabla}^{ij} \cZ
	- \frac{1}{2} L \cZ \eta_{1A} {\bm\nabla}^{ij} Y^A
	+ \frac{1}{2} L \bar \cZ \eta_{1A} \bar{\bm\nabla}^{ij} Y^A
	\eol&\quad
	- L \eta_{1A} {\bm\nabla}^{\a(i} \cZ {\bm\nabla}_\a^{j)} Y^A
	+ L \eta_{1A} \bar{\bm\nabla}_\ad^{(i} \bar \cZ \bar{\bm\nabla}^{\ad j)} Y^A
	\eol&\quad
	- 2 \cZ \eta_{1A} {\bm\nabla}^{\a(i} L {\bm\nabla}_\a^{j)} Y^A
	+ 2 \bar \cZ \eta_{1A} \bar{\bm\nabla}_\ad^{(i} L \bar{\bm\nabla}^{\ad j)} Y^A
	\eol&\quad
	- \cZ \eta_{AB} Y^A {\bm\nabla}^{ij} Y^B
	+ \bar \cZ \eta_{AB} Y^A \bar{\bm\nabla}^{ij} Y^B
	\eol&\quad
	- 2 \eta_{AB} Y^A {\bm\nabla}^{\a(i} \cZ {\bm\nabla}_\a^{j)} Y^B
	+ 2 \eta_{AB} Y^A \bar{\bm\nabla}_\ad^{(i} \bar \cZ \bar{\bm\nabla}^{\ad j)} Y^B
	\eol&\quad
	- 2 \cZ \eta_{AB} {\bm\nabla}^{\a(i} Y^A {\bm\nabla}_\a^{j)} Y^B
	+ 2 \bar \cZ \eta_{AB} \bar{\bm\nabla}_\ad^{(i} Y^A \bar{\bm\nabla}^{\ad j)} Y^B~.
\end{align}

A further dramatic simplification occurs if we consider the real part
of this constraint. One finds the equation
\begin{align}\label{VTConstraint2R}
0 = \cZ {\bm\nabla}^{ij} L + 2 {\bm\nabla}^{\a(i} L {\bm\nabla}_\a^{j)} \cZ
	+ \frac{1}{2} L {\bm\nabla}^{ij} \cZ + \textrm{c.c.} \ ,
\end{align}
which is \emph{completely linear} in $L$.
Together with \eqref{VTConstraint1}, we have two linear constraints in $L$;
the imaginary part of \eqref{VTConstraint2} gives an additional \emph{nonlinear} constraint.

Before discussing the nonlinear constraint, we should make one further observation.
The superfield $Y^A$ was constructed by taking the imaginary part of
the combination $\cW^A / \cZ$. Since $\cW^A$ and $\cZ$ are both vector multiplets,
their respective Bianchi identities should imply some sort of constraint
on $Y^A$. Indeed, it is easy to check that
\begin{align}\label{WConstraint2}
0 = \cZ {\bm\nabla}^{ij} Y^A + 2 {\bm\nabla}^{\a(i} Y^A {\bm\nabla}_\a^{j)} \cZ
	+ \frac{1}{2} Y^A {\bm\nabla}^{ij} \cZ + \textrm{c.c.} \ ,
\end{align}
which is exactly the same form as \eqref{VTConstraint2R}! Furthermore, one can easily
check that
\begin{align}\label{WConstraint1}
{\bm\nabla}_\a^{(i} \bar{\bm\nabla}_\bd^{j)} Y^A = -\bar{\bm\nabla}_\bd^{(j} {\bm\nabla}_\a^{i)}  Y^A  = 0~,
\end{align}
which is the same form as \eqref{VTConstraint1}.
This suggests that we ought to consider the superfield $L$ as $Y^1$ and
consider the set of real superfields $Y^I = (L, Y^A)$ obeying the
linear constraints \eqref{WConstraint2} and \eqref{WConstraint1}.
It must be emphasized that Claus {\it et al.} already observed
at the component level that it is profitable to interpret the one-forms
within $L$ and $\cW^A$ on the same footing.
Our observation lifts their consideration to superspace.

Finally, let us take the imaginary part of the constraint \eqref{VTConstraint2}.
The equation one finds is equivalent to
\begin{align}\label{VTConstraint2I}
0 = \eta_{IJ} G^{IJ \,ij}~,
\end{align}
where the quantity $G^{IJ\, ij}$ is defined by
\begin{align}\label{Gij}
G^{IJ\, ij} &:=
	\ri \cZ {\bm\nabla}^{\a(i} Y^I {\bm\nabla}_\a^{j)} Y^J
	+ \frac{\ri}{4} \cZ Y^I {\bm\nabla}^{ij} Y^J + \frac{\ri}{4} \cZ Y^J {\bm\nabla}^{ij} Y^I
	\eol & \quad
	+ \frac{\ri}{2} Y^I {\bm\nabla}^{\a(i} Y^J {\bm\nabla}_\a^{j)} \cZ
	+ \frac{\ri}{2} Y^J {\bm\nabla}^{\a(i} Y^I {\bm\nabla}_\a^{j)} \cZ
	+ \HC
\end{align}
and the numeric coefficients $\eta_{IJ}$ are given by
\begin{align}
\eta_{IJ} = \left(
\begin{array}{cc}
\eta_{11} & \eta_{1A} \\
0 & \eta_{AB}
\end{array}
\right)~,
\end{align}
with $\eta_{A1}$ defined to vanish.
The equation \eqref{VTConstraint2I} is bilinear in $Y^I = (L, Y^A)$.
The imaginary part of the constraint \eqref{VTConstraint2} arises by solving \eqref{VTConstraint2I} 
for ${\bm\nabla}^{ij} L - \bar {\bm\nabla}^{ij} L$, with non-linearities
arising when $\eta_{11} \neq 0$.

There is a simple 5D interpretation of these results, at least in the
Minkowski limit, due to Kuzenko and Linch \cite{KL}.\footnote{Kuzenko
and Linch \cite{KL} considered the nonlinear case 
with $\eta_{1A} = \eta_{AB}=0$, but the generalization is completely straightforward.}
The nonlinear vector-tensor multiplet in 4D Minkowski space naturally arises
from dimensional reduction of a 5D Chern-Simons theory. Within 5D superspace,
the real 5D vector multiplet $Y^I$ is governed by two Bianchi
identities, which can be written as \eqref{WConstraint2} and \eqref{WConstraint1}.
If one constructs a 5D Chern-Simons action in superspace, its \emph{equation of motion}
matches \eqref{VTConstraint2I} in the Minkowski limit. In dimensionally
reducing this action to 4D, this equation of motion is reinterpreted as a 4D
constraint on $Y^1 = L$. 

A second observation we would like to make is how the above reformulation
clarifies an observation made in \cite{Claus2}. There it was noted that for general
VT multiplets with non-linearities ($\eta_{11} \neq 0$), it was possible
to make certain redefinitions of the components of the VT multiplet so
that $\eta_{1A}$ vanishes. In the reformulation given above, this
property is manifest. First, we observe that since both
the superfields $L$ and $Y^A$ obey the same linear constraints
-- \eqref{VTConstraint1} and \eqref{VTConstraint2R} for $L$
and \eqref{WConstraint1} and \eqref{WConstraint2} for $Y^A$ --
it is possible to \emph{redefine} $L$ by adding to it a certain
linear combination of $Y^A$,
\begin{align}\label{L'}
L = L' + c_A Y^A~,
\end{align}
where $c_A$ is some real constant. Due to the bilinearity of
the remaining constraint \eqref{VTConstraint2I}, one can easily
show that this transformation can be countered by a redefinition
of the parameters $\eta_{IJ}$
\begin{align}\label{eta'}
\eta_{11} = \eta_{11}'~,\qquad
\eta_{1A} = \eta_{1A}' - 2 \eta_{11}' c_A, \qquad
\eta_{AB} = \eta_{AB}' - \eta_{1(A}' c_{B)} + \eta_{11}' c_A c_B
\end{align}
so that the form of the constraint remains unchanged.
In particular, the choice $c_A = - \eta_{1A} / 2\eta_{11}$ will
set $\eta'_{1A} = 0$, as noted in \cite{Claus2}. This property
implies that for the nonlinear VT multiplet, the most general
case can always be mapped to the case $\eta_{1A} = 0$ and
vice-versa by a superfield redefinition.

There is a third, technical observation which we wish to make about the
constraints.  In a series of papers, formulations of the VT multiplet have been worked out 
in the case of conventional $\N =2$ superspace \cite{DT, HOW, GHH, BHO} and $\cN = 2$ 
harmonic superspace \cite{DKT, DK, IS, DIKST}
by making use of consistency conditions. A recent paper \cite{KN}, using a similar 
approach has found a superspace formulation in terms of superfields for the cases of the
linear and non-linear VT multiplets within $\rm SU(2)$ superspace.
One of the consistency conditions has an interesting relationship
to the constraints when written in projective superspace.
Since $L$ is independent of the isotwistors, we have
\be\label{isoDL}
0 = D^{--} L = D^{++} L = D^0 L \ .
\ee
Recalling that
${\bm\nabla}_\a^{\pm} := u^\pm_i {\bm\nabla}_\a^i$ and
$\bar{{\bm\nabla}}_\ad^{\pm} :=  u^\pm_i \bar{{\bm\nabla}}_\ad^i$
in the language of section \ref{ProjectiveComponents}, we may apply successive
gauged covariant derivatives to the first condition of \eqref{isoDL}
to derive a consistency condition:
\begin{align}
0 &= {\bm\nabla}^{\a+} {\bm\nabla}_\a^{+} \bar{{\bm\nabla}}_\ad^{+} \bar{{\bm\nabla}}^{\ad +} D^{--} L \non\\
&= D^{--} {\bm\nabla}^{\a+} {\bm\nabla}_\a^{+} \bar{{\bm\nabla}}_\ad^{+} \bar{{\bm\nabla}}^{\ad +} L 
+ 8 \ri {\bm\nabla}^{\a\ad} {\bm\nabla}_\a^{+} \bar{{\bm\nabla}}^{+}_\ad L 
	\eol & \quad
- 2 {\bm\nabla}^{\a-} {\bm\nabla}_\a^{+} \bar{{\bm\nabla}}_\ad^{+} \bar{{\bm\nabla}}^{\ad+} L
- 2 \bar{{\bm\nabla}}_\ad^{-} \bar{{\bm\nabla}}^{\ad+} {\bm\nabla}^{\a+} {\bm\nabla}_\a^{+} L \non\\
& \quad - 4 \D \Big( ({\bm\nabla}^{+})^2 (\cZ L) + (\bar{{\bm\nabla}}^{+})^2 (\bar{\cZ} L) 
 - \hf L ({\bm\nabla}^{+})^2 \cZ - \hf L (\bar{{\bm\nabla}}^{+})^2 \bar{\cZ} \Big) \ .
\end{align}
This places restrictions on the possible constraints for $L$. For instance, 
if we impose the constraint \eqref{VTConstraint1}, we have
\begin{align}
0 =& \D \Big( ({\bm\nabla}^{+})^2 (\cZ L) + (\bar{{\bm\nabla}}^{+})^2 (\bar{\cZ} L) - L ({\bm\nabla}^{+})^2 \cZ \Big) ~.
\end{align}
This is guaranteed by the (slightly stronger) constraint
\begin{align}
0= {\bm\nabla}^{ij} (\cZ L) + \bar{{\bm\nabla}}^{ij} (\bar{\cZ} L) 
-  L {\bm\nabla}^{ij} \cZ ~,
\end{align}
which is equivalent to \eqref{VTConstraint2R}. It is remarkable that
the second linear constraint on $L$ may be motivated from the first
solely on the basis of consistency.

\subsection{Action formulation for the vector-tensor multiplet}

The action principle for the VT multiplet requires a composite real
linear multiplet $L^{ij}$,
\begin{align}
{\bm\nabla}^{(i}_\a L^{jk)} = \bar{\bm\nabla}^{(i}_\ad L^{jk)} = 0~,
\end{align}
which transforms under the central charge. If this linear multiplet
is coupled to conformal supergravity using the vector multiplet
$\cZ$ that gauges the central charge, an invariant action can
be constructed at the component level (see {\it e.g.} the
discussion in \cite{Claus2, Claus3} or the original literature
\cite{BS, sct_rules, sct_structure}). This action works even if the
linear multiplet is invariant under the central charge, and
coincides exactly in that case with \eqref{PsiWAction} and
\eqref{GVprojFinalResult}, which we derived directly from superspace.
It would be useful to have a direct projective superspace formulation of this
action for non-vanishing central charge, but as yet no such
formulation exists.\footnote{In harmonic superspace,
one can construct the analogue to \eqref{Projective_GV}.
See the discussion in \cite{DIKST} for the globally supersymmetric
case and \cite{KT} for the local case.}

We may nevertheless consider the lift to superspace of the linear multiplet
$L^{ij}$ constructed by Claus {\it et al.} \cite{Claus3} and look for
simplifications inspired by the observations made in the previous
subsection. The linear multiplet $L^{ij}$ was based on an ansatz which
reads, in superfield notation,
\begin{align}
L^{ij} &= \cZ \cA {\bm\nabla}^{\a i} L {\bm\nabla}_\a^j L
	+ \bar \cZ \bar\cA \bar{\bm\nabla}_\ad^i L \bar{\bm\nabla}^{\ad j} L
	+ \cZ \cB_I {\bm\nabla}^{\a(i} X^I {\bm\nabla}_\a^{j)} L 
	+ \bar \cZ \bar\cB_I \bar {\bm\nabla}_\ad^{(i} \bar X^I \bar{\bm\nabla}^{\ad j)} L 
	\eol & \quad
	+ \cC_{IJ} {\bm\nabla}^{\a i} X^I {\bm\nabla}_\a^j X^J
	+ \bar\cC_{IJ} \bar{\bm\nabla}_\ad^i \bar X^I \bar{\bm\nabla}^{\ad j} \bar X^J
	-\frac{1}{2} \cG_I {\bm\nabla}^{ij} X^I~.
\end{align}
The index $I$ runs $0, 2, \cdots, n$ with $X^I = (\cZ, \cW^A)$.
The coefficient functions $\cA, \cB_I, \cC_{IJ}$ and $\cG_{I}$
fell into three distinct cases. The first case was given by
\begin{gather}
\cA = \eta_{11} (L + \ri \zeta) - \frac{1}{2} g~, \quad
\cB_I = -\frac{1}{2} (L + \ri\zeta) \partial_I g - 2 \ri \partial_I b~, \quad
\cC_{IJ} = -\frac{\ri}{2} (L + \ri \zeta) \partial_I \partial_J (Z b)~, \eol
\cG_I = \mathrm{Re\,} \left(
	\left[\frac{1}{3} \eta_{11} (L + \ri \zeta)^3
	- \frac{\ri}{2} \zeta (L + \ri \zeta) g\right] \delta_I{}^0
	+ \frac{1}{2} (L + \ri \zeta) Z \partial_I (g L + 4 \ri b)
	\right)~,
\end{gather}
where the parameter $\zeta$ was defined as
\begin{align}
\zeta = \frac{\mathrm{Im\,} (L g + 4 \ri b)}{2 \eta_{11} L - \mathrm{Re\, } g}~.
\end{align}
This led to general couplings of the VT multiplet to conformal
supergravity and the abelian multiplets $\cW^A$.

The second case turned out to give a total derivative and corresponded to
\begin{align}
\cA &= \ri \eta_{11} \zeta' - \ri \alpha~, \qquad
\cB_I = -\frac{\ri}{2} \zeta' \partial_I g - 2 \ri \partial_I \gamma~,\qquad
\cC_{IJ} = \frac{1}{2} \zeta' \partial_I \partial_J (Z b)~, \eol
\cG_I &= \mathrm{Re\,} \left(2 \ri Z L \partial_I \gamma + \frac{\ri}{2} \zeta' Z L \partial_I g
	- 2 \zeta' \partial_I (Z b) \right)~, \eol
\gamma &= \frac{\ri}{4} \alpha_A \cW^A / Z~,\qquad
\zeta' = \frac{2 \alpha L + 4 \mathrm{Re\, } \gamma}{2 \eta_{11} L - \mathrm{Re\,} g}~,
\end{align}
where the parameters $\alpha$ and $\alpha_A$ were arbitrary real numbers.
This solution can be included into the first solution by the redefinitions
\begin{align}\label{gbShift}
g \rightarrow g + 2\ri \alpha~,\qquad
b \rightarrow b + \gamma~.
\end{align}

The third case was
\begin{align}
\cA &= \cB_I = 0~,\qquad
\cC_{IJ} = -\frac{\ri}{8} \partial_I \partial_J (f(X) / \cZ)~,\qquad
\cG_I = -\frac{1}{2} \mathrm{Im\,} \partial_I (f(X) / \cZ)~,
\end{align}
where $f(X)$ is a holomorphic function of $X^I = (\cZ, \cW^A)$ of degree 2;
this corresponded to an alternative formulation for general
vector multiplet couplings.

Each of these tensor multiplets can be rewritten in a way which makes their
properties somewhat more transparent. Let us begin with the first
case. Making use of the constraint obeyed by $L$, one can rewrite
$L^{ij}$ as
\begin{align}\label{ClausLag1}
L^{ij} = \bm\nabla^{ij} (\cZ \G) + \bar{\bm\nabla}^{ij} (\bar{\cZ} \G)
	- \frac{1}{2} \left(\frac{\cW^A}{\cZ} + \frac{\bar \cW^A}{\bar \cZ}\right)
		\left(\hf \eta_{1A} G^{11\, ij} + 2 \eta_{AB} G^{1B\, ij} \right) \ ,
\end{align}
where $G^{IJ\, ij}$ is given by \eqref{Gij} and we have defined
\begin{align}
\G &= \frac{1}{6} \eta_{11} L^3 + \frac{1}{4} \eta_{1A} Y^A L^2 + \hf \eta_{AB} Y^A Y^B L~.
\end{align}
The remarkable feature of this action is that it is completely
\emph{tri-linear} in the fields $L$, $\cW^A$, and $\bar \cW^A$. Moreover, except
for the single factor $\cW^A/\cZ + \bar \cW^A / \bar\cZ$ appearing in the third term,
the action depends only on the combination $Y^A$.

The second action can be rewritten in a similar way as
\begin{align}\label{ClausLag2}
L^{ij} = -\alpha G^{11\, ij} + \alpha_A G^{1A\, ij}~.
\end{align}
Note that the action is purely \emph{bilinear}
in the fields $L$, $\cW^A$ and $\bar \cW^A$.
It is immediately apparent that a term of the form
\eqref{ClausLag2} arises by making the replacement
$\cW^A \rightarrow \cW^A + m^A \cZ$ for a real constant $m^A$
in \eqref{ClausLag1}, consistent with the shift symmetry
\eqref{gbShift}. From a superspace perspective, it is unclear
why \eqref{ClausLag2} should lead to a total derivative.

Finally, the last case of Lagrangian can be rewritten
\begin{align}
L^{ij} = -\frac{\ri}{8} \nabla^{ij} \left(\frac{f(X)}{\cZ} \right) + \HC
\end{align}
In this case, $L^{ij}$ is invariant under the central charge and
we may utilize the projective superspace action with $\cL^{++} = \cV L^{++}$
(where $\cV$ is the prepotential for the central charge multiplet
$\cZ$) to rewrite the action as \cite{Kuzenko:2008ry}
\begin{align}
S = \frac{\ri}{2} \int \rd^4x\, \rd^4\q\, \cE \, f(X) + \HC
\end{align}

Using these superfield reformulations, it is possible to show that
the symmetry of the constraints generated by the redefinitions
\eqref{L'} and \eqref{eta'} can be lifted to symmetries of the action
itself. Combining the first case \eqref{ClausLag1} and the third case
with $f(X) = f_{ABC} \cW^A \cW^B \cW^C$ for real $f_{ABC}$ leads to
a linear multiplet
\begin{align}
L^{ij} &= \bm\nabla^{ij} (\cZ \G) + \bar{\bm\nabla}^{ij} (\bar{\cZ} \G)
	\eol & \quad
	- \frac{1}{2} \left(\frac{\cW^A}{\cZ} + \frac{\bar \cW^A}{\bar \cZ}\right)
		\left(\hf \eta_{1A} G^{11\, ij} + 2 \eta_{AB} G^{1B\, ij}
	- 3 f_{ABC} G^{BC \, ij}\right)~, \eol
\Gamma &= \frac{1}{6} \eta_{11} L^3 + \frac{1}{4} \eta_{1A} Y^A L^2 + \hf \eta_{AB} Y^A Y^B L
	- \hf f_{ABC} Y^A Y^B Y^C~,
\end{align}
which is symmetric under the shifts \eqref{L'} and \eqref{eta'} along with
\begin{align}\label{f'}
f_{ABC} = f'_{ABC} + \frac{1}{3} \eta_{11}' c_A c_B c_C 
	- \frac{1}{2} \eta_{1 (A}' c_B c_{C)}
	+ \eta_{(AB}' c_{C)}~.
\end{align}
One can straightforwardly prove that $\Gamma$ is invariant
while the the term involving $G^{IJ\, ij}$ is invariant when
one uses the constraint \eqref{VTConstraint2I}.

\subsection{Comparison with existing superspace approaches}

Recently, two superspace formulations of the VT
multiplet, the linear and nonlinear VT multiplet,
were presented in AdS and in supergravity \cite{KN}.
Both of these cases correspond to special choices for the
coefficients $\eta_{IJ}$, which we can now explicitly demonstrate.
The nonlinear VT multiplet obeys the constraint\footnote{We
have rewritten the results of \cite{KN} to use the conformal
superspace geometry of \cite{Butter4D}.}
\begin{align}
{\bm \nabla}^{ij} L =
	- \frac{2}{\cZ} {\bm \nabla}^{\a (i} \cZ {\bm \nabla}_\a^{j)} L
	+ \frac{1}{\cZ L} \left(
	- \cZ {\bm \nabla}^{\a i} L {\bm \nabla}_{\a}^j L
	+ \bar \cZ \bar{\bm \nabla}_\ad^i L \bar{\bm \nabla}^{\ad j} L
	- \frac{L^2}{2} {\bm \nabla}^{ij} \cZ
	\right)
\end{align}
as well as the constraint \eqref{VTConstraint1}. Similarly,
the Lagrangian $L^{ij}$ constructed in \cite{KN} is simply
\begin{align}
L^{ij} = \frac{1}{24} {\bm\nabla}^{ij}(\cZ L^3) + \frac{1}{24} \bar{\bm\nabla}^{ij}(\bar\cZ L^3)~.
\end{align}
It is straightforward to compare these equations to
\eqref{VTConstraint2} and \eqref{ClausLag1} and observe agreement
(up to an overall normalization of the Lagrangian) for the
case $\eta_{1A} = \eta_{AB} = 0$.

Similarly, the linear VT multiplet, which couples not only to
the central charge multiplet $\cZ$ but also an additional
vector multiplet $\cW$, obeys \eqref{VTConstraint1} along with
\begin{align}
{\bm\nabla}^{ij} L &= \frac{2 \bar \cW}{\bar \cZ \cW - \cZ \bar \cW}
	\left({\bm\nabla}^{\a (i} \cZ {\bm\nabla}_\a^{j)} L + \bar{\bm\nabla}_\ad^{(i} \bar \cZ \bar{\bm\nabla}^{\ad j)} L
		+ \frac{1}{2} L {\bm\nabla}^{ij}\cZ \right)
	\eol & \quad
	- \frac{2 \bar \cZ}{\bar \cZ \cW - \cZ \bar \cW}
	\left({\bm\nabla}^{\a (i} \cW {\bm\nabla}_\a^{j)} L + \bar{\bm\nabla}_\ad^{(i} \bar \cW \bar{\bm\nabla}^{\ad j)} L
		+ \frac{1}{2} L {\bm\nabla}^{ij}\cW \right)~.
\end{align}
This constraint matches \eqref{VTConstraint2} for
the case $\eta_{11} = \eta_{AB} = 0$.
Similarly, the Lagrangian $L^{ij}$ for this case, eq. (6.7) in \cite{KN},
can be shown to match \eqref{ClausLag1} up to an overall constant.

It should be mentioned that the nonlinear case with Chern-Simons
terms ({\it i.e.} setting only $\eta_{1A} = 0$), which was also considered
in \cite{KN}, is equivalent to the most general case by a simple superfield
redefinition. Given a VT multiplet with $\eta_{1A}=0$, one can redefine
$L$ by
\begin{align}
L = L' + c_{A} Y^A~.
\end{align}
This leads to a new set of $\eta$ coefficients
\begin{align}
\eta'_{11} = \eta_{11}~, \qquad \eta'_{1A} = 2 \eta_{11} c_A~, \qquad
\eta'_{AB} = \eta_{AB} + \eta_{11} c_A c_B~.
\end{align}
The choice of $c_A$ then sets the new value $\eta'_{1A}$, and
the most general case is recovered.

%%%%%%%%%%%%%%%%%%%%%%%%%%%%%%%%%%%%%%%%%%%%%%%%%%%%%%
%%%%%%%%%%%%%%%%%%%%%%%%%%%%%%%%%%%%%%%%%%%%%%%%%%%%%%

\section{Discussion}

In this paper, we have studied component reductions from superspace
using the newly-constructed conformal superspace \cite{Butter4D}
as well as the conventional formulation based on $\rm SU(2)$
superspace \cite{Grimm}.
In section 3, we demonstrated how to construct
component actions of vector multiplets coupled to supergravity,
both for the superconformal case (which reproduces results originally
constructed using the superconformal tensor calculus) and for the
non-superconformal case. We also briefly discussed the component
construction of tensor multiplet self-couplings.

For these applications, only conventional $\cN=2$ superspaces were used,
but off-shell hypermultiplets (and even the most natural formulation
of tensor multiplets) require projective superspace. For this
reason, we demonstrated in section 4 explicitly how to
perform component reductions in the local projective superspace
formulated in \cite{KLRT-M08}. Our
action was a relatively simple one, but we were able to
demonstrate explicitly how the auxiliary contour integral
selects out the physical degrees of freedom for a simple class
of models involving (possibly composite) vector and tensor multiplets.
To our knowledge, this is the first direct application of the
local projective superspace component reduction rule derived in
\cite{Kuzenko:2008ry}.

It would be interesting to calculate the component action involving general
two-derivative self-couplings of arctic multiplets. It is
believed that the elimination of the infinite number of 
auxiliary fields must lead to a model of on-shell hypermultiplets
coupled to conformal supergravity with a target space geometry of
a hyperk\"ahler cone. However, this has never been demonstrated by
an explicit calculation. In analogy to flat projective superspace,
one might construct an action in $\cN=1$ superspace where the
arctic multiplet is decomposed as an infinite number of $\cN=1$
superfields. Algebraic elimination of the auxiliaries followed
by a duality transformation would then lead to a model involving
purely $\cN=1$ chiral superfields. The resulting $\cN=1$ action
would exhibit one manifest supersymmetry as well as the requisite
target space geometry and could be reduced completely to components.
However, there is currently no way of handling the supergravity
couplings in such an approach. It is possible that this action
could instead be handled by direct reduction to components
(as we have done for a far simpler model). Whether the auxiliary
fields can be eliminated (even formally) in this approach remains
to be seen.

In section \ref{VectorTensorSection}, we considered an inverse procedure. Rather than
deriving component results from superspace, we showed how
superspace results may be obtained from components. Taking a
particularly intricate component multiplet -- the
vector-tensor multiplet constructed by Claus {\it et al.} --
we showed how a superfield formulation can simplify its
constraint structure.\footnote{The authors of \cite{Claus3}
indeed pointed out the ``need for a suitable superspace formulation''
due to the complexity of their component results.} We
found that the nonlinear constraints can be understood
using the three relatively simple relations \eqref{VTConstraint1},
\eqref{VTConstraint2R} and \eqref{VTConstraint2I}. This
last constraint in particular hints that the off-shell VT multiplet
considered by Claus {\it et al.} could have a five-dimensional
origin, as shown in flat superspace by Kuzenko and Linch
for the nonlinear VT multiplet \cite{KL}.

A number of unanswered questions remain regarding the VT multiplet.
The most prevalent question is whether it can be derived from five
dimensions when coupled to conformal supergravity. Recently, a detailed
connection between off-shell conformal supergravity in 5D
and 4D has been developed \cite{BdeWK:4d/5d}. It would be interesting
to see if the VT multiplet can indeed be derived directly
in this formalism.

Secondly, the superspace geometry of the 1-form and 2-form
of the VT multiplet have not yet been constructed in supergravity.
As reviewed earlier, both the vector and tensor multiplets
can be naturally encoded, respectively, in 1-form and
2-form superspace geometries with suitable superspace constraints.
The VT multiplet should also have an elegant
reformulation involving coupled superforms.
In fact, the geometry for the linear VT multiplet in Minkowski
superspace has already been worked out in a number of
references \cite{HOW, GHH, BHO}.
Generalizing these results to supergravity and to the
more general VT multiplet considered here would be of interest.

Another interesting question is whether the off-shell VT multiplet
constructed in \cite{Claus3} can be generalized.
In 2001, a new off-shell formulation of the VT multiplet
in flat superspace was constructed by Theis \cite{Theis2}
(see also \cite{Theis1}).
The novel feature of this formulation is that the VT multiplet \emph{itself} provides
the vector field necessary for gauging the central charge.
It would be interesting to understand if such a VT multiplet
can be formulated in supergravity, and what modifications
are necessary to the component procedure of Claus {\it et al.}
to accommodate this.
On a related note, there exists a geometric approach to the VT
multiplet in supergravity constructed using Free Differential
Algebra \cite{ADS, ADST}, which seems to be quite general. It would be
interesting to understand explicitly how the off-shell formulations
in superspace and in tensor calculus relate to this approach.

Although there is much to be learned about the VT multiplet, there is
one immediate application for this formulation: the construction of
higher derivative actions.
It was demonstrated in \cite{Butter:2010jm} that higher derivative
actions involving vector and tensor multiplets may be constructed
based on the action \eqref{PsiW} by treating one or both of the multiplets
as composite. Building composite vector multiplets out of
composite tensor multiplets naturally leads to a hierarchy of
higher derivative actions.
It was argued in \cite{KN} that this same principle can naturally
be applied to higher derivative actions involving the VT multiplet,
since there is no requirement that the multiplets $\cW^A$ be \emph{fundamental};
we can just as easily consider them to be composite and built out of
other vector multiplets or tensor multiplets along the lines
discussed in \cite{Butter:2010jm}. This quite economically
accommodates higher derivative interactions into the VT Lagrangian.
It would be interesting to see if such interactions can be
found within string calculations which inspired the renewal
of interest in the VT multiplet.

\vspace{1cm}

\noindent
{\bf Acknowledgements:}\\
DB and JN would like to thank Sergei Kuzenko for suggesting the project, for
numerous discussions, and for reading the manuscript; and Richard Grimm for
correspondence. JN would also like
to thank Gabriele Tartaglino-Mazzucchelli for useful discussions.
The work  of DB  is supported by the Australian Research Council
(grant No. DP1096372)  and by a UWA Research Development Award.
The work of JN is supported by an Australian Postgraduate Award.

\appendix

%%%%%%%%%%%%%%%%%%%%%%%%%%%%%%%%%%%%%%%%%%%%%%%%%%%%%%

\section{Notations and conventions} \label{NC}

Our notations and conventions follow mostly those in \cite{Ideas}. We briefly summarize them here.

We use two-component notation where
dotted and undotted spinor indices are raised and lowered by $\ve$ tensors
\be
\psi_\a = \ve_{\a\b} \psi^\b \ , \quad \bar{\chi}^\ad = \ve^{\ad\bd} \bar{\chi}_\bd \ ,
\ee
obeying
\be \ve_{\a\b} = - \ve_{\b\a} \ , \quad \ve_{\a\b} \ve^{\b\g} = \d_\a^\g \ , \quad \ve_{\ad\bd} \ve^{\bd\gd} = \d_\ad^\gd \ , \quad \ve^{12} = 1 \ . \non\\
\ee
Similarly ${\rm SU}(2)$ indices are raised and lowered by $\ve_{ij}$ and $\ve^{ij}$ having the same properties as $\ve_{\a\b}$.
Spinor indices are contracted as
\begin{align}
\psi \chi &:= \psi^\a \chi_\a \ , \quad \bar{\psi} \bar{\chi} = \bar{\psi}_\ad \bar{\chi}^\ad~.
\end{align}
For spinors which are also isospinors, we define
\begin{align}
\psi \chi &= \psi^\a_i \chi_\a^i \ , \quad \bar{\psi} \bar{\chi} = \bar{\psi}_\ad^i \bar{\chi}^\ad_i \ .
\end{align}

The metric is $\eta_{ab} = \textrm{diag}(-1, 1,1,1)$. The sigma matrices are defined as
\be
(\s^a)_{\a\ad} = (1, \vec{\s}) \ , \quad (\tilde{\s}^a)^{\ad\a} = \ve^{\ad\bd} \ve^{\a\b} (\s^a)_{\b\bd} = (1 , - \vec{\s}) 
\ee
and have the properties
\begin{align}
(\s_a)_{\a\bd} (\tilde{\s}_b)^{\bd\b} = - \eta_{ab} \d^\b_\a - 2 (\s_{ab})_\a{}^\b \ , \\
(\tilde{\s}_a)^{\ad\b} (\s_b)_{\b\bd} = - \eta_{ab} \d^\ad_\bd - 2 (\tilde{\s}_{ab})^\ad{}_\bd \ ,
\end{align}
together with the following useful identities
\begin{subequations}
\begin{align*}
(\s^a)_{\a \ad} (\s_a)_{\b \bd} &= - 2 \ve_{\a\b} \ve_{\ad \bd}~, \\
(\s_{ab})_{\a\b} (\s^{ab})_{\g \d} &= -2 \ve_{\g (\a} \ve_{\b) \d}~, \\
(\s_{ab})_{\a\b} (\tilde{\s}^{ab})_{\gd \dd} &= 0~, \\
\tr(\s_{ab} \s_{cd}) &= (\s_{ab})_\a{}^\b (\s_{cd})_\b{}^\a = - \eta_{a[c} \eta_{d] b} - \frac{\ri}{2} \ve_{abcd}~, \\
(\s_{[a})_{\a \bd} (\s_{bc]})_{\g\d} &= \frac{\ri}{3} \ve_{abcd} \ve_{\a (\g} (\s^d)_{\d)\bd}~, \\
\ve^{abcd} \ve_{a' b' c' d'} &= - 4! \d^a_{[a'} \d^b_{b'} \d^c_{c'} \d^d_{d']}~, \\
(\s_{ab} \s_c)_\a{}^\ad &= (\s_{ab})_\a{}^\b (\s_c)_\b{}^\ad = - \eta_{c[a} (\s_{b]})_\a{}^\ad - \frac{\ri}{2} \ve_{abcd} (\s^d)_\a{}^\ad~, \\
\ve_{abcd} (\s^{cd})_{\a \b} &= -2 i (\s_{ab})_{\a\b} \ , \qquad \ve_{abcd} (\tilde{\s}^{cd})_{\ad \bd} = 2 i (\tilde{\s}_{ab})_{\ad\bd} \ .
\end{align*}
\end{subequations}
The antisymmetric tensor is
\be \ve^{0123} = - \ve_{0123} = 1 \ ,
\ee
and (anti-)symmetrization includes a normalization factor, for example
\be V_{[ab]} = \frac{1}{2!} (V_{ab} - V_{ba}) \ , \quad \psi_{(\a\b)} = \frac{1}{2!} (\psi_{\a\b} + \psi_{\b\a}) \ .
\ee
For superform indices, we introduce graded antisymmetrization, {\it e.g.}
\begin{align}
V_{[AB\}} = \frac{1}{2!} (V_{AB} - (-)^{ab} V_{BA})~.
\end{align}
When an index is not included, we separate it with vertical bars, {\it e.g.}
\begin{align}
T_{[AB}{}^D F_{|D|C\}} &= \frac{1}{3!} \Big(T_{AB}{}^D F_{DC}
	- (-)^{ab} T_{BA}{}^D F_{DC}
	+ (-)^{ca+cb} T_{CA}{}^D F_{DB}
	\eol & \quad
	- (-)^{cb} T_{AC}{}^D F_{DB}
	+ (-)^{ab+ac} T_{BC}{}^D F_{DA}
	- (-)^{ab+ac+cb} T_{CB}{}^D F_{DA}\Big)~.
\end{align}

A vector $V_a$ can be rewritten with spinor indices as
\be
V_{\a\bd} = (\s^a)_{\a\bd} V_a \ , \quad V_a = - \hf (\tilde{\s}_a)^{\bd\a} V_{\a\bd} \ .
\ee
A real antisymmetric tensor, $F_{ab} = - F_{ba}$ is converted to spinor indices as
\be
F_{\a\b} = \hf (\s^{ab})_{\a\b} F_{ab} \ , \quad \bar{F}_{\ad\bd} = - \hf (\tilde{\s}^{ab})_{\ad\bd} F_{ab} \ ,\quad
F_{ab} = (\s_{ab})^{\a\b} F_{\a\b} - (\tilde{\s}_{ab})_{\ad\bd} \bar{F}^{\ad\bd} \ .
\ee

%%%%%%%%%%%%%%%%%%%%%%%%%%%%%%%%%%%%%%%%%%%%%%%%%%%%%%

\section{Solution to constraints for the tensor multiplet} \label{TensorBI}

In this section we summarize the analysis for solving the Bianchi
identities associated with the tensor multiplet field strength in
the supergravity formulation of \cite{Butter4D}.

The field strength,
\be
H = \rd B \implies H_{ABC} = 3 (\nabla_{[A} B_{BC\}} - T_{[AB}{}^D B_{|D|C\}}) \ , \label{defH'}
\ee
satisfies the Bianchi identity
\be \nabla_{[A} H_{BCD\}} - \frac{3}{2} T_{[AB}{}^E H_{|E|CD\}} = 0 \ .\ee
We impose the mass dimension-$\frac{3}{2}$ constraints
\be
H_\a^i{}_\b^j{}_\g^k = H^\ad_i{}^\bd_j{}^\gd_k
= H_\a^i{}_\b^j{}^\gd_k = H^\ad_i{}^\bd_j{}_\g^k = 0
\ee
and proceed to analyze the consequences of the Bianchi identities at each mass dimension.

\noindent \emph{Mass dimension-2} \\
At mass dimension 2, we must analyze two independent cases.
Putting $A = {}_\a^i$, $B = {}_\b^j$, $C = {}_\g^k$, $D = {}^\dd_l$ gives
\be \d^i_l H_{\a \dd}{}^j_\b{}^k_\g + \d^j_l H_{\b \dd}{}_\a^i{}_\g^k + \d^k_l H_{\g \dd}{}_\a^i{}_\b^j = 0 \ .
\ee
One can easily check that the only solution to this equation is $H_{\a \ad}{}_\b^i{}_\g^j = 0$. 
Similarly $H_{\a\ad}{}^\bd_i{}^\gd_k = 0$.

Putting $A = {}_\a^i$, $B = {}_\b^j$, $C = {}^\ad_k$, $D = {}^\bd_l$ gives
\be \d^i_j H_{\a\ad}{}_\b^k{}_{\bd l} + \d^k_l H_{\b\bd}{}_\a^i{}_{\ad j} + \d^i_l H_{\a\bd}{}^k_\b{}_{\ad j} + \d^k_j H_{\b\ad}{}^i_\a{}_{\bd l} = 0 \ .
\ee
This is solved by
\be H_{\a\ad}{}^k_\b{}_{\bd l} = -4 \ve_{\a\b} \ve_{\ad\bd} \cG^k{}_l + \d^k_l (\ve_{\ad\bd} H_{\a\b} + \ve_{\a\b} H_{\ad\bd}) \ ,
\ee
where $\cG_{ij}$ is a {\em traceless hermitian} superfield of vanishing $\rm U(1)$ weight and we make the identifications
\be \cG^{i j} = - \frac{1}{16} H^{\a \ad}\,{}_\a^{(i}{}_{\ad}^{j)}
	= - \frac{1}{16} H^{\a \ad}\,{}_\a^{i}{}_{\ad}^{j}\ ,\ \
	H_{\a\b} = -\frac{1}{4} H_{(\a}{}^\gd{}_{\b) i}{}_\gd^{i} \ , \ \ H_{\ad\bd} = -\frac{1}{4} H_{\a(\ad}\,{}^{\a i}{}_{\bd) i} \ .
\ee
Now from the definition of $H$ we see that $H_a{}_\a^i{}^\bd_i$ contains a term proportional to $(\s^b)_\a{}^\bd B_{ab}$ and thus by redefining $B_{ab}$ we can make $H_{\a\b} = 0$ 
(similarly we impose $H_{\ad\bd} = 0$). Hence we end up with the {\em constraints}:
\begin{subequations}
\begin{align}
H_\a^i{}_\b^j{}_\g^k &= H^\ad_i{}^\bd_j{}^\gd_k
= H_\a^i{}_\b^j{}^\gd_k = H^\ad_i{}^\bd_j{}_\g^k = 0 ~, \\
H_a{}_\a^i{}_\b^j &= 0 \ , \ \ H_a{}^\ad_i{}^\bd_j = 0 \ , \quad
H_a{}^i_\a{}^\ad_j = 2 (\s_a)_\a{}^\ad \cG^i{}_j \ .
\end{align}
\end{subequations}

\noindent{\em Mass dimension-$\frac{5}{2}$:} \\
Putting $A = a$, $B = {}_\b^j$, $C = {}_\g^k$ and $D = {}_\d^l$ yields no additional constraints. \\
Putting $A = a$, $B = {}_\b^j$, $C = {}_\g^k$ and $D = {}^\dd_l$ gives
\be (\s_a)_\g{}^\dd \nabla_\b^j \cG^k{}_l + (\s_a)_\b{}^\dd \nabla_\g^k \cG^j{}_l + \ri \d^j_l (\s^b)_\b{}^\dd H_{ab}{}^k_\g + \ri \d^k_l (\s^b)_\g{}^\dd H_{ab}{}^j_\b = 0 \ .
\ee
Converting to spinor indices
\be -2 \ve_{\a \g} \ve_{\ad\bd} \nabla_\b^j \cG^k{}_l - 2 \ve_{\a\b} \ve_{\ad\bd} \nabla_\g^k \cG^j{}_l + \ri \d^j_l H_{\a\ad\b\bd}{}^k_\g + \ri \d^k_l H_{\a\ad\g\bd}{}_\b^j = 0 \ .
\ee
Taking the symmetric part in $j$, $k$ and $l$ gives
\be \nabla^{(i}_\a \cG^{jk)} = 0 \ .
\ee
Alternatively contracting $k$ and $l$ gives
\be - 2 \ve_{\a\b} \ve_{\ad\bd} \nabla_\g^k \cG^j{}_k + \ri H_{\a\ad\b\bd}{}^j_\g + 2 \ri H_{\a\ad\g\bd}{}_\b^j = 0 \ .
\ee
Now taking the symmetric part in $\ad$ and $\bd$ we deduce
$H_{\a(\ad}{}^\b{}_{\bd)\,}{}_\g^j = 0$.
On the other hand contracting  $\ad$ and $\bd$ and taking the symmetric and antisymmetric parts in $\b$ and $\g$ gives
\be H_{\a\ad}{}_{(\b}{}^{\ad}{}_{\g)j} = \frac{4 \ri}{3} \ve_{\a ( \b} \nabla_{\g)}^k \cG_{kj} \ , \quad
H_{\a\ad}{}_{[\b}{}^\ad{}_{\g]j} = 2 \ri \ve_{\b\g} \nabla_\a^k \cG_{kj} \ .
\ee
Then using the fact that $H_{ab}{}_\g^k = H_{[ab]}{}_\g^k$ and the above results we find
\be H_{\a\ad\b\bd}{}_\g^j = \frac{4 \ri }{3} \ve_{\ad\bd} \ve_{\g ( \a} \nabla_{\b )}^k\cG^j{}_k \ , \ \ \nabla^{(i}_\a \cG^{jk)} = 0 \ .
\ee
Similarly we deduce
\be H_{\a\ad\b\bd}{}^\gd_j =  \frac{4 \ri}{3} \ve_{\a\b} \d^\gd_{( \ad} \bar{\nabla}_{\bd ) k} \cG^k{}_j \ , \ \ \bar{\nabla}^{(i}_\ad \cG^{jk)} = 0 \ .
\ee

\noindent{\em Mass dimension-3:} \\
Putting $A = a$, $B = b$, $C = {}_\a^i$, $D = {}^\ad_j$ gives
\begin{align} -2 \ri (\s^c)_\a{}^\ad \d^i_j H_{abc} &= - 4 (\s_{[a})_\a{}^\ad \nabla_{b]} \cG^i{}_j + \frac{2 \ri}{3} (\ts_{ab})^\ad{}_\bd \nabla_\a^i \bar{\nabla}^\bd_k \cG^k{}_j + \frac{2 \ri}{3} (\s_{ab})_\a{}^\b \bar{\nabla}^\ad_j \nabla^k_\b \cG^i{}_k \ .
\end{align}
Now contracting $i$ and $j$ gives
\be - 4 \ri (\s^c)_\a{}^\ad H_{abc} = \frac{2 \ri}{3} (\ts_{ab})^\ad{}_\bd \nabla_\a^j \bar{\nabla}^\bd_k \cG^k{}_j + \frac{2 \ri}{3} (\s_{ab})_\a{}^\b \bar{\nabla}^\ad_k \nabla_\b^j \cG^k{}_j \ .
\ee
Then rearranging and simplifying the above gives
\be H_{abc} = \frac{\ri}{24} \ve_{abcd} (\s^d)^\a{}_\bd [\nabla_\a^j, \bar{\nabla}^\bd_k] \cG^k{}_j \equiv \ve_{abcd} \tilde{H}^d \ .
\ee
 
 All other Bianchi identities yield no further constraints, however the following useful identities can be found:
\begin{align} \nabla_\a^i \tilde{H}^a &= \frac{2}{3} (\s^{ab})_\a{}^\b \nabla_b \nabla_\b^k \cG^i{}_k + \frac{2}{3} (\tilde{\s}^{ab})^\bd{}_\gd T_b{}^i_\a{}^j_\bd \nabla^\gd_k \cG^k{}_j 
+ \ve^{abcd} (\s_b)_\a{}^\bd T_{cd}{}^j_\bd \cG^i{}_j \ , \non\\
\nabla_a \tilde{H}^a &= - \frac{2}{3} T_{\a\b}{}^\a_i \nabla^{\b k} \cG^i{}_k + {\rm c.c.}
\end{align}

Having solved the constraints for the three-form, we now turn to finding a solution for the gauge two-form under some appropriate constraints. We make use of the Yang-Mills constraints
\be B_\a^i{}_\b^j = 4 \ri \ve_{\a\b} \ve^{ij} \bar{\Psi} \ , \quad B^\ad_i{}^\bd_j = 4 \ri \ve^{\ad\bd} \ve_{ij} \Psi \ , \quad B_\a^i{}_\bd^j= 0 \ .
\ee
We can then solve \eqref{defH} for $B$ under these conditions. We again proceed to analyze the consequences by mass dimension.

\noindent {\em Mass dimension-$\frac{3}{2}$:} \\
Letting $A = {}_\a^i$, $B = {}_\b^j$ and $C = {}_\g^k$ gives
\be \nabla_\a^i B_\b^j{}^k_\g + \nabla_\b^j B_\a^i{}_\g^k + \nabla_\g^k B_\a^i{}_\b^j = 0 \ .
\ee
Substituting $B_\a^i{}_\b^j = 4 \ri \ve_{\a\b} \ve^{ij} \bar{\Psi}$ and contracting $\b$ with $\g$ and $j$ with $k$ gives
\be \nabla_\a^i \bar{\Psi} = 0 \ .
\ee
Similarly we also have $\bar{\nabla}^\ad_i \Psi = 0$.

Now letting $A = {}_\a^i$, $B = {}_\b^j$ and $C = {}^\ad_k$ gives
\be \nabla_\a^i B_\b^j{}^\ad_k + \nabla_\b^j B_\a^i{}^\ad_k + \bar{\nabla}^\ad_k B_\a^i{}^j_\b + 2 \ri \d^i_k (\s^a)_\a{}^\ad B_a{}^j_\b + 2 \ri \d^j_k (\s^a)_\b{}^\ad B_a{}^i_\a = 0 \ .
\ee
Substituting our constraints leads to
\be 4 \ri \ve_{\a\b} \ve^{ij} \bar{\nabla}_{\ad k} \bar{\Psi} + 2 \ri \d^i_k B_{\a\ad}{}_\b^j + 2 \ri \d^j_k B_{\b\ad}{}_\a^i = 0 \ .
\ee
Contracting $j$ and $k$
\be 4 \ri \ve_{\a\b} \bar{\nabla}_\ad^i \bar{\Psi} + 2 \ri B_{\a\ad}{}_\b^i + 4 \ri B_{\b\ad}{}_\a^i = 0 \ ,
\ee
and taking the symmetric part in $\a$ and $\b$ gives
$B_{(\a}{}^\ad{}_{\b)i} = 0$.
On the other hand taking the antisymmetric part in $\a$ and $\b$ gives
\be B_{[\a}{}^\ad{}_{\b]i} = 2 \ve_{\a\b} \bar{\nabla}_i^\ad \bar{\Psi} \ .
\ee
Thus we have
\be B_{\a\ad}{}_{\b}^i = 2 \ve_{\a\b} \bar{\nabla}^i_\ad \bar{\Psi} \ , \ \ B_{\a\ad}{}^\bd_i = 2 \d^\bd_\ad \nabla_{\a i} \Psi \ .
\ee

\noindent {\em Mass dimension-2:} \\
Letting  $A = a$, $B = {}_\b^j$ and $C = {}_\g^k$ yields no new information. Setting  $A = a$, $B = {}_\a^i$ and $C = {}^\bd_j$ gives
\be 2 (\s_a)_\a{}^\bd \cG^i{}_j = \nabla_a B_\a^i{}^\bd_j - \nabla_\a^i B_a{}^\bd_j - \bar{\nabla}^\bd_j B_a{}^i_\a - 2 \ri \d^i_j (\s^b)_\a{}^\bd B_{ab} - T_a{}_\a^i{}_\gd^k B^\bd_j{}^\gd_k- T_a{}^\bd_j{}^\g_k B_\a^i{}_\g^k \ .
\ee
Substituting our constraints and torsion and simplifying leads to
\begin{align} 4 \eta_{ab} \cG^{ij} &= \eta_{ab} \nabla^{ij} \Psi + \eta_{ab} \bar{\nabla}^{ij} \bar{\Psi} - \ve^{ij} (\s_{ab})^{\a\b} \nabla_{\a\b} \Psi - \ve^{ij} (\tilde{\s}_{ab})_{\ad\bd} \bar{\nabla}^{\ad\bd} \bar{\Psi} + 4 \ri \ve^{ij} B_{ab} \non\\
& + 4 \ve^{ij} (\s_{ab})^{\a\b} W_{\a\b} \bar{\Psi} + 4 \ve^{ij} (\s_{ab})_{\ad\bd} \bar{W}^{\ad\bd} \Psi \ .
\end{align}
From which it follows that
\begin{align}
\cG^{ij} &= \frac{1}{4} \nabla^{ij}\Psi + \frac{1}{4} \bar{\nabla}^{ij} \bar{\Psi}~, \\
B_{ab} &= - \frac{\ri}{4} (\s_{ab})^{\a\b} (\nabla_{\a\b} \Psi - 4 W_{\a\b} \bar{\Psi}) - \frac{\ri}{4} (\tilde{\s}_{ab})_{\ad\bd} (\bar{\nabla}^{\ad\bd} \bar{\Psi} - 4 \bar{W}^{\ad\bd} \Psi) \ .
\end{align}
All other Bianchi identities are identically satisfied.

%%%%%%%%%%%%%%%%%%%%%%%%%%%%%%%%%%%%%%%%%%%%%%%%%%%%%%

\section{Component results from superspace} \label{CompIdentities}
In this appendix, we briefly describe certain component results in
the conformal superspace formulation of \cite{Butter4D} as well as
the supergravity formulation developed in \cite{KLRT-M08},
which was based on $\rm SU(2)$ superspace \cite{Grimm}.

\subsection{Conformal superspace in components}

The component structure of $\cN=2$ conformal superspace is fully described
in \cite{Butter4D}. We summarize here the results necessary for this
paper, adapted to the notation of \cite{Ideas}.

The superspace derivative $\nabla_a$, when projected to lowest
components, yields a fully supercovariant derivative, which
we can write as
\begin{align}
e_m{}^a \nabla_a \lc &= \partial_m
	- \hf \psi_m{}^\g_k \nabla_\g^k\lc - \hf \bar{\psi}_m{}^\gd_k \bar{\nabla}_\gd^k\lc
	+ \hf \omega_m{}^{bc} M_{bc}+ \phi_m{}^{ij} J_{ij} + \ri A_m Y + b_m \mathbb{D}
	\eol & \quad
	+ \frak{f}_m{}^b K_b + \hf \phi_m{}^i_\a S^\a_i + \hf \phi_m{}_i^\ad \bar{S}_\ad^i~.
\end{align}
Relative to the definitions used in \cite{Butter4D}, we have made
several sign flips in the definitions of the connections to match the
conventions of \cite{Ideas}. The connections
$\omega_m{}^{ab}$, $\phi_m{}^{ij}$, $b_m$, and $\frak{f}_m{}^a$ differ
by a sign from the corresponding objects in \cite{Butter4D}, while
$A_m$ and $\phi_m{}^i_\alpha$ are the same. Similarly, the superfield
$W_{\a \b}$ used here differs by a sign while $\bar W_{\ad \bd}$ is
the same.

It is useful to introduce the partially degauged derivative
\begin{align}
\nabla_a' = e_a{}^m \nabla_m' := e_a{}^m \left(\partial_m + \frac{1}{2} \omega_m{}^{ab} M_{ab}
		+ \phi_m{}^{ij} J_{ij}
		+ \ri A_m Y
		+ b_m \mathbb D\right)
\end{align}
in terms of which the supercovariant derivative can be written
\begin{align}
\nabla_a\lc = \nabla_a'
	- \hf \psi_a{}^\g_k \nabla_\g^k\lc - \hf \bar{\psi}_a{}^\gd_k \bar{\nabla}_\gd^k\lc
	+ \frak{f}_a{}^b K_b + \hf \phi_a{}^i_\a S^\a_i + \hf \phi_a{}_i^\ad \bar{S}_\ad^i~.
\end{align}

The connections $\omega_m{}^{ab}$, $\frak{f}_m{}^a$ and $\phi_m{}^i_\a$ are composite
fields built out of the other components of the Weyl multiplet. The spin connection
is defined in terms of the vierbein and the gravitino in the usual way.\footnote{See
{\it e.g.} \cite{Butter4D} for the relation, which holds here except for an overall
sign flip in the definition of the spin connection.}
The $S$-supersymmetry connection is given in two-component notation by
\begin{align}
\phi_{\beta \dbeta\,}{}_\alpha^j &=
     \frac{\ri}{12} \bar\Psi_\beta{}^\gd{}_{\alpha \gd\,}{}_\dbeta^j
     + \frac{\ri}{6} \bar\Psi_\alpha{}^\gd{}_{\beta \gd\,}{}_\dbeta^j
     + \frac{\ri}{12} \eps_{\beta \alpha} \bar\Psi_\dbeta{}^\g{}_{\g \gd}{}^{\gd j}
     \eol & \quad
     - \frac{1}{6} \bar W_\dbeta{}^\gd \psi_{\beta \gd\,}{}_\alpha^j
     - \frac{1}{3} \bar W_\dbeta{}^\gd \psi_{\alpha \gd}{}_\beta^j
	+ \frac{\ri}{2} \eps_{\beta \alpha} \bar\S_\bd^j~,
	\\
\bphi_{\beta \dbeta \, \dalpha j} &=
     -\frac{\ri}{12}  \Psi_\dbeta{}^\g{}_{\g \dalpha \,\beta j}
     - \frac{\ri}{6} \Psi_\dalpha{}^\g{}_{\g \dbeta \,\beta j}
     - \frac{\ri}{12} \eps_{\dbeta \dalpha} \Psi_\beta{}^\gd{}_{\g \gd\,}{}^\g_j
     \eol & \quad
     + \frac{1}{6} W_\beta{}^\g \bpsi_{\g \dbeta \dalpha j}
     + \frac{1}{3} W_\beta{}^\g \bpsi_{\g \dalpha \dbeta j}
	+ \frac{\ri}{2} \eps_{\dbeta \dalpha} \S_{\beta j}~.
\end{align}

Only the trace of the special conformal connection $\frak{f}_m{}^a$ is required
for our calculations:
\begin{align}
\frak{f}_a{}^a &= e_a{}^m \frak{f}_m{}^a = -D -\frac{1}{12} \poin \cR
     + \frac{1}{24} \eps^{mnpq} (\bar{\psi}_m{}^j \ts_n \nabla'_p \psi_{q j})
     - \frac{1}{24} \eps^{mnpq} (\psi_{m j} \sigma_n \nabla'_p \bar{\psi}_q{}^j)
     \eol & \quad
     + \frac{\ri}{8} (\psi_{a j} \sigma^a \bar \S^j)
     - \frac{\ri}{8} (\bar{\psi}_a{}^j \ts^a \S_j)
     + \frac{1}{12} W^{ab +} (\bar{\psi}_a{}^j \bar{\psi}_{bj})
     - \frac{1}{12} W^{ab-} (\psi_{a j} \psi_b{}^j)~.
\end{align}
Here $\poin\cR = \poin\cR(e, \omega)$ corresponds to the Poincar\'e version of the Lorentz
curvature, constructed in the usual way from the spin connection.

\subsection{$\rm SU(2)$ superspace in components}
Now for $\rm SU(2)$ superspace we introduce
\be
\cD_a' = e_a{}^m \cD_m| \equiv e_a{}^m \big( \partial_m + \hf \omega_m{}^{bc} M_{bc} + \phi_m{}^{ij} J_{ij} \big) \ ,
\ee
with the algebra
\be [\cD_a', \cD_b'] = \cT_{ab}{}^c(x) \cD_c' + \hf R_{ab}{}^{cd}(x) M_{cd} + R_{ab}{}^{ij}(x) J_{ij}\ ,
\ee
where
\begin{align}
\cT_{ab}{}^c &:= e_a{}^m e_b{}^n T_{mn}{}^c| \ ,\\
\cR_{ab}{}^{cd} &:= e_a{}^m e_b{}^n R_{mn}{}^{cd}| \ , \\
\cR_{ab}{}^{ij} &:= e_a{}^m e_b{}^n R_{mn}{}^{ij}| \ .
\end{align}
Now considering the projection of the covariant derivative algebra, $[\cD_m, \cD_n]$ we find the following relations for the 
torsion and curvatures:
\begin{align}
T_{mn}{}^c| &= -2 \cD'_{[m} e_{n]}{}^c = - 2 \partial_{[m} e_{n]}{}^c + 2 \omega_{[mn]}{}^c \ ,\\
T_{mn}{}^\g_k| &= - \cD_{[m}' \psi_{n]}{}^\g_k \ , \\
R_{mn}{}^{cd}| &= 2 \partial_{[m} \omega_{n]}{}^{cd} + 2 \omega_{[m}{}^{cb} \omega_{n]}{}_b{}^d \ , \\
R_{mn}{}^{ij}| &= 2 \partial_{[m} \phi_{n]}{}^{ij} + 2 \phi_{[m}{}^i{}_k \phi_{n]}{}^{k j}~.
\end{align}
The projection of their corresponding covariantized versions can then be found to be
\begin{align}
T_{ab}{}^c| &= 0 = \cT_{ab}{}^c - \ri \psi_{[a}{}^\g_k \psib_{b]}{}^k_\dd (\s^c)_\g{}^\dd \ , \\
T_{ab}{}^\g_k| &= - \Psi_{ab}{}^\g_k + 2 \psi_{[a}{}^\a_i T_{b]}{}^i_\a{}^\g_k| + 2 \psib_{[a}{}^i_\ad T_{b]}{}^\ad_i{}^\g_k| \ , \\
R_{ab}{}^{cd}| &= \cR_{ab}{}^{cd} + \psi_{[a}{}^\g_k R_{b]}{}^k_\g{}^{cd}| + \psib_{[a}{}^k_\gd R_{b]}{}^\gd_k{}^{cd}| + \frac{1}{4} \psi_{[a}{}^\g_k \psi_{b]}{}^\d_l R_\g^k{}^l_\d{}^{cd}|  \eol
&\qquad + \frac{1}{4} \psib_{[a}{}^k_\gd \psib_{b]}{}^l_\dd R^\gd_k{}^\dd_l{}^{cd}| + \hf \psi_{[a}{}^\g_k \psib_{b]}{}^l_\dd R^k_\g{}^\dd_l{}^{cd}| \ , \\
R_{ab}{}^{kl}| &= \cR_{ab}{}^{kl} + \psi_{[a}{}^\g_j R_{b]}{}^j_\g{}^{kl}| + \psib_{[a}{}^j_\gd R_{b]}{}^\gd_k{}^{kl}| + \frac{1}{4} \psi_{[a}{}^\g_i \psi_{b]}{}^\d_j R^i_\g{}^j_\d{}^{kl}|
	\non\\&\qquad
	+ \frac{1}{4} \psib_{[a}{}^i_\gd \psib_{b]}{}^j_\dd R^\gd_i{}^\dd_j{}^{kl}|
	+ \hf \psi_{[a}{}^\g_i \psib_{b]}{}^j_\dd R^i_\g{}^\dd_j{}^{kl}| \ ,
\end{align}
where
\begin{align}
\Psi_{ab}{}^\g_k &:= 2 \cD_{[a}' \psi_{b]}{}^\g_k - \cT_{ab}{}^c \psi_c{}^\g_k \ , \quad \psi_a{}^\g_k := e_a{}^m \psi_m{}^\g_k \ .
\end{align}

From the above results we can derive the following useful relations:
\begin{align}
\cT_{ab}{}^c &= \ri \psi_{[a}{}^\g_k \psib_{b]}{}^k_\dd (\s^c)_\g{}^\dd \ , \\
\cD_\g^k W_{\a\b}| &= - \frac{2}{3} \eps_{\g (\a} \cD_{\b) j} \bar{S}^{jk}| - (\s^{ab})_{\a\b} \psi_{ab}{}^k_\g - 2 \ri \eps_{\g (\a} \psi^a{}^k_{\b )} G_a| \non\\
&\quad
+ 4 \ri \eps_{( \g| (\a} (\s^{ab})_{\b ) | \d )} \psi_a{}^{\d k} G_b| - \ri \eps_{\g (\a} (\s^a)_{\b ) \ad} \psib_a{}^\ad_i \bar{S}^{ik}| + \ri (\s^a)_{(\a}{}^\ad \psib_a{}^k_\ad W_{\b ) \g}| \non\\
&\quad
- \ri \eps_{\g ( \a} (\s^a)_{\b ) \bd} \psib_a{}^k_\ad \bar{Y}^{\ad \bd}| \ , \\
\bar{\cD}^\gd_k Y_{\a\b}| &= (\s^{ab})_{\a\b} \psib_{ab}{}^\gd_k - 2 i (\s^a)_{( \a}{}^{(\ad} (\tilde{\s}^b)^{\gd )}{}_{\b )} \psib_a{}_{\ad k} G_b| \non\\
&\quad
+ \ri (\s^a)_{(\a}{}^{\gd} \psi_a{}^i_{\b)} S_{ik}| + \ri (\s^a)_{(\a \dd} \psi_a{}_{\b ) k} \bar{W}^{\gd \dd}| + \ri (\tilde{\s}^a)^\gd{}_{( \a} \psi_a{}^\g_k Y_{\b) \g}| \ .
\end{align}

\begin{footnotesize}

\end{footnotesize}

\end{document}